%% file: Liu.tex
\input {epsf}
\input {def}

\documentstyle {ucsdthesis}
\begin{document}
\pagenumbering{roman}
\setlength{\baselineskip}{24pt}
\include{int1}
\tableofcontents
\include{int2}
\pagenumbering{arabic}
\setlength{\baselineskip}{24pt}
\setlength{\jot}{24pt}
\setlength{\belowdisplayskip}{12pt}
\setlength{\abovedisplayskip}{12pt}
\setlength{\abovedisplayshortskip}{12pt}
\setlength{\belowdisplayshortskip}{12pt}
\renewcommand{\arraystretch}{2}
\include{c1}
\include{c2}

\include{c3}
\include{c4}
\include{c5}
\include{c6}
\include{c7}
%
%
%
\end{document}

%% file: def.tex
\newcount\dx  \newcount\dy  
\def\dim_in_pt#1{#1 true pt}            
\def\psneg#1{#1 neg}
\def\InsertPlot#1#2#3#4#5{
  \relax\ifdraft
    {\message {  Draft mode only for plot file in #1.ps}
    \vskip 10pt}
  \else {
    \message {  Inserting plot file in #1.ps}
    \dx=#4 \advance\dx by -#2
    \dy=#5 \advance\dy by -#3          
    \centerline{ 
      \hbox to \dim_in_pt{\number\dx}{\hfil
        \vbox to \dim_in_pt{\number\dy}{\vfil
          \special{ps::[asis,begin] 
            0 SPB
            /InsertPlot save def
            Xpos Ypos translate
            \psneg{\number\dx} 0 translate
            \psneg{#2} \psneg{#3} translate
          }
          \special{ps: plotfile #1.ps asis}
          \special{ps::[asis,end]
            InsertPlot restore
            0 SPE
          }
        }
      }        
    }
  }
  \fi
}                     
\def\be{\begin{equation}} 
\def\ee{\end{equation}} 
\def\ba{\begin{eqnarray}} 
\def\ea{\end{eqnarray}}

\def\bk{{\bf k}}

%% file: int1.tex
\begin{titlepage}
\null \vfil
\begin{center}
{UNIVERSITY OF CALIFORNIA, SAN DIEGO}\\
\vspace{28 pt}
 {\large\bf Strongly Interacting Higgs Sector Without Technicolor} \\
\vspace{44 pt}
A dissertation submitted in partial satisfaction of the\\
\vspace{16 pt}
requirements for the degree Doctor of Philosophy\\
\vspace{16 pt}
in Physics\\
\vspace{32 pt}
by\\
\vspace{32 pt}
Chuan Liu\\
\vspace{52 pt}
\end{center}
\addtolength{\baselineskip}{-2 ex}
\begin{tabbing}
Comm\=ittee in charge:\\
\\
\>Professor Julius~G.~Kuti, Chairman\\
\>Professor Aneesh~V.~Manohar\\
\>Professor James~G.~Branson \\
\>Professor Bruce~K.~Driver \\
\>Professor Jeffrey~M.~Rabin\\
\end{tabbing}
\addtolength{\baselineskip}{2 ex}
\vfil
\begin{center}
1994
\end{center}
\end{titlepage}

\newpage
\setcounter{page}{0} 
\null \vfill
\begin{center}
Copyright\\
Chuan Liu, 1994\\
All rights reserved.
\end{center}

\newpage
\setcounter{page}{3}
\addcontentsline{toc}{schapter}{Signature Page}
\thispagestyle{plain}
\null \vfil
\begin{center}
The dissertation of Chuan Liu is approved, and it is\\
acceptable in quality and form for publication on\\
microfilm:\\
\vspace{12 pt}
\rule{4.0 in}{0.010 in}\\
\vspace{12 pt}
\rule{4.0 in}{0.010 in}\\
\vspace{12 pt}
\rule{4.0 in}{0.010 in}\\
\vspace{12 pt}
\rule{4.0 in}{0.010 in}\\
\vspace{12 pt}
\rule{4.0 in}{0.010 in}\\
\vspace{-12 pt} \hspace*{\fill}Chairman\makebox[1.0 in]{ } \\
\vspace{24 pt}
University of California, San Diego\\
\vspace{12 pt}
1994\\
\end{center}
\vfil \null

\newpage
\setcounter{page}{4}
\addcontentsline{toc}{schapter}{Dedication Page}
\thispagestyle{plain}
\null \vfil
\begin{center}
{\it To my wife Dan\\
my parents and grandparents\\
}
\end{center}
\vfil \null

%% file: int2.tex
\chapter*{Acknowledgements}

It has been a very pleasant and fruitful six years for me to stay at
Department of Physics at UCSD, during which many friends helped me in
many ways. Here I want to express my appreciation to them. 

First, I would like to thank my advisor, Professor Julius Kuti, for his
continued energy and enthusiasm 
under all circumstances.
He encourages me to enter the  exciting fields of high energy 
physics and field theory and keeps  
stimulating my interests in the field.
He has given me not only important directions on physics research
but also precious advices on how to be a physicist. 
It was from many intriguing discussions with him
that I gained so much knowledge of physics.
Without his patient guidance, it would be much more 
difficult for me to complete this work 
satisfactorily. 

I would also like to thank the other members of my committee
(Professor Aneesh Manohar, Professor  
James Branson, Professor Bruce Driver and Professor 
Jeffrey M. Rabin) for taking the time to review my work  
and giving me helpful suggestions.
I should also thank the following Professors at UCSD Physics
Department from whom I benefit so much in their lectures:
Professor D. Arovas, Professor P. Diamond, 
Professor R. Dashen, Professor H. Levine and 
Professor H. Paar.

Special thanks go to Dr. Karl Jansen, who helped me so much
in many ways. He has been a collaborator with me for
quite some time  and I have benefited a great deal from
discussions with him. It has always been a great pleasure
for me to work with him.

Bonnie Horstmann and Debra Bomar 
in the Physics Office have helped 
a lot. I would especially thank 
Mary Ann for her help on the administrative works and her patient
reading the thesis and correcting the language mistakes.
 
Many friends have supported and encouraged me throughout this journey,
among them, Y. Shen, L. Lin, X. Che, X. Hong, G. Sun, Z, Guralnik,
M. Schmaltz, C. Chen, X. Zou, W. Zhang and Y. Liang.
 
I would also like to thank my  parents and my grandmother 
for their understanding and  support.  
 
Finally, I would like to thank my wife Dan Wei, not only for her 
moral support , but also for her patience of  
reading the 
thesis and even debugging the programs. Without her continued   
encouragement and help, it would have been very difficult for me
to accomplish this goal successfully.
\chapter*{Vita}
\vfil
\begin{tabbing}
1111111111111111111\=   \kill
 
14 February 1966 \> Born, Beijing, P. R. China\\
\\
1988     	 \> B.S.~in Physics,\\ 
		 \> Peking University, Beijing, P. R. China\\
\\
1988-1994        \> Research/Teaching Assistant,\\
                 \> University of California, San Diego\\
\\
1991    	 \> M.S.~in Physics,\\ 
		 \> University of California, San Diego\\
\\
1994 		 \> Ph.D.~in Physics,\\
                 \> University of California, San Diego\\
\end{tabbing}

\chapter*{Publications}
\begin{enumerate}
\item K. Jansen, J. Kuti and C. Liu, Phys. Lett. B 309, (1993) 119. 
\item K. Jansen, J. Kuti and C. Liu, Phys. Lett. B 309, (1993) 127. 
\item K. Jansen, J. Kuti and C. Liu, 
Nucl. Phys. B 30 (Proc. Suppl.),(1993) 681. 
\item C. Liu, K. Jansen and J. Kuti,
Nucl. Phys. B 34 (Proc. Suppl.),(1994) 635. 
\end{enumerate}

\begin{abstract}
\begin{center}
\vspace{12 pt}
{\large\bf Strongly Interacting Higgs Sector Without Technicolor  
}\\
\vspace{24 pt}
by\\
\vspace{24 pt}
Chuan Liu\\
Doctor of Philosophy in Physics\\
University of California, San Diego, 1994\\
Professor Julius Kuti, Chairman\\
\end{center}
\vspace{24 pt}
 \addtolength{\baselineskip}{2 ex}

The theoretical  framework 
for the higher derivative $O(N)$ scalar field
theory is established and the theory  is 
shown to be finite and unitary
with the indefinite metric quantization. 
It has been shown that if the ghost states  are
represented by a complex conjugate pair, the theory 
is free of  any logical inconsistencies and the ghost pair can
easily evade the experimental  tests. 

With an underlying  hypercubic lattice structure,
the higher derivative $O(4)$ model is
studied nonperturbatively in computer simulations. 
The Higgs mass bound problem is also studied within 
the framework  of higher derivative theory. A much
higher Higgs mass value in the TeV range  is found 
with the ghost pair well-hidden in the multi-TeV range.
Therefore, the higher  derivative $O(4)$ model can
incorporate a strongly interacting Higgs sector  
without introducing
more complicated structures, like technicolor, which
was impossible   for the conventional lattice scalar model. 
This means that, although the added   higher derivative term 
is a higher dimensional operator,
it changes   the fundamental features (metric, energy spectrum, 
strength of interaction, etc.) of the theory so
much that we can no longer  view it as an irrelevant operator
in the Lagrangian. Moreover,   due to the strong interaction
of the theory, it would be impossible to   meaningfully define
the scaling violation in the higher derivative $O(4)$ model.
This implies that  we will not be able to set up
the Higgs mass bound in this theory unless  a
new nonperturbative   interpretation of the
Higgs mass bound is    developed.

\end{abstract}

%% file: c1.tex
\chapter{Introduction}
\label{ch:INTR}

\section{The Higgs Sector of the Minimal Standard Model}

The Standard Model  was first
introduced in the late 1960's to unify the electromagnetic and
weak interactions \cite{wein1,salam1,glash1}. The symmetry group of
the Standard Model is 
$SU(2)_{L}\times U(1)_{Y}$. 
The Minimal Standard Model corresponds to taking only
one Higgs doublet in the basic representation of $SU(2)$.  
The action of the Standard Model consists of several sectors
which are coupled together.
The action  of the Higgs sector for the theory can be written
as: 
\be 
S_{H}= \int d^4x \left\{ {1\over2} (D_{\mu}\phi)^{\dagger}
                                   (D^{\mu}\phi) - V(\phi) \right\},
\ee 
where $\phi$ is a $SU(2)$-doublet  Higgs field, 
\be
\phi(x)= \left( \begin{array}{ll}
                \phi_1(x)+i \phi_2(x) \\
                \phi_3(x)+i \phi_4(x) 
                \end{array} \right) ,
\ee 
and the potential can be written as:
\be
V(\phi(x))=-{1\over2}m^2 \phi(x)^{\dagger}\phi(x)
           + \lambda [ \phi(x)^{\dagger}\phi(x) ]^2  .
\ee
In the limit of small gauge coupling and Yukawa
coupling, the Higgs sector decouples from the rest and becomes a 
$\phi^4$ type scalar field theory with a global symmetry $O(4)$. 
This limit is also referred to as the $O(4)$ limit of the
Minimal Standard Model.
In the Standard Model, the symmetry $SU(2)_L\times U(1)_Y$ is
spontaneously broken to $U(1)_{em}$, which, in the $O(4)$ limit, 
corresponds to the symmetry breaking $O(4)\rightarrow O(3)$. 
In this limit, the  Higgs mass and the vacuum expectation value
$v$ are related by
\be
m_H=\sqrt{8\lambda} v ,
\ee
where $\lambda$ is the renormalized coupling constant.  
The experimental value for $v$  is fixed  to be $v\approx 250$ GeV.
Therefore, the ratio $m_H/v$ also  characterizes
the strength of the quartic self-interaction.

The $O(4)$ limit of the Standard Model is a very interesting 
limit to study for two  reasons. First of all, the
$SU(2)$ gauge coupling is found to be very small,     
$g^2 \approx 0.4$. Therefore, the effects of the gauge fields
on the Higgs sector is perturbative. 
Although the mass of the top quark remains to be
determined, it is unlikely that the top mass will be much higher than
$200$ GeV. All the quark masses are 
rather light compared with the weak scale, therefore the effects of
the fermion sector can also be evaluated within perturbation theory.
In other words, the symmetry breaking mechanism is almost completely
determined by the Higgs sector alone, plus some perturbative corrections.
Secondly, on the phenomenology side, people have shown a so-called
Equivalence Theorem \cite{quigg1} which says: when the center of mass energy
$\sqrt{s}$ is much higher than the $W$ boson mass, the scattering
amplitude of the $W$ bosons in the full Standard Model is equal to
the scattering amplitude of the corresponding channel in the
$O(4)$ model with ${\cal O}(m_W/\sqrt{s})$ corrections.
One of the  methods in the Higgs search experiment is utilizing
the $WW$ or $ZZ$ boson scattering channel.
Therefore, if we consider the energy range for the Higgs   
search, assuming that  the Higgs mass
is above the vev scale or even in the TeV range, the $O(4)$ limit
would be a good approximation for the $WW$ scattering in the
Standard Model. 
Thus, we conclude that the $O(4)$ limit of the
Standard Model would  be a very good laboratory for the investigation
of the symmetry breaking mechanism and the mass of the Higgs particle. 

\section{ The Triviality Higgs Mass Bound }
     
Despite the successes of the Standard Model, two types of
particle that are important in this model remain unconvinced by the 
experiments, namely, the top quark and the Higgs particle. The missing
of the Higgs particle is very problematic because the Higgs plays such an
important role in the spontaneous symmetry breaking which   
gives rise to all
the masses of gauge bosons and fermions. 
In the past decade, there have been many efforts to put an upper bound 
on the Higgs mass. The early works utilized the tree level unitarity 
and the unitarity bound was found to be around $1$ TeV  \cite{quigg1}.
Later, it was then realized that the $O(N)$ scalar field theory is a trivial
field theory and this implies an upper bound on the Higgs mass 
\cite{maiani1,dash1}. 
 
The triviality picture of field theory
was first encountered by Landau et. al. 
long ago when studying the renormalization properties of Quantum
Electrodynamics (QED) \cite{landau1}. 
They discovered that, if the
cutoff was brought to infinity in QED, the renormalized coupling constant
of QED (the electric charge)
 was driven to zero logarithmically. Therefore, in order to
have an interacting theory, a large but finite cutoff had to remain
in the theory. Thus, QED has a built-in
cutoff parameter.
This implies that 
every quantity  calculated in QED depends on this arbitrary
cutoff parameter.  As the energy scale  gets closer and
closer to the cutoff, there is  more  dependence on this
arbitrary cutoff parameter. It seems then that we will
lose the predictability of the theory.
In fact, this is not a problem at all for QED. The built-in
cutoff scale, also know as the Landau ghost scale, is enormous
(typically $\Lambda \sim 10^{137}$ MeV)  for  
QED and therefore the dependence of the physical quantities on this cutoff
scale is negligible. Furthermore, before this energy scale is reached, 
new physics (weak interactions, strong interactions) 
will set in and QED must be modified.
However, one thing becomes clear from the above discussion, namely,
we cannot calculate to arbitrary accuracy in  a trivial field theory
due to the existence of the arbitrary cutoff parameter. 

The triviality scenario of the Higgs sector
is quite similar to that of QED, except that
in the  Higgs sector we do not
know the mass of the Higgs and the coupling constant.
Therefore, the built-in
cutoff for the Higgs sector could be as low as a few TeV, or as
high as the Planck scale, depending on the value of the Higgs mass.
Also, we do not know the nature of the new physics that lies between
the built-in cutoff and the weak scale, if there is any.

The triviality of the Higgs sector can be easily seen 
in either perturbation theory or
in the $1/N$ expansion \cite{einh1} of the model.  Extensive
nonperturbative studies have also been performed on this 
model with a lattice regulator \cite{kuti1,lusc1,hase1}.
All nonperturbative simulation results suggest that the 
triviality scenario found in perturbation theory 
is a feature of the full theory. In these studies, the upper 
bound of the Higgs particle was found to be about
$640$ GeV  under some well defined conditions which we now come to.  

With the lattice regulator, the theory is made finite and the momentum
cutoff is given by $\Lambda=\pi/a$. The continuum limit is achieved by
taking  $\Lambda \rightarrow \infty$, or equivalently, taking the
lattice correlation length $\xi \rightarrow \infty$. For very large
$\Lambda/m_H$, triviality of the theory forces the renormalized 
coupling constant $\lambda_R$ to go to zero logarithmically. Since 
the vacuum expectation value $v$ is fixed in physical units, this
would mean the Higgs mass is  also going to zero in this limit 
like $m_H \sim (\log(\Lambda/m_H))^{-1/2}$. Making a larger
Higgs mass is therefore equivalent to bringing down the cutoff $\Lambda$
relative to the Higgs mass. Of course, this will generate larger cutoff
dependent terms (scaling violation) in the physical scattering cross section. 
In the case of the lattice cutoff, the scaling violation is represented
by the violation of Euclidean invariance.
The old triviality Higgs mass bound was
obtained by  demanding that in a  Goldstone scattering process 
(which is equivalent to $WW$ scattering in the Standard Model
according to the Equivalence Theorem), 
there was not more than a few percent Euclidean invariance violation in the 
scattering cross section \cite{lusc1}. 
It is evident from the above discussion
 that two things are crucial to set up the triviality
mass bound of the Higgs particle. First, one has to know what 
the scaling violation will be when a certain type of regulator is introduced.
Second, one has to have a well-defined method to calculate this
scaling violation for a given set of parameters. 
In the case of the lattice Higgs bound study, the scaling violation
is the Euclidean invariance violation and the method to calculate 
it is perturbation theory.  Perturbation theory is a valid approach
for the Higgs sector, 
because for all Higgs mass values below the bound,
the coupling is weak enough for meaningful perturbative expansion.

The old triviality Higgs mass
bound was  rather low because even at the upper
bound value the renormalized coupling constant of the theory remains
perturbative. In terms of the Higgs mass over
vev ratio, $R$ is only about 3. 
Further increase to the Higgs mass  results in 
a scattering amplitude with large lattice effects and can no longer represent
the low energy continuum theory.  Therefore, 
if the hypercubic lattice will not be the new
physics, then the existence of
a strongly interacting Higgs sector is excluded in a lattice regulated
scalar field theory. 
There has been great concern that this finding was
an artifact of the lattice regulator itself which breaks Euclidean
invariance.  This concern is reasonable if we consider   
the analogue in QCD. We know that the linear sigma model,
which is nothing but the $O(4)$ model in the broken phase, 
will generate the right physics 
of QCD at low energies (low energy theorems, PCAC, etc.).
However, the corresponding ratio $m_{\sigma}/f_{\pi} \sim 7$
is much higher than in the Higgs case. 
Based on this analogy, technicolor models have been introduced which offers
a possibility of strongly interacting Higgs sector.
Due to the strong interacting nature of the technicolor at low
energies, perturbation theory breaks down. Most of the analytic
calculations are therefore performed using the effective chiral
Lagrangian methods. A complete nonperturbative simulation of the
technicolor theory including the dynamical fermions is very costly.
Therefore, it would be nice to have a scalar model that can incorporate
a strongly interacting Higgs sector.
People have tried to perform the lattice calculation
with better Euclidean invariance for the scalar models.
The first significant increase of the Higgs mass
bound ($750$ GeV) was reported 
\cite{zimm1} within the Symanzik improvement program
on a hypercubic lattice structure \cite{syman1}. Similar results on
different lattice structures, with higher dimensional lattice operators
in the interaction term, have also been  reported \cite{neub1}.     

In this thesis, I will study 
the scalar sector of the Minimal Standard Model and the Higgs
mass problem by adding a higher derivative term in the kinetic
energy of the Higgs Lagrangian. With the higher derivative term,
we have a finite $O(N)$ scalar field theory interacting via a 
quartic coupling constant.     
 
\section{Higher Derivative Field Theory and Indefinite Metric 
         Quantization} 
   
There have been serious concerns about the potential difficulties
in higher derivative field theories 
\cite{ostro1,podo1,pais1,lee1,polk1,ghost1,gross1,simon1,slav1,hawk1,stelle1,tomb1}. 
I will briefly mention some of these difficulties in
this section, and the detailed
study will be  the subject of the subsequent chapters. 

First of all, as we will see in Chapter~(\ref{ch:QUAN}), the conventional
quantization procedure does not offer a meaningful theory because
the spectrum is neither bounded below nor above. So, finding new ways of 
quantizing the higher derivative theory is necessary. One of the 
choices is the indefinite metric quantization \cite{pauli1,simon1,ghost1}. By 
doing this the theory has a unique vacuum but, in the meantime, the
positivity of the norm in the Hilbert space is lost. 
Therefore, one has
to identify a subspace in the full Hilbert space as the physical
space and maintain all the physical principles. 

Unitarity is one of
these principles that one would like to maintain because this is
at the heart of any quantum theory for which Born's probability
description still applies. Before any  meaningfully
interpretation of  negative probability is found,
unitarity should be preserved in any physical theory.  This is
 a big challenge for the higher derivative theories simply 
because the full Hilbert space is not positively normed, and negative
normed states, also called ghost states, may violate unitarity.  
This is one of the main reasons why many people have 
abandoned the higher derivative field theories. However, 
I will demonstrate that, there could be a scenario in which the ghost particles
are represented as a complex conjugate pair, 
and unitarity is maintained \cite{lee1,ghost1,scat1}.
This possibility was first pointed out by T. D. Lee in the late
sixties. There have been a lot of discussions on this issue and
it still remains quite controversial.

Causality is another principle of the physical theory. As has been
pointed out earlier by Lee \cite{lee1}, with the complex conjugate ghost pair,
only microscopic causality is violated, and macroscopically it is 
very difficult to detect in the experiments (see Chapter~(\ref{ch:NUNI}) for
further information). 

\section{ Higgs Mass Problem in Higher Derivative Scalar Field Theory}

It is very interesting to study the Higgs mass bound problem in
this higher derivative scalar field theory. 
There have always been several ways of viewing this theory.
The first and most conventional 
way is to view it as  the Pauli-Villars regulated Higgs
theory \cite{pv1,ghost1,hhiggs1,dallas1}. 
The second  is to view  it as a stand-alone, finite,
well-defined theory with ghosts.
The third is to view it as some truncated expansion of the
effective low energy theory after the degrees of freedom  representing
the new physics have been integrated out.  
The original full theory probably has no ghost states, but after the
truncation of the full series, the model may contain
ghost excitations.
Obviously, the distinction between the second and the third view
is ambiguous since we do not know what the full theory should be.
The first view, however, should be taken very carefully. 
Strictly speaking, this view is only valid in the limit of
small $m/M$ ratio, where $m$ is the Higgs mass and $M$ is the
Pauli-Villars mass parameter. If the mass of the Higgs is getting
close to the Pauli-Villars mass parameter, we have to take the 
second view and treat the theory as a finite theory with ghosts.
In the limit of $m/M \rightarrow 0$, 
this finite theory  coincides with
the conventional $O(N)$ scalar field theory with a Pauli-Villars
cutoff. When the Higgs mass scale is comparable with the ghost
parameter, the higher derivative field theory becomes a theory with
complicated particle contents.  

To study the Higgs mass bound problem in the higher derivative
$O(N)$ model, we have to answer the same two questions. First, what
is the scaling violation; second, how does one calculate it?

The answer to the first question is not easy in the case of the
higher derivative theory. Naively thinking, one would expect there should
be some ghost effects. However, despite the negative metric ghost
states in the theory, it remains unitary and the scattering cross
section of ordinary particles looks perfectly normal (see
Chapter~(\ref{ch:NUNI}) for details). 
The only unusual effect 
found for the higher derivative $O(N)$ model is the
violation of microscopic causality. As has been mentioned above, 
this type of acausal effect is extremely difficult to detect. 
That is to say, introducing the higher derivative terms to the
theory makes the theory finite,  only at the cost of violating
microscopic causality, which is invisible for practical reasons.
One might still worry that, in this theory, all the results will depend on
the ghost mass parameter and this is some sort of scaling violation.
This leads us to the second fundamental
question of the problem, namely, how to calculate the scaling
violations.

Obviously, if the Higgs particle remains light and the theory
is still in the perturbative regime, we can do the perturbative
calculation and find out how the scattering amplitude depends on
the new parameter $M$. Whether to call it the scaling violation
is still a question. It is some deviation from the Minimal Standard
Model in the perturbative range. 
However, if the Higgs is heavy and the
interaction is getting stronger, we will not be able to find out
the scaling violation simply because we have nothing to compare with. 
In a strong interacting theory, we have no idea what the universal 
scattering amplitude will look like. In fact, we do not know how to
define such a quantity meaningfully. 
A new nonperturbative 
interpretation of the Higgs mass bound therefore becomes necessary.

From the above discussion, we can see that there are several
major differences between the higher derivative theory and the
conventional theory on the lattice in regards to the Higgs
mass bound problem. First, the scaling violation in the conventional
theory with the lattice regulator is unambiguously defined, both
perturbatively and nonperturbatively. Even without the help of the
perturbation theory, we can quantify the violation of the Euclidean
invariance meaningfully \cite{lang1}. In the higher derivative case,
however, the scaling violation is not well-defined, at least not
nonperturbatively. One can try to search the $M$ dependence of the
theory only in perturbation theory. 
 
Although the higher derivative theory is a finite theory, it still
has infinite degrees of freedom. In order to carry out
a nonperturbative simulation of the model, one must make the number
of degree of freedom finite. This can be done by introducing an
underlying hypercubic lattice structure to the model. The lattice 
spacing $a$ introduces a new short distance energy scale with
the associated lattice momentum cutoff $\Lambda=\pi/a$. In order to
recover the higher derivative theory in the continuum, one would have
to work towards the limit $\Lambda/M \rightarrow \infty$ with a fixed
ratio of $M/m_H$. In so doing, the higher derivative $O(N)$ model
has the same scaling violation as the conventional model, that is,
it violates Euclidean invariance. In the lattice higher derivative
model, in order to recover the corresponding continuum model, one
only has to eliminate the scaling violation that is associated with
the lattice. 
For the higher derivative model on the lattice, one can view it
as the conventional model on the lattice plus some so-called
higher dimensional (or irrelevant) operators. 

Recently, Neuberger et. al. \cite{neub1} 
reported a new Higgs mass bound based
on the systematic search in all the dimension $6$ operators added 
to the conventional Higgs model on the $F_4$ lattice. 
 Based on their study, they claim that
the triviality Higgs mass bound is
$m_H=710 \pm 40$ GeV, and this bound value is
universal in the sense that no other higher dimensional
operators will change it. However, our model discussed herein
contradicts their  conclusion. Our model can be viewed as
the conventional model plus one dimension $8$ operator, which
is supposed to be irrelevant according to their study. However, from
all our simulation results, we can easily drive the Higgs mass
value into the TeV range (see Chapter~(\ref{ch:SIMU}) 
for more details). We believe that the so-called ``irrelevant
operators'' are not irrelevant at all, at least not for the Higgs 
mass bound problem. After all, by adding new dimension $6$ irrelevant
operators, Neuberger et. al. have found a rather different bound.
Therefore, the notion of irrelevant operators is a very misleading
one as far as the Higgs mass bound problem is concerned.
As we discussed above, in our model, 
it is impossible to set up a precise Higgs mass bound  
due to the strong interaction. However, the model is
 able to accommodate a Higgs particle which is heavier than the
old Higgs mass bounds with no lattice scaling violations.

My thesis is organized as follows: in Chapter 2 , 
the quantization of the higher derivative theory is
established using indefinite metric quantization.
In Chapter 3, the higher derivative $O(N)$ model is 
studied within the framework of $1/N$ expansion and 
the important issue of
unitarity and causality are also discussed
. In Chapter 4, the lattice version of the 
higher derivative field theory is presented and the possibility 
of nonperturbative studies using Monte Carlo simulation  is discussed,
and the symmetry breaking mechanism in the finite
volume is studied within the Born-Oppenheimer approximation.
In Chapter 5,
numerical results of the simulation are presented and analyzed.
These simulation results demonstrate 
that the interaction of  the higher derivative
scalar field theory is much stronger than the conventional
scalar field theory. Therefore, a heavy Higgs particle in the
TeV range becomes a real possibility in the theory. 
Chapter 6  discusses  the extraction of the resonance
parameters of the unstable Higgs particle in the finite volume
using  finite size techniques. This method, first suggested by
L\"uscher \cite{luscf1,luscwolf1}, has  proved to work very well for the
conventional $O(N)$ model \cite{zimmph11,zimmph21}. 
We demonstrate that this  also
works in the higher derivative $O(N)$ model after 
appropriate adjustments. In fact, we believe this is the
only sensible way to extract the mass parameter in the simulation
of a strongly interacting theory.

\vfill\eject

%% file: c2.tex
%
    \newcommand{\appi}{a_{i, \bf p}^{(+)}}
    \newcommand{\apmi}{a_{i, \bf p}^{(-)}}
    \newcommand{\aabar}{ \overline{a_{0, \bf p}^{(-)}} }
    \newcommand{\bbbar}{ \overline{a_{1, \bf p}^{(-)}} }
    \newcommand{\ccbar}{ \overline{a_{2, \bf p}^{(-)}} }
    \newcommand{\aam}{a_{0, \bf p}^{(-)}}
    \newcommand{\aap}{a_{0, \bf p}^{(+)}}
    \newcommand{\bbm}{a_{1, \bf p}^{(-)}}
    \newcommand{\bbp}{a_{1, \bf p}^{(+)}}
    \newcommand{\ccm}{a_{2, \bf p}^{(-)}}
    \newcommand{\ccp}{a_{2, \bf p}^{(+)}}
    \newcommand{\ompi}{\omega _{i, \bf p}}
    \newcommand{\cm}{{\cal M}}
    \newcommand{\cmb}{\overline{{\cal M}}}
    \newcommand{\bp}{{\bf p}} 
    \newcommand{\bq}{{\bf q}} 
    \newcommand{\bei}{{\bf e_{\rm i}}} 
    \newcommand{\bx}{{\bf x}} 
    \newcommand{\by}{{\bf y}} 
    \newcommand{\bM}{{\bf M}} 
    \newcommand{\phial}{|\phi_{\alpha}\rangle}
    \newcommand{\psialp}{|\psi_{\alpha}^{(+)}\rangle}
    \newcommand{\psialm}{|\psi_{\alpha}^{(-)}\rangle}
    \newcommand{\psialpm}{|\psi_{\alpha}^{(\pm)}\rangle}
    \newcommand{\gzpmal}{{1 \over E_{\alpha}-H_0 \pm i \epsilon}}
    \newcommand{\gpmal}{{1 \over E_{\alpha}-H \pm i \epsilon}}
    \newcommand{\gzpmbe}{{1 \over E_{\beta}-H_0 \pm i \epsilon}}
    \newcommand{\gpmbe}{{1 \over E_{\beta}-H \pm i \epsilon}}
    \newcommand{\gzpal}{{1 \over E_{\alpha}-H_0 + i \epsilon}}
    \newcommand{\gpal}{{1 \over E_{\alpha}-H + i \epsilon}}
    \newcommand{\gzpbe}{{1 \over E_{\beta}-H_0 + i \epsilon}}
    \newcommand{\gpbe}{{1 \over E_{\beta}-H + i \epsilon}}
    \newcommand{\fmab}{ {\cal M}_{\alpha\beta} } 
\chapter{Higher Derivative Field Theories 
         and Indefinite Metric Quantization}
\label{ch:QUAN}

\section{Higher Derivative Oscillator}
\label{sec:hdos}

\subsection{Classical Hamiltonian} 

Many important features of higher derivative field theories can
be illustrated by their simple quantum mechanical counterparts. As   
an example, let us first  study a higher derivative oscillator 
\cite{simon2} 
given by the following Lagrangian
\be 
L = {1\over2}(1+2{m^2 \over M^2}\cos2\Theta)
      \dot{x}^2 
      -({\cos2\Theta \over M^2}+ {m^2 \over 2M^4})\ddot{x}^2  
      + {1\over 2M^4}{\stackrel{\cdots}{x}}^2
      -{m^2 \over 2}x^2.  
\ee 
This Lagrangian describes a simple harmonic oscillator of
frequency $m$ with second and third derivative terms added.
For simple interpretation of the spectrum, the coefficients of
the  derivative terms are given in terms of $M$ and $\Theta$;
the only restrictions imposed are $m/M <1 $ and $0< \Theta <\pi/2$.
With the higher derivative terms added,
this Lagrangian produces new features that are not present in the conventional
theory. Classically, one can look at the time evolution of the
position $x(t)$ which is a solution of the 
corresponding Euler-Lagrange equation  
\be
(1+2{m^2 \over M^2}\cos2\Theta) {d^2 x \over dt^2}
+({2 \cos2\Theta \over M^2}+ {m^2 \over M^4}) {d^4 x \over dt^4}
+ M^{-4} {d^6 x \over dt^6} +m^2 x = 0 .
\ee 
Some of the new features of the higher derivative theory
already appear at the classical level. For example, in order
to specify the solution, one has to know more initial conditions
than in the usual theory. In this particular example, one needs to
know $x^{(n)}(0),n=0,1,\cdots,5$ to specify a unique solution,
where $x^{(n)}$ denotes the $n$-th time derivative of 
the variable $x$.
This in fact tells us that the higher derivative theory has
more degrees of freedom than the conventional theory.
Another new feature is that
there are runaway solutions to this classical
equation of motion \cite{podo2}. 
The Hamiltonian of a higher derivative Lagrangian 
was  worked out long time ago by
Ostrogradsky \cite{ostro2}. In the Hamiltonian formalism, new degrees
of freedom show up explicitly due to the higher derivative terms. In this 
particular example, there are three independent coordinates and their
corresponding conjugate momenta,  given by 
\ba
q_1&=&x,\;\;\;\;\; q_2=\dot{x}, \;\;\;\;\; q_3=\ddot{x} , \nonumber\\
p_1&=&{1\over2}(1+2{m^2 \over M^2}\cos2\Theta)\dot{x}
      +({\cos2\Theta \over M^2}
      + {m^2 \over 2M^4}) \stackrel{\cdots}{x}
      +{1 \over 2M^4} \stackrel{\cdots\cdot\cdot}{x}, \nonumber\\
p_2&=&
      -({\cos2\Theta \over M^2}+ {m^2 \over 2M^2})\ddot{x} 
      -{1 \over 2M^4} \stackrel{\cdots\cdot}{x}, \\
p_3&=&{1 \over 2M^4} \stackrel{\cdots}{x}. \nonumber
\ea 
Notice that $p_1$ is not proportional to $\dot{x}$ any more.  
Instead, both $\dot{x}$ and $\ddot{x}$ become independent variables.
In terms of these variables the Hamiltonian reads
\be
\label{eq:ch2.hamosci}
H = p_1q_2+p_2q_3+{M^4 \over 2}{p_3}^2
      -{1\over2}(1+2{m^2 \over M^2}\cos2\Theta){q_2}^2
      +({\cos2\Theta \over M^2}+ {m^2 \over 2M^2}) {q_3}^2
         +{m^2 \over 2} {q_1}^2 . 
\ee 
The classical equation of motion can be written out in the
Hamiltonian form 
\be
{d \over dt} q_i =  {\partial H \over \partial p_i},
\;\;\;\;\;\;\;
{d \over dt} p_i = -{\partial H \over \partial q_i},
\ee
where $i$ runs from $1$ to $3$. It is easy to verify that
the Hamilton equations of motion are equivalent to the Euler-Lagrange form, 
 once we have expressed everything in terms of $q_1(t) \equiv x(t)$ 
and its time derivatives. 
Note , however, that this Hamiltonian does not look like the conventional
Hamiltonian at all. The limit of small $m/M$ is a singular limit, and
we will not be able to recover the standard oscillator Hamiltonian by
taking this limit.

\subsection{Quantization} 

Let us now try to quantize this Hamiltonian with the
conventional canonical method. We will treat $q_1$, $q_2$ and 
$q_3$ as independent variables and they have the usual
commutator with the corresponding momenta
\be
[q_i,p_j]=i\delta_{ij} \;\;\; i,j=1,2,3 ,
\ee
with other commutators vanishing. This already 
says something unusual about this quantum theory, namely that   
the position of a particle and its velocity 
are {\em independent} variables and can be measured 
simultaneously; while in conventional quantum mechanics
they form a conjugate pair and cannot be measured simultaneously.
In the higher derivative theory, it is the quantity $p_1$ that
cannot be simultaneously measured with $q_1$. 
From the expression of $p_1$, 
it implies that the  
measurement of $x$ together with $\stackrel{\cdots\cdot\cdot}{x}$, 
$\dot{x}$ together with $\stackrel{\cdots\cdot}{x}$ and 
$\ddot{x}$ together with $\stackrel{\cdots}{x}$ are impossible. 
    
It is not very easy to see that 
the quadratic Hamiltonian in
Equation~(\ref{eq:ch2.hamosci}) still
represents the oscillator spectrum. 
In fact, using a linear transformation, the quadratic part
of the Hamiltonian can be diagonalized exactly 
\be
H_0 = (a^{\dagger}a+{1\over2})m 
     -(b^{\dagger}b+{1\over2}){\cal M} 
     +(c^{\dagger}c+{1\over2}) \overline{{\cal M}} ,
\ee
with ${\cal M}=Me^{i\Theta}$ and $\overline{{\cal M}}
=Me^{-i\Theta}$. The creation and
annihilation operators appeared in the
above equation are linear combinations
of $q_i$ and $p_i$ and satisfy the 
following standard commutation relations:
\be
[a,a^{\dagger}]=[b,b^{\dagger}]=[c,c^{\dagger}]= 1 .
\ee
The other commutators all vanish.
This type of spectrum has many problems \cite{simon2}.
It is not bounded below, not even the real part.
Therefore, no ground state exists in this theory. This unboundedness is
a very common feature to all higher derivative quantum theories.
It is a direct reflection of the the ``wrong sign'' in front of
one of the quadratic terms in the Hamiltonian. 
One way of dealing with these problems is to try another 
quantization procedure and this is where indefinite metric 
quantization \cite{pais2,lee2,pauli2} comes in.

The idea of using negative metric in the quantization procedure 
was introduced long ago, especially for the 
quantization of gauge fields \cite{pauli2}.
In this framework, the full Hilbert space is too large
for physical interests. It contains negative normed states which are
necessary for the consistent quantization.
The negative normed states must  be removed from  the 
physical subspace to maintain the probability interpretation
of the theory. We will apply the same idea here \cite{ghost2}.

First,  notice that by appropriate scaling: $q_1 \rightarrow 
\rho q_1$ and $q_2 \rightarrow q_2/\rho$, with  
$\rho^2=1+2m^2\cos2\Theta/M^2$ and by the  change
$(p_2,q_2) \rightarrow (-q_2,p_2)$,  we can rewrite the 
Hamiltonian into the following form
\be
H=p_1p_2-{p_2^2 \over 2}+{p_3^2 \over 2}
  -{M^2 \over \rho}q_2q_3 +{1\over2}(m^2+2M^2\cos2\Theta)q_3^2
    +{1 \over 2}m^2 \rho^2 q_1^2 .
\ee
Now make the substitution
\be 
p_2  \rightarrow +ip_2, \;\;\;\; 
q_2  \rightarrow -iq_2  .
\ee
This will not change the commutator of $q_2$ and $p_2$ and we may
write the Hamiltonian as
\be 
H={1\over2}P_1^2+{1\over2}P_2^2+{1\over2}P_3^2 
    +{1\over2}Q^{T}{\bf M}Q ,
\label{eq:ch2.hammat} 
\ee 
where the P's and Q's are related to original variables 
by the following table
\ba
P_1=p_1, \;\;\; P_2=p_2+ip_1, &&  P_3=p_3, \nonumber\\
Q_1=q_1-iq_2, \;\;\; Q_2=q_2, && Q_3=q_3 .
\ea
We have used the matrix notation  $Q$ and $Q^{T}$ and 
the mass matrix ${\bf M}$ is 
\be
{\bf M}= \left( \begin{array}{ccc}
           m^2 \rho^2  &  im^2 \rho^2 & 0  \\ 
           im^2 \rho^2 &  -m^2 \rho^2 & i{M^2 \over \rho}  \\ 
           0  &  i{M^2 \over \rho} & m^2+2M^2\cos2\Theta   
           \end{array} \right) .
\ee 
Negative metric quantization corresponds to 
demanding that the $p$'s and $q$'s 
are hermitian, so the Hamiltonian~(\ref{eq:ch2.hammat}) 
itself is not hermitian. Rather, it is self-adjoint with respect
to a metric operator $\eta$ satisfying
\ba
\eta H^{\dagger} \eta &=& H ,  \nonumber \\   
\eta q_2 \eta = -q_2, && 
\eta p_2 \eta = -p_2,  \\
\eta^2=1 , && \eta=\eta^{\dagger} . \nonumber  
\ea 

In this indefinite Hilbert space, the inner product of any two
states, $|\psi\rangle$ and $|\phi \rangle$, is defined to be
$\langle \psi|\eta |\phi \rangle$. It is easy to show that
the expectation value of any self-adjoint operator is real  
in any states.
 Therefore, the expectation value of the
Hamiltonian in any state is real, although the eigenvalues 
of the Hamiltonian may be complex.     
This immediately implies that the complex energy
eigenstates have zero norm.
The dynamics of any state 
vector are still governed by the Schr\"odinger equation
\be
i{\partial \over \partial t}|\Psi(t) \rangle
= H |\Psi(t)\rangle .
\ee
It is easy to show that the norm of a state is still preserved 
under time evolution. 

\subsection{Diagonalization}

We can now perform  transformation  of the variables 
$Q$ and $P$ according to a ``rotation'' $A$ 
\ba
\tilde{Q}=AQ, && \tilde{P}=AP, \nonumber \\
A^{T}A=AA^{T}&=& 1 ,
\ea
and diagonalize the mass matrix ${\bf M}$. The eigenvalues of this matrix
are simply given by $m^2$, ${\cal M}^2 =M^2e^{2i\Theta}$  
and ${\overline{\cal M}}^2=M^2e^{-2i\Theta}$.
 This is why we chose complicated parametrization of the Lagrangian.  
Therefore, we can define the creation and annihilation operators as
\ba
a^{(\pm)} &=& {1\over\sqrt{2}}(\sqrt{m}\tilde{Q_1}
              \mp {i \tilde{P_1} \over \sqrt{m}} ), \nonumber \\
b^{(\pm)} &=& {1\over\sqrt{2}}(\sqrt{{\cal M}}\tilde{Q_2}
              \mp {i \tilde{P_2} \over \sqrt{{\cal M}}}) ,\\  
c^{(\pm)} &=& {1\over\sqrt{2}}(\sqrt{\overline{{\cal M}}}\tilde{Q_3}
              \mp {i \tilde{P_3} \over \sqrt{\overline{{\cal M}}}}  ) .
              \nonumber 
\ea 
These operators satisfy the standard commutation relation
\be
[a^{(-)},a^{(+)}]=[b^{(-)},b^{(+)}]= [c^{(-)},c^{(+)}]=1 ,
\ee
and the Hamiltonian has the standard oscillator form 
\be
H=(a^{(+)}a^{(-)}+{1\over2})m 
 +(b^{(+)}b^{(-)}+{1\over2}) {\cal M} 
 +(c^{(+)}c^{(-)}+{1\over2}) {\overline{{\cal M}}} .
\ee
The ground state is defined to be the state which is simultaneously
annihilated by $a^{(-)}$, $b^{(-)}$ and $c^{(-)}$. 
We assume that the
ground state is positively normed to $1$. Negative metric is seen from
the adjoint relations among the creation and annihilation operators 
\be
\!\!\!\!\!\!\!\eta {a^{(-)}}^{\dagger} \eta = a^{(+)}, \;\;\;
\eta {b^{(-)}}^{\dagger} \eta = c^{(+)}, \;\;\;
\eta {c^{(-)}}^{\dagger} \eta = b^{(+)}.
\ee 
We can then build up our full Hilbert space by applying the various
creation operators to the ground state. The eigenvalues of the 
Hamiltonian can, in general, be complex if the complex ghost pair
is not evenly excited. All the eigenstates with
complex energy have zero norms. This is a common feature for all
self-adjoint Hamiltonians. The excited states are constructed and 
normalized according to
\ba
|n_a,n_b,n_c\rangle &=& {(a^{(+)})^{n_a} \over \sqrt{ n_a!}}
                        {(b^{(+)})^{n_b} \over \sqrt{ n_b!}}
        {(c^{(+)})^{n_c} \over \sqrt{ n_c!}}
         |0,0,0\rangle , \nonumber \\
\langle n_a',n_b',n_c'|\eta |n_a,n_b,n_c\rangle &=& \delta_{n_a,n_a'}
                              \delta_{n_b,n_c'}
                               \delta_{n_c,n_b'} .
\ea

\subsection{Ground State Wave Function} 

\begin{figure}[htb]
\vspace{12mm}
\centerline{ \epsfysize=3.0cm 
             \epsfxsize=5.0cm \epsfbox{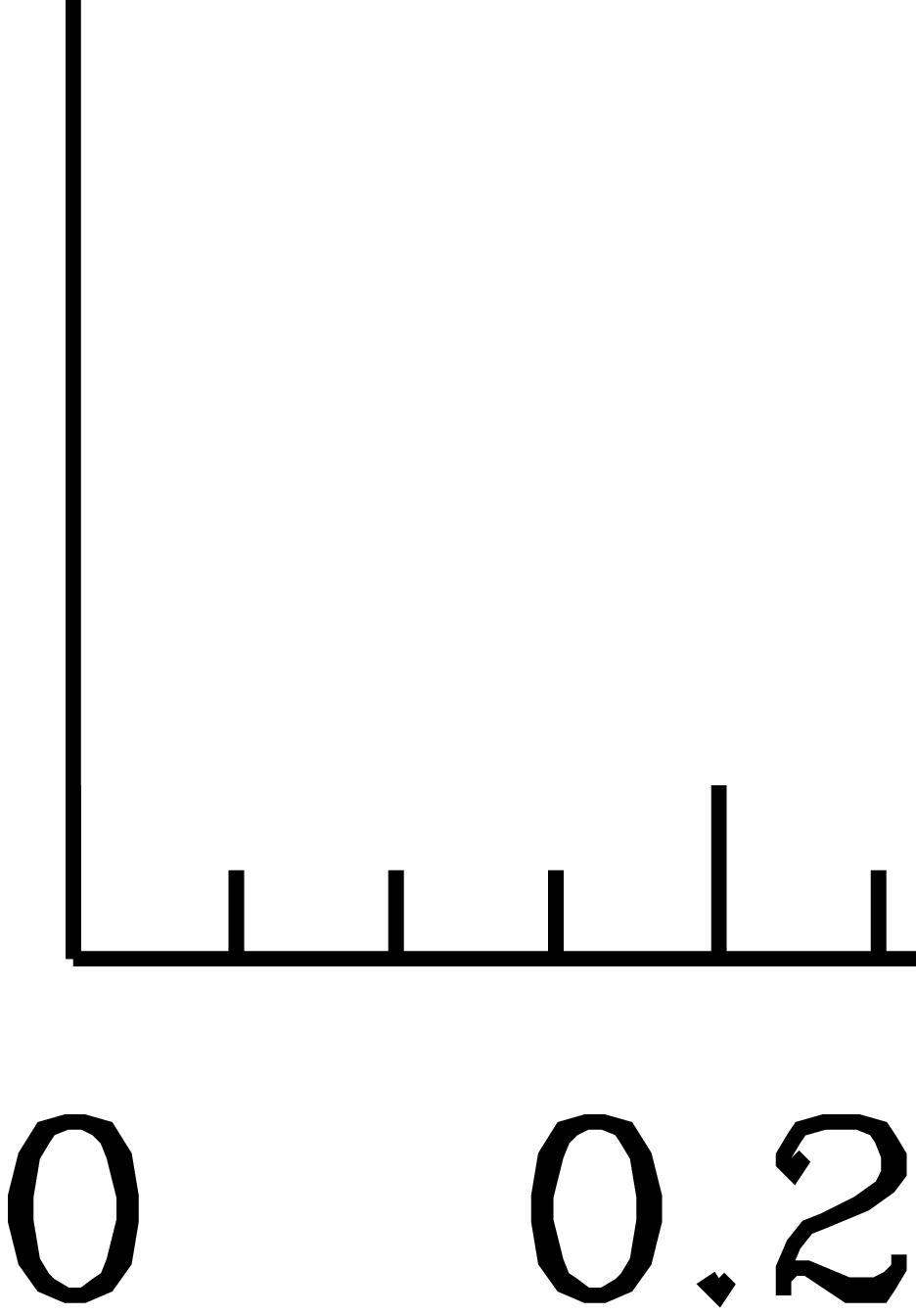} }
\vspace{12mm}
\caption{ The expression 
$f_0(x,\Theta)f_0((1/x),\Theta)-f_1(x,\Theta)^2$,  
as given in the above equations  
for various values of $x$ is plotted versus the variable
$2\Theta /\pi$. This combination is always positive for 
a given value of $x$ as long as $\Theta$ is less than some 
critical value $\Theta_c(x)$. }
\label{fig:ch2.theta}
\end{figure}

\begin{figure}[htb]
\vspace{12mm}
\centerline{ \epsfysize=3.0cm 
             \epsfxsize=5.0cm \epsfbox{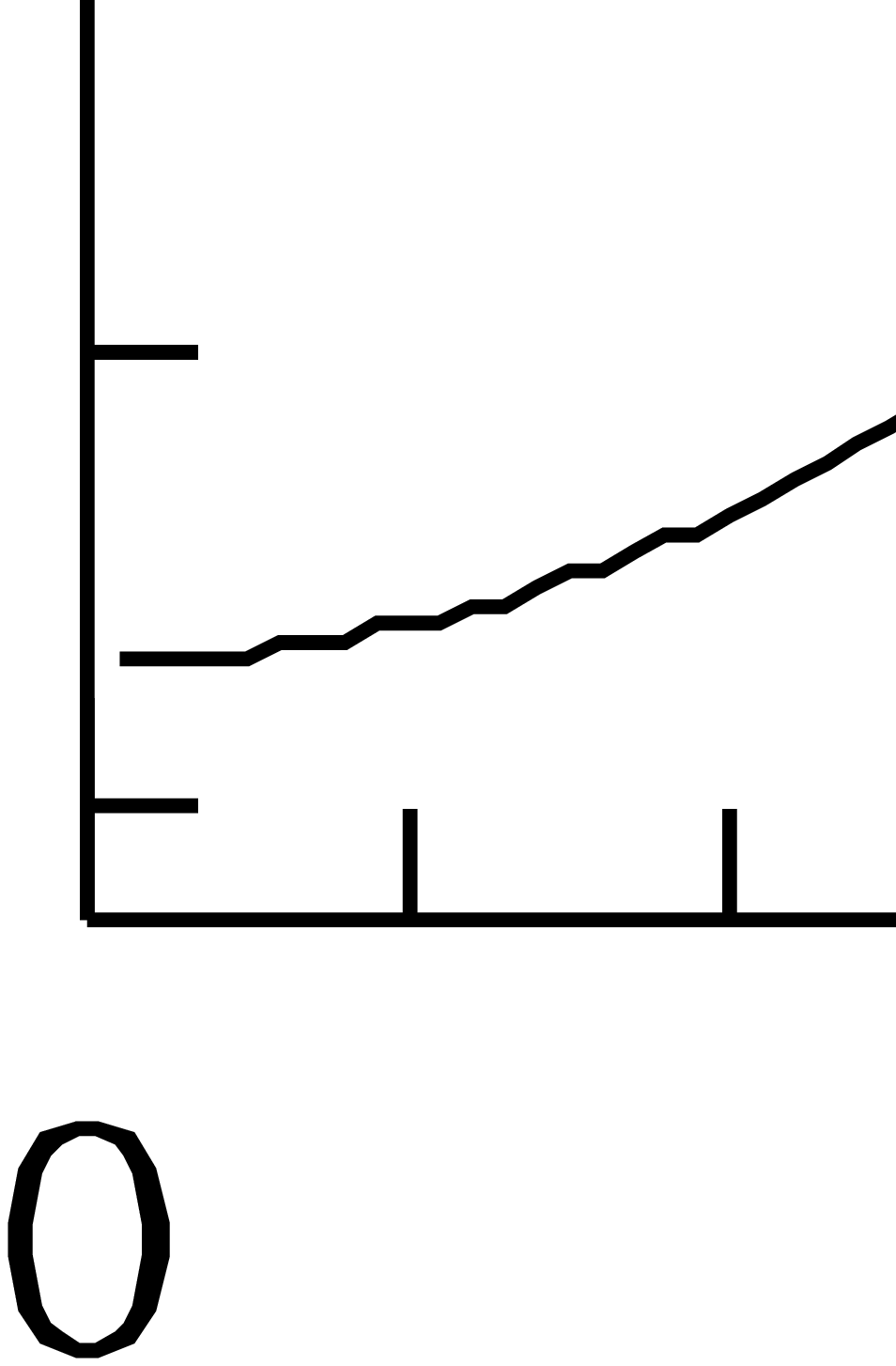}}
\vspace{16mm}
\caption{ The critical value $\Theta_c(x)$ is plotted as
a function of $x=m/M$. All the parameter pairs $(x,\Theta)$
below this curve will ensure the normalizability of the
ground state wave function. }
\label{fig:ch2.thetac}
\end{figure}

We can work out the coordinate space wavefunction for the ground
state  by substituting the old variables. We get
\ba   
\Psi (q_1,q_2,q_3)&=&N_{000} \exp \left(
             -{m \over 2} { 1-{m^3 \over M^3}
                            {\sin5\Theta \over \sin2\Theta}
                           +{m^5 \over M^5}
                            {\sin3\Theta \over \sin2\Theta}
                         \over
                        { 1-2{m^2 \over M^2} \cos2\Theta
                           +{m^4 \over M^4} } } q_1^2  \right. \nonumber \\
             &-&{m \over 2} { {({m \over M}+{M \over m})
                            {\sin\Theta \over \sin2\Theta}
                           -1} 
                         \over
                        { 1-2{m^2 \over M^2} \cos2\Theta
                           +{m^4 \over M^4} } } q_2^2  \nonumber \\
             &-&{m \over 2}({m^4 \over M^4}) 
                           { 1-{M^3 \over m^3}
                            {\sin5\Theta \over \sin2\Theta}
                           +{M^5 \over m^5}
                            {\sin3\Theta \over \sin2\Theta}
                         \over
                        { 1-2{m^2 \over M^2} \cos2\Theta
                           +{m^4 \over M^4} } } q_3^2  \nonumber \\
             &+&{m} 
                           { 1-{m \over M}
                            {\sin3\Theta \over \sin2\Theta}
                           -{m^3 \over M^3}
                            {\sin\Theta \over \sin2\Theta}
                         \over
                        { 1-2{m^2 \over M^2} \cos2\Theta
                           +{m^4 \over M^4} } } (iq_1q_2)  \nonumber \\
             &-&{m^3 \over M^3} 
                           { 1+{M^3 \over m^3}
                            {\sin\Theta \over \sin2\Theta}
                           -{M \over m}
                            {\sin3\Theta \over \sin2\Theta}
                         \over
                        { 1-2{m^2 \over M^2} \cos2\Theta
                           +{m^4 \over M^4} } } (iq_2q_3)  \nonumber \\
             &+& \left. {m^3 \over M^3} { { 1-({m \over M}+{M \over m})
                            {\sin\Theta \over \sin2\Theta}
                           } 
                         \over
                        { 1-2{m^2 \over M^2} \cos2\Theta
                           +{m^4 \over M^4} } } (q_1q_3) \right)  . 
\ea 
In order for the ground state to be normalizable, some constraints must
be put on the parameters $M/m$ and $\Theta$. 
First of all, the normalization condition is somewhat different in the
case of indefinite metric quantization. The condition is
\be
\langle 0|\eta |0 \rangle \equiv
 \langle 0|\eta |q_1,q_2,q_3 \rangle \langle q_1,q_2,q_3|0\rangle =1 ,
\ee 
where we have omitted the sum (integration) over the $q_i$'s.
The ground state wave function given above is just
$\langle q_1,q_2,q_3|\eta |0 \rangle$. Therefore, due to the
existence of $\eta$ which flips the sign of $q_2$, the normalization 
condition for the ground state wave function is written as
\be
\int dq_1dq_2dq_3 \Psi^{*}(q_1,-q_2,q_3)\Psi(q_1,q_2,q_3)=1 .
\ee
Now we can write down the sufficient condition for
this Gaussian type integral to converge.
Since the quantity 
$1-2(m^2/M^2)\cos 2\Theta + m^4/M^4$
is always positive, the condition for normalizability 
reduces to the following
\ba
\label{eq:normal}
f_0(m/M,\Theta)>0,  \;\;\;\;\;\;\; f_0(M/m,\Theta)>0 ,  
\nonumber \\
f_1(m/M,\Theta)>0, \;\;\;\;\;\;\; 
f_0(m/M,\Theta) f_0(M/m,\Theta)- f_1(m/M,\Theta)^2 >0 ,
\nonumber \\
f_0(x,\Theta)=1-x^3{\sin 5\Theta \over \sin 2\Theta} 
 +x^5{\sin 3\Theta \over \sin 2\Theta} ,
\nonumber \\
f_1(x,\Theta)=(x+{1 \over x}){\sin \Theta \over \sin 2\Theta} -1 .
\ea
The condition $f_1(m/M, \Theta)>0$ is equivalent to the
condition $0< \Theta <\pi/2$. In order to fulfill the other
conditions the parameter pair $(m/M, \Theta)$ has to be
in some range. In Figure~(\ref{fig:ch2.theta}),
the function 
$f_0(x,\Theta)f_0((1/x),\Theta)-f_1(x,\Theta)^2$ is plotted
as a function of $\Theta$ for some values of $x=m/M$. Since this
combination is symmetric with respect to the change $x\rightarrow
(1/x)$, it is sufficient to study 
the behavior in the parameter range $0<x<1$.
It is seen from this figure that for any value of the parameter $x$, there
exists a critical value $\Theta_c(x)$ below which the
ground state normalizability is preserved. In 
Figure~(\ref{fig:ch2.thetac}), this function is plotted in
the whole range $0<x<1$. As a result, if we restrict
the angle $\Theta$ to be less than about $\pi/3$, the ground
state wave function is normalizable for all values of $m/M$.
      
It is useful to have the expression of $q_1$ in terms of the
creation and annihilation operators:
\ba
\label{eq:ch2.q1exp}
q_1&=&{a^{(+)} +a^{(-)} \over \sqrt{2m}
                      \sqrt{(1-{m^2 \over {\cal M}^2})
                            (1-{m^2 \over \overline{{\cal M}}^2})}
                       }
    +{b^{(+)} +b^{(-)} \over \sqrt{2{\cal M}}
                      \sqrt{(1-{m^2 \over {\cal M}^2})
                            (-1+{{\cal M}^2 \over \overline{{\cal M}}^2})}
                       } \nonumber\\
    &+&{c^{(+)} +c^{(-)} \over \sqrt{2\overline{{\cal M}}}
                      \sqrt{(1-{m^2 \over \overline{{\cal M}}^2})
                            (-1+{\overline{{\cal M}}^2 \over {\cal M}^2})}
                       } .
\ea
This concludes our discussion of the higher derivative oscillator.

Note that, if we have an extra term $-\lambda_0 x^4$ in the starting
Lagrangian, then our Hamiltonian would consist of two parts,
$H=H_0 + H_1$, where $H_0$ is just the oscillator Hamiltonian
discussed above and $H_1=\lambda_0 q_1^4$ with $q_1$ given by
Equation~(\ref{eq:ch2.q1exp}) . Thus, the oscillator 
gives us a good starting point
for perturbation theory. 

\subsection{ Euclidean Path Integral }

Now let us  evaluate the partition function of the
higher derivative theory defined by 
\be
{\cal Z} = Tr e^{-\beta H} \equiv 
\sum_{s} \langle \bar{s} |\eta e^{-\beta H}|s \rangle ,
\ee
where the summation is over all states $|s\rangle$
such that they are complete:
\be
\sum_{s} |s \rangle \langle \bar{s} | \eta = 1 .
\ee
One convenient choice for the states is $|q_1,q_2,q_3\rangle$.
Then one can make use of the derivative forms of the momentum
operators and derive a path integral form for the partition
function, just like in the usual theory. First one 
has to slice the Euclidean time $\beta$ into small intervals and, then, 
the partition function is written in terms of the integration 
of the intermediate positions. This path integral form of
the partition function is exactly the 
Euclidean path integral that one would naively write down
when not concerned with the canonical quantization 
procedure \cite{gross2,hawk2}  
\ba
\!\!\!\!{\cal Z}[J]&=& \int Dq \exp \left( -\int^{\beta}_0 d \tau
      L_{E}[q(\tau)] + J(\tau)q(\tau) \right) ,
\nonumber \\
\!\!\!\!L_{E} &=& {1\over2}(1+2{m^2 \over M^2}\cos2\Theta)
      \dot{q}^2 
      +({\cos2\Theta \over M^2}+ {m^2 \over 2M^4})\ddot{q}^2  
      + {1\over 2M^4}{\stackrel{\cdots}{q}}^2
      +{m^2 \over 2}q^2 .  
\ea
The Euclidean propagator of the variable $q(\tau)$ 
can be found by differentiating the partition functional
with respect to the external source $J(\tau)$. In Fourier
space, it is given by
\be
D_{E}(E)={M^4 \over (E^2+m^2)(E^2+M^2e^{2i\Theta})(E^2+M^2e^{2i\Theta})}
.
\ee
The multiple pole structure in the propagator is a manifestation of
the spectrum of the theory. As can be seen clearly, the poles are
located exactly at three types of energy gaps  of the theory.       

There is a big difference here in the
higher derivative theory as compared with the usual theory.
The Minkowski path integral \cite{ghost2,slav2} 
is not well defined. In fact, due
to the complex ghost energy, it has runaway modes at large
temporal separation. Also, 
we cannot do a wick rotation from the Euclidean to the Minkowski
because of the complex ghost pole on the first sheet.
We should emphasize that the Euclidean path integral
is still well defined. This is the object that we will
be using in our numerical simulation of the theory. 
Also, the Euclidean path integral in principle contains all
the information about the higher derivative theory.  
By measuring the Euclidean propagator of the theory, one can
extract the energy excitations of the higher derivative theory
and, hence, the eigenvalues of the Hamiltonian.

\section{Higher Derivative Free Field Theory}
\label{sec:freeft}

Having discussed the quantum mechanical oscillator, 
let us now turn
to the simplest higher derivative field theory, 
free field theory. 
Since most of the procedures are quite similar to the
quantum mechanical case, we will be very brief in this section.
Consider the one component higher derivative scalar field theory
parametrized by the
Lagrangian 
\be
{\cal L}={1 \over 2} \phi(x)(-\Box-m_0^2)
                  (1+{\Box \over {\cal M}^2})
                  (1+{\Box \over \overline{{\cal M}}^2})\phi(x) ,
\ee
where the $\Box$ is the Minkowski d'Alambert operator. 
The Hamiltonian density can be obtained in the same way as in
the quantum mechanical example 
\ba
{\cal H}&=&\pi_1 \phi_2+\pi_2 \phi_3+{M^4 \over 2}{\pi_3}^2
        -{1 \over 2} \phi_2 (\rho_1-2\rho_2 \nabla^2 +3 \rho_3 \nabla^4
                        ) \phi_2
        +{1 \over 2} \phi_3 (\rho_2 -3 \rho_3 \nabla^2
                        ) \phi_3
\nonumber \\
      &&+{1 \over 2} \phi_1 (-\rho_1 \nabla^2+\rho_2 \nabla^4 
            -\rho_3 \nabla^6
                        +m^2_0) \phi_1 .
\ea
Again, we can interchange the role of $\pi_2$ and $\phi_2$, which 
amounts to $\phi_2 \rightarrow \pi_2$ and 
$\pi_2 \rightarrow -\phi_2$ .
We also impose negative metric on $\pi_2$ and $\phi_2$ by doing the
substitution $\phi_2 \rightarrow -i\phi_2$ and
$\pi_2 \rightarrow +i\pi_2$ , and after these changes our Hamiltonian density
is,                                           
\ba
{\cal H} &=& i\pi_1\pi_2 +{1 \over 2\rho_3}{\pi_3}^2
           +{1 \over 2} \pi_2 (\rho_1 - 2\rho_2 \nabla^2
                              +3 \rho_3 \nabla^4) \pi_2
        +{1 \over 2} \phi_3 (\rho_2 
                            -3\rho_3 \nabla^2)\phi_3
\nonumber \\
         &&+{1 \over 2} \phi_1 (-\rho_1 \nabla^2 + \rho_2 \nabla^4
                              -  \rho_3 \nabla^6 +m^2_0) \phi_1
         +i\phi_2 \phi_3 .
\ea 
Negative metric quantization then corresponds to making $\phi_i,
i=1,2,3$ and $\pi_i,i=1,2,3$ hermitian operators. Then the 
Hamiltonian itself is not hermitian but still self-adjoint 
with respect to the negative metric $\eta$ which flips the 
sign of $\pi_2$ and $\phi_2$. We have   
\be
\eta {\cal H}^{\dagger} \eta = {\cal H} .
\ee
Introducing  the Fourier modes 
\be
\phi_i({\bf x}) = \bar{\phi_i}
          +\sum_{{\bf k} >0}{1 \over \sqrt{V}} 
           [ \phi_{i,\bf k} e^{i{\bf k}\cdot{\bf x}}
           + \phi_{i,\bf k}^{*} e^{-i{\bf k}\cdot{\bf x}} ] ,
\ee
where the index $i$ runs from $1$ to $3$.  
We can also write the similar expression for $\pi_i$
\be
\pi_i({\bf x}) = \bar{\pi_i}
          +\sum_{{\bf k} >0}{1 \over \sqrt{V}} 
           [ \pi_{i,\bf k} e^{-i{\bf k}\cdot{\bf x}}
           + \pi_{i,\bf k}^{*} e^{i{\bf k}\cdot{\bf x}} ] .
\ee
To ensure basic commutation relations,  we must have 
\be
[\phi_{i,\bf k}, \pi_{j,\bf k'}]=
[\phi_{i,\bf k}^{*}, \pi_{j,\bf k'}^{*}]=i \delta_{ij}\delta_{\bf k k'}
,
\;\; \;\;
[\bar{\phi_{i}}, \bar{\pi_{j}}]=i \delta_{ij} .
\ee
We can then write the Hamiltonian as 
\ba
H&=& {1 \over V} ( i\bar{\pi_1}\bar{\pi_2}
              +{\bar{\pi_3}\bar{\pi_3} \over 2 \rho_3}
          +{\rho_2 \over 2} \bar{\pi_2}^2 )
       +V( {\rho_2 \over 2} \bar{\phi_3}^2 +{m^2 \over 2} \bar{\phi_1}^2
         +i \bar{\phi_2}\bar{\phi_3} )
\nonumber \\
&& + \sum_{\bf k >0} i (\pi_{1,\bf k}\pi_{2,\bf k}^{*}
                     +\pi_{1,\bf k}^{*}\pi_{2,\bf k})
              +{\pi_{3,\bf k}\pi_{3,\bf k}^{*} \over \rho_3}
            +(\rho_1 +2 \rho_2 {\bf k}^2 +3 \rho_3 {\bf k}^4)
             \pi_{2,\bf k}\pi_{2,\bf k}^{*}
\nonumber \\
&&    +(\rho_1 {\bf k}^2 +\rho_2 {\bf k}^4 +\rho_3 {\bf k}^6 +m^2_0)
             \phi_{1,\bf k}\phi_{1,\bf k}^{*}
      +(\rho_2 +3 \rho_3 {\bf k}^2 )
             \phi_{3,\bf k}\phi_{3,\bf k}^{*}
\nonumber \\
&&    +i( \phi_{2,\bf k}\phi_{3,\bf k}^{*}
        +\phi_{2,\bf k}^{*}\phi_{3,\bf k}) .
\ea
After some rescaling of the variables 
we can bring the Hamiltonian into  similar form as in the 
quantum mechanical oscillator case
\be
H=H_0+\sum_{\bk >0} \pi^{*}_{i \bk}\pi_{i \bk}
   +\phi^{*}_{i \bk}\bM_{ij}\phi_{j \bk} .
\ee
We then perform the same ``rotation'' transformation 
as in the oscillator case, and then the above Hamiltonian is
diaganolized to
\be
H=H_0+\sum_{\bk >0} \Pi^{*}_{i \bk}\Pi_{i \bk}
   +\Phi^{*}_{i \bk} \omega^2_{i \bk} \Phi_{j \bk} ,
\ee
where the frequency  
$\omega_{0 \bk}=\sqrt{ m^2_0 +\bk^2}$,
$\omega_{1 \bk}=\sqrt{ {\cm}^2 +\bk^2}$ and
$\omega_{2 \bk}=\sqrt{ {\cmb}^2 +\bk^2}$.
The creation and annihilation operators are given by
\ba
a^{(-)}_{i  \bk}&=& {1 \over \sqrt{2}} (
             \sqrt{\omega_{i \bk}} \Phi_{i \bk}
            +{i \over \sqrt{\omega_{i \bk}}} \Pi^{*}_{i \bk} ) ,
\nonumber \\
a^{(+)}_{i  \bk}&=& {1 \over \sqrt{2}} (
             \sqrt{\omega_{i \bk}} \Phi_{i \bk}
            -{i \over \sqrt{\omega_{i \bk}}} \Pi^{*}_{i \bk} ) ,
\nonumber \\
a^{(-)}_{i -\bk}&=& {1 \over \sqrt{2}} (
             \sqrt{\omega_{i \bk}} \Phi^{*}_{i \bk}
            +{i \over \sqrt{\omega_{i \bk}}} \Pi_{i \bk} ) ,
\nonumber \\
a^{(+)}_{i -\bk}&=& {1 \over \sqrt{2}} (
             \sqrt{\omega_{i \bk}} \Phi^{*}_{i \bk}
            -{i \over \sqrt{\omega_{i \bk}}} \Pi_{i \bk} ) ,
\ea 
with $i=1,2,3$. The Hamiltonian finally looks like
\be
H=\sum_{\bk}(a_{i \bk}^{(+)}a_{i \bk}^{(-)}+{1\over2})\omega_{i \bk} ,
\label{eq:frees}
\ee
where the summation is over all the momentum modes and   
three types of excitations. The
creation and annihilation operators have the standard
commutation relations
\be
[a^{(-)}_{i \bk},a^{(+)}_{j \bp}]=\delta_{\bk \bp} \delta_{ij} .
\ee 
Similarly, the field $\phi(x)$ can be expressed as a linear combination
of the creation and annihilation operators which will be
given explicitly in the next section. The particle contents of this
free Hamiltonian is now clear. One has three types of excitations for
each three-momentum $\bk$. The operator 
$a^{(+)}_{0 \bk}$
creates an ordinary particle of mass $m_0$, momentum $\bk$ and
energy $\omega_{0 \bk}=\sqrt{m^2_0+\bk^2}$. 
The operator 
$a^{(+)}_{1 \bk}$
creates a ghost particle of mass ${\cal M}$, momentum $\bk$ and
energy $\omega_{1 \bk}=\sqrt{{\cal M}^2+\bk^2}$. 
The operator 
$a^{(+)}_{2 \bk}$
creates an antighost particle.
 
\section{Higher Derivative $O(N)$ Model in the Symmetric Phase}
\label{sec:symm}

The higher derivative field theory can be easily generalized
to an $O(N)$-symmetric scalar field theory with a quartic coupling.
In the symmetric phase it is convenient to parametrize the Lagrangian
as 
\ba  
{\cal L} &=& -{1 \over 2}(1+2{m_0^2 \over M^2}\cos 2\Theta) 
                  \phi^{a}\Box\phi^{a} \nonumber \\
         &+& ({\cos 2\Theta \over M^2}+ {m_0^2 \over 2M^4})  
                  \phi^{a}\Box^2\phi^{a} 
        - {1 \over M^4} 
            \phi^{a}\Box^3\phi^{a} \nonumber \\
         &-& {m_0^2 \over 2}\phi^{a}\phi^{a}   
        -{\lambda_0}(\phi^{a}\phi^{a})^2 .
\ea
The Hamiltonian of the theory,  
after indefinite metric quantization, can be expressed in terms of
creation and annihilation operators, $H=H_{0}+H_{\rm int}$,  
where the free part of the Hamiltonian 
is given by Equation~(\ref{eq:frees}).
The interaction part of the Hamiltonian is the conventional one, namely
$H_{\rm int}=\int d^3{\bf x}{\lambda_0}(\phi^{a}\phi^{a})^2$,
where the field $\phi^{a}$ can be written as a linear combination
of the creation and annihilation operators, 
\ba 
\phi^{a}\!\!\!\!&=&\!\!\!\!\sum_{\bf p} 
    \sqrt{{c_0 \over 2V\omega_{0\bf p}}}
    \left(a^{(-)a}_{0\bf p} e^{i{\bp \cdot x}}  
    +a^{(+)a}_{0{\bf p}} e^{-i{\bp \cdot x}}\right) \nonumber\\  
&+&\!\!\!\!
    \sqrt{{c_1 \over 2V\omega_{1 \bf p}}}
    \left(a^{(-)a}_{1\bf p} e^{i{\bp \cdot x}}  
    +a^{(+)a}_{1{\bf p}} e^{-i{\bp \cdot x}}\right) \\  
&+&\!\!\!\!
    \sqrt{{c_2 \over 2V\omega_{2 \bf p}}}
    \left(a^{(-)a}_{2\bf p} e^{i{\bp \cdot x}}  
    +a^{(+)a}_{2{\bf p}} e^{-i{\bp \cdot x}}\right) , \nonumber  
\ea
where the 
values for $c_i$ are given by the following list:
\ba  
c_0&=&M^{-4}[(m_0^2-{\cm}^2)
      (m_0^2-{\cmb}^2)]^{-1} , \nonumber\\
c_1&=M^{-4}&[({\cm}^2-m_0^2)
       ({\cm}^2-{\cmb}^2)]^{-1} , \\
c_2&=M^{-4}&[({\cmb}^2-m_0^2)
      ({\cmb}^2-{\cm}^2)]^{-1} . \nonumber
\ea
The negative metric is seen from the
adjoint relations among the creation and annihilation operators
\ba 
\overline{a^{(-)a}_{0{\bf p}}}
&\equiv& \eta  a^{(-)a \dagger}_{0{\bf p}} \eta 
= a^{(+)a}_{0{\bf p}} , \nonumber\\  
\overline{a^{(-)a}_{1{\bf p}}}
&\equiv& \eta  a^{(-)a \dagger}_{1{\bf p}} \eta 
= a^{(+)a}_{2{\bf p}} , \\  
\overline{a^{(-)a}_{2{\bf p}}}
&\equiv& \eta  a^{(-)a \dagger}_{2{\bf p}} \eta 
= a^{(+)a}_{1{\bf p}} , \nonumber  
\ea 
where $\eta$ is the metric operator satisfying $\eta=\eta^{\dagger}$ and
$\eta^2=1$. It is clear that the
Hamiltonian itself is self-adjoint with respect to the metric $\eta$, i.e.
$\overline{H}\equiv\eta H^{\dagger} \eta=H$. 

\section{Higher Derivative $O(N)$ Model in the Broken Phase}
\label{sec:broken}

  
One starts with the general higher derivative Lagrangian which 
has a global $O(N)$ symmetry
\be 
{\cal L} = {1 \over 2} \phi^a (-\rho_1 \Box
          -\rho_2 \Box^2-\rho_3 \Box^3 ) \phi^a 
          +{1 \over 2} \mu_0^2 \phi^a\phi^a 
           -\lambda_0  (\phi^a \phi^a)^2  ,
\ee
where $\Box=\partial^2_t -\nabla^2$ 
is the Minkowski space d'Alambert operator and 
the coefficients are parametrized as
\be
\rho_1=1+{m^2_0 \over {\cal M}^2}+{m^2_0 \over \bar{{\cal M}}^2} ,
\;\;\;
\rho_2={1 \over {\cal M}^2}+{1 \over \bar{{\cal M}}^2}
       +{m^2_0 \over {\cal M}^2\bar{{\cal M}}^2} ,
\;\;\;
\rho_3={1 \over {\cal M}^2\bar{{\cal M}}^2} .
\ee
After the usual steps of indefinite metric quantization, the
Hamiltonian has the form
\ba
\label{eq:ch2.hambro}
{\cal H}&=& i \pi^a_1\pi^a_2 + {1 \over 2\rho_3}\pi^a_3\pi^a_3
           +{1 \over 2} \pi^a_2 (\rho_1 -2\rho_2\nabla^2
                       +3 \rho_3 \nabla^4) \pi^a_2
\nonumber \\
\!\!\!\! &+&{1 \over 2} \phi^a_1 (-\rho_1\nabla^2 -\rho_2\nabla^4
                       -\rho_3 \nabla^6) \phi^a_1
           +{1 \over 2} \phi^a_3 (\rho_2
                       -3\rho_3 \nabla^2) \phi^a_3
           +i \phi^a_2\phi^a_3
\nonumber \\
         &-&{1 \over 2} \mu_0^2 \phi^a_1\phi^a_1 
           + \lambda_0  (\phi^a_1 \phi^a_1)^2  .
\ea 
The corresponding $O(N)$ generators are given by
\be
Q^{ab}=\sum_{i,\bx} \phi^{a}_i(\bx)\pi^{b}_i(\bx)
                 -\phi^{b}_i(\bx)\pi^{a}_i(\bx) ,
\ee
which obviously commute with the Hamiltonian.

Next, the Fourier modes are introduced for each variable
\ba 
\phi^a_i(\bx) &=& \bar{\phi}^a_i +
            {1 \over \sqrt{V}} \sum_{\bk > 0}
 \phi^a_{i,\bk} e^{i\bk \cdot \bx}
           +\phi^{a*}_{i,\bk} e^{-i\bk \cdot \bx} ,
\nonumber \\
\pi^a_i(\bx) &=& {(-i) \over V}{\partial \over \partial \bar{\phi}^a_i}
           +{(-i) \over \sqrt{V}} \sum_{\bk > 0}
 e^{-i\bk \cdot \bx}{\partial \over \partial \phi^a_{i,\bk}}
    +e^{+i\bk \cdot \bx}{\partial \over \partial \phi^{a*}_{i,\bk}} .
\ea
The Hamiltonian is brought into the following form:
\ba
\label{eq:ch2.hambrof}
H &=& {1 \over V}(i \pi^a_{10}\pi^a_{20}
      +{1 \over 2 \rho_3} \pi^a_{30}\pi^a_{30}
      +{\rho_1 \over 2} \pi^a_{20}\pi^a_{20} )
      +V ( {\rho_2 \over 2} \bar{\phi}^a_3\bar{\phi}^a_3
         + i \bar{\phi}^a_2\bar{\phi}^a_3 )
\nonumber \\
      && + 
        \sum_{\bk >0} i \pi^a_{1\bk}\pi^{a*}_{2\bk}
                    +i \pi^{a*}_{1\bk}\pi^{a}_{2\bk}
               +{1 \over \rho_3} \pi^{a}_{3\bk}\pi^{a*}_{3\bk}
               +(\rho_1+2\rho_2\bk^2+3\rho_3\bk^4)
                \pi^{a}_{2\bk}\pi^{a*}_{2\bk}
\nonumber \\
      &&   
      + (\rho_1\bk^2 + \rho_2\bk^4 + \rho_3\bk^6 ) 
                \phi^{a}_{1\bk}\phi^{a*}_{1\bk}
      + ( \rho_2\bk^4 +3\rho_3\bk^2 ) 
                \phi^{a}_{3\bk}\phi^{a*}_{3\bk}
       + i \phi^a_{2\bk}\phi^{a*}_{3\bk}
                    +i \phi^{a*}_{2\bk}\phi^{a}_{3\bk}
\nonumber \\
      &&   
          -\sum_{\bx}{1 \over 2} \mu_0^2 \phi^a_1\phi^a_1 
           +\sum_{\bx}\lambda_0  (\phi^a_1 \phi^a_1)^2  .
\ea 
We will single out the direction of the $\bar{\phi^a_1}$ variable
and fix it in some direction in the $O(N)$ space. 
This treatment is only valid in the limit of infinite volume.
Strictly speaking, in a finite volume, the symmetry is not
broken. Therefore, the description of symmetry breaking in the finite
volume needs more careful study. As we will show in Chapter~(\ref{ch:BOA}),
in a very large but finite volume, one can apply the adiabatic 
approximation (or Born-Oppenheimer Approximation) 
 to the direction of the zeromode. 
We find that  the direction of the zeromode rotates very slowly and
decouples from the other modes in the theory.
Therefore,  if the volume is very large,  
it is legitimate to assume that the direction of the
zeromode is frozen in the $O(N)$ space.
With this in mind, we can then decompose
\be
\phi^a_1 =v n^a + h(\bx) n^a +\tilde{\phi}^a_{1T}(\bx) ,
\ee
and similarly for the $\phi_2$ and $\phi_3$ variables.
The value of $v$ is set to $\sqrt{\mu^2_0/4\lambda_0}$.
The Hamiltonian is then written as sum of three types of
terms:
\ba
\label{eq:ch2.hampiece}
H &=& H_0+H_{\bk \neq 0}+H_{\rm int}  ,
\nonumber \\
H_0&=&
      {1 \over V}(i \pi^a_{10}\pi^a_{20}
      +{1 \over 2 \rho_3} \pi^a_{30}\pi^a_{30}
      +{\rho_1 \over 2} \pi^a_{20}\pi^a_{20} )
      +V ( {\rho_2 \over 2} \bar{\phi}^a_3\bar{\phi}^a_3
         + i \bar{\phi}^a_2\bar{\phi}^a_3+{m^2_0 \over 2}\sigma^2 ) ,
\nonumber \\
H_{\bk \neq 0}&=&
        \sum_{\bk >0} i \pi^a_{1\bk}\pi^{a*}_{2\bk}
                    +i \pi^{a*}_{1\bk}\pi^{a}_{2\bk}
               +{1 \over \rho_3} \pi^{a}_{3\bk}\pi^{a*}_{3\bk}
               +(\rho_1+2\rho_2\bk^2+3\rho_3\bk^4)
                \pi^{a}_{2\bk}\pi^{a*}_{2\bk}
\nonumber \\
    &+&(\rho_1\bk^2+\rho_2\bk^4+\rho_3\bk^6+m^2_0)
       \phi^{a}_{1\bk L}\phi^{a*}_{1\bk L}
     + (\rho_1\bk^2+\rho_2\bk^4+\rho_3\bk^6)
       \phi^{a}_{1\bk T}\phi^{a*}_{1\bk T} 
\nonumber \\
    &+&(\rho_2+3\rho_3\bk^2)
       \phi^{a}_{3\bk}\phi^{a*}_{3\bk}
     +i\phi^{a}_{2\bk}\phi^{a*}_{3\bk}
     +i\phi^{a*}_{2\bk}\phi^{a}_{3\bk} ,
\nonumber \\
H_{\rm int}&=& \sum_{\bx}
      {4 \lambda_0 v }h(h^2+
           \tilde{\phi}^{a}_{1T}\tilde{\phi}^{a}_{1T})
     +{\lambda_0 }(h^2+
           \tilde{\phi}^{a}_{1T}\tilde{\phi}^{a}_{1T})^2 ,
\ea 
where $m^2_0=2\mu^2$.
We will examine  each piece separately.

The $\bk \neq 0$ piece can be diagonalized the same way
as in section~(\ref{sec:symm}). The interaction piece is also
expressed as the creation and annihilation operators through
the field variables. The $H_0$ piece is the only one that
is new in the broken phase. For convenience we use the rescaled
variables given by
\ba
p^a_1&=&(\rho_1 V)^{-1/2}\pi^a_{10},
\;\;\;\;\;
p^a_2=\sqrt{{\rho_1 \over V}}\pi^a_{20},
\;\;\;\;\;
p^a_3=(\rho_3 V)^{-1/2}\pi^a_{30} ,
\nonumber \\
q^a_1&=&(\rho_1 V)^{1/2}\bar{\phi}^a_{1},
\;\;\;\;\;
q^a_2=\sqrt{{V \over \rho_1}}\bar{\phi}^a_{2},
\;\;\;\;\;
q^a_3=(\rho_3 V)^{1/2}\bar{\phi}^a_{3} ,
\ea
and express the radial variables  $q^a_1$ as
\be
q^a_1=\sqrt{\rho_1 V}(v+\sigma) n^a= \rho n^a .
\ee
The derivatives for the $q^a_1$ are now substituted by
\be
{\partial \over \partial q^a_1}=n^a 
   {\partial \over \partial \rho} 
\ee
where the index $a$ runs from $1$ to $N$.
We have assumed that the volume is practically infinite and 
the direction $n^a$ is really a constant unit vector in
$O(N)$ space. As we will see in Chapter~(\ref{ch:BOA}), this is
only approximately true in a finite volume.
Use the following identity
\be
i p^a_1p^a_2 = i p_{2L}p_{1 \rho} ,
\ee
$H_0$ is further decomposed into
two parts
\ba
H_0&=& H_{0L}+H_{0T} ,
\nonumber \\
H_{0L}&=& i p_{2L}p_{y} +{1 \over 2}p^2_{2L}
        +{1 \over 2}p^2_{3L}+{\rho_2 \over 2\rho_3}q^2_{3L}
        +i \sqrt{{\rho_1 \over \rho_3}}q_{2L}q_{3L}
        +{m^2_0 \over 2\rho_1} y^2 ,
\nonumber \\
H_{0T}&=& {1 \over 2}p^a_{2T}p^a_{2T}
        +{1 \over 2}p^a_{3T}p^a_{3T}
        +{\rho_2 \over 2\rho_3}q^a_{3T}q^a_{3T}
        +i \sqrt{{\rho_1 \over \rho_3}}q^a_{2T}q^a_{3T} .
\ea 
The longitudinal part has the same form as the simple oscillator
and can be easily diagonalized. The transverse part $H_{0T}$ can also be   
diagonalized with the transformation
\be
q_T=A Q_T ,  \;\;\;\; A A^T=A^T A=1 ,
\ee
where $A$ is a two by two matrix
\be
A= \left( \begin{array}{cc}
         {-1 \over (1-e^{-4i\theta_g})^{1/2}} &
        {1 \over (1-e^{+4i\theta_g})^{1/2}} \\
         {-ie^{-2i\theta_g} \over (1-e^{-4i\theta_g})^{1/2}} &
        {ie^{2i\theta_g} \over (1-e^{+4i\theta_g})^{1/2}} 
         \end{array}  \right) .
\ee
The angle $\theta_g$ is the complex phase of the
Goldstone ghost mass parameter 
${\cal M}_g=|{\cal M}_g|e^{i\theta_g}$, which is given by
\be
{\cal M}^2_g={m^2_0+\cm^2+\cmb^2 \over 2}
     + i \cm \cmb \sqrt{ \rho_1-{1 \over4} 
       ( {m^2_0 \over \cm \cmb}+ {\cm \over \cmb}+ {\cm \over \cmb})^2
         } .
\ee
The transverse part of the Hamiltonian is then diagonalized to
\be
H_{0T}=\sum_{i \neq 0,a}a^{(+)a}_{i0T}a^{(-)a}_{i0T} \omega_{i0T}  ,
\ee
where the summation of $a$ is from $1$ to $N$ and the 
energy gap is $\omega_{10T}=\cm_g$ and 
$\omega_{20T}=\cmb_g$. 
In terms of these operators we can write out the explicit
form of $p^a_{2T}$ 
\be
p^a_{2T}=\sum_{i\neq 0} \sqrt{\omega_{i0T} \over 2} \epsilon_{i} ,
          (a^{(-)a}_{i0T}-a^{(+)a}_{i0T})
\ee
where the polarization factor $\epsilon_i$ is given by
$\epsilon_1=\epsilon^{*}_2=i/(1-e^{-4i\theta_g})^{1/2}$.

To summarize, in the broken phase we would have the following
Hamiltonian
\ba
H&=& H_0 + H_{\rm int} , 
\nonumber \\
H_0&=& \sum_{i,\bk,\lambda} a^{(+)a}_{i\bk\lambda}a^{(-)a}_{i\bk\lambda}
                           \omega_{i\bk\lambda} ,
\nonumber \\
H_{\rm int}&=& \sum_{\bx}
      {4 \lambda_0 v }h(h^2+
           \tilde{\phi}^{a}_{1T}\tilde{\phi}^{a}_{1T})
     +{\lambda_0  }(h^2+
           \tilde{\phi}^{a}_{1T}\tilde{\phi}^{a}_{1T})^2 .
\ea
The
index $\lambda$ takes the value $L$ and $T$ respectively. All the 
operators can be expressed in terms of the creation and annihilation
operators as
\ba
h(\bx)&=& \sum_{i\bk} {c_{iL} \over \sqrt{2\omega_{i\bk L}V}}
      \left( n^a a^{(-)a}_{i\bk L} e^{i\bk\cdot\bx}
           +n^a a^{(+)a}_{i\bk L} e^{-i\bk\cdot\bx} \right) ,
\nonumber \\
\tilde{\phi}^a_{T}(\bx)&=& \sum_{i\bk \neq 0} 
       {c_{iT} \over \sqrt{2\omega_{i\bk T}V}}
      \left(  a^{(-)a}_{i\bk T} e^{i\bk\cdot\bx}
           + a^{(+)a}_{i\bk T} e^{-i\bk\cdot\bx} \right) ,
\\
\rho&=&\sqrt{\rho_1 V}(v+\sigma)=\sqrt{\rho_1 V}\left(v
          +\sum_{i}{c_{iL} \over \sqrt{2\omega_{i0L}V}}
          (a^{(-)}_{i0L}+a^{(+)}_{i0L}) \right) ,
\nonumber
\ea
where the form factors $c_{i\lambda}$ are given by the following
table
\ba
c_{0L}&=& \sqrt{ \cm^2 \cmb^2 \over (m^2_0-\cm^2)(m^2_0-\cmb^2)},
\;\;\;\;\;c_{1L}=c^{*}_{2L}=\sqrt{ \cm^2 \cmb^2 \over (\cm^2-m^2_0)(\cm^2-\cmb^2)} ,
\nonumber \\
c_{0T}&=& 1,
\;\;\;\;\;c_{1T}=c^{*}_{2T}=\sqrt{\cmb^2_g \over (\cm^2_g-\cmb^2_g)} .
\ea
The creation and annihilation operators enjoy the following
commutation relations
\be
[ a^{(-)a}_{i\bk\lambda} , a^{(+)b}_{j\bp\lambda^{'}} ] 
= \delta_{ij}\delta_{\bk \bp}\delta_{\lambda \lambda^{'}}P^{ab}_{\lambda}.
\ee

\vfill\eject

%% file: c3.tex
\chapter{ Unitarity and Large $N$ Expansion }
\label{ch:NUNI}

\section{Lippmann-Schwinger Equation And Unitarity}
 
In this section, we will try to answer one of the most important
questions about higher derivative theories, namely, the unitarity
problem \cite{pais3,lee3,polk3,ghost3,scat3}. 
In the first part of the discussion, 
 we will set up the general formalism of scattering
matrix in the higher derivative theory and  argue that the
$S$-matrix defined within the 
physical subspace can be made unitary. 
In the second part, we will present 
a concrete example of the unitary scattering amplitude in the
large $N$ limit of the $O(N)$ model  which involves the ghost
states as intermediate states.

\subsection{General Formalism and Unitarity}
 
Let us imagine that our Hilbert space is built up by all the states
generated from the vacuum by successive operations of creation operators
 as described in Chapter~(\ref{ch:QUAN}). 
Some states will have negative norm and complex energy
components. We assume that 
all states available to build the initial state
contain only real
energy components of the free Hamiltonian in all Lorentz frames 
\cite{lee3}. We will call these states ``normal states'' or
``physical states''.  Denote
the eigenstate of the free Hamiltonian by $\phial$ such that
\ba
H_0 \phial &=& E_{\alpha} \phial , \\
E_{\alpha} &\in&  \Re .   \nonumber
\ea 
Then one can construct two states, denoted as $\psialp$ and $\psialm$, 
from the Lippmann-Schwinger equation
\ba
\psialpm &=& \phial +\gzpmal V \psialpm , \nonumber \\ 
\psialpm &=& \phial +\gpmal V \phial .
\ea 
It is then easy to show that the states $\psialpm$ are eigenstates of
the full Hamiltonian with corresponding energy $E_{\alpha}$. If we now form
wavepackets from these states, one can see that they correspond to
incoming and outgoing waves in the past or future. Therefore, they give us
a good description of the scattering process. We can rewrite the
above equation as 
\begin{equation}
\psialpm = {\pm i \epsilon} \gpmal \phial .
\end{equation}
In this form, it is clear that only energy conserving components of
$\phial$ survive the scattering since, if the components are of different
energy, they will make the operator $(E_{\alpha}
-H\pm i\epsilon)^{-1}$ nonsingular in the $\epsilon$
goes to zero limit, hence are killed by the $\epsilon$ in front. The $S$-
matrix between any two normal states $\alpha$ and $\beta$ is then
defined to be 
\ba
S_{\beta \alpha}\!\!\!&\equiv& \langle\psi_{\beta}^{(-)} |\eta \psialp \\
\!\!\!\!&=&\!\!\!\! \langle\phi_{\beta} | \eta 
            {i \epsilon \over E_{\beta}-H+ i \epsilon} 
            {i \epsilon \over E_{\alpha}-H+ i \epsilon} 
          \phial . \nonumber 
\ea 
Using the perturbative expansion of the Green's function one can show
that the $S$-matrix element defined above is related to the so called
$R$-matrix (or $T$-matrix) element by
\ba
\label{eq:ch3.rmat}
S_{\beta \alpha}(E_{\alpha}) &=& \delta_{\beta\alpha}
   -2\pi i \delta(E_{\alpha}-E_{\beta})
    R_{\beta\alpha}(E_{\alpha}) , \nonumber \\
R_{\beta\alpha}(E_{\alpha})&=&\langle \phi_{\beta} 
   |R(E_{\alpha}) \phial , \nonumber\\
R(E)&=& V +V {1 \over E-H_0 +i \epsilon}R(E) , \\
R(E)&=& V +V {1 \over E-H_0 +i \epsilon}V+\cdots  , \nonumber \\
R(E)&=& V +V {1 \over E-H +i \epsilon}V . \nonumber
\ea 
To show unitarity, we write the Lippmann-Schwinger equation in a
special way 
\ba
\psialpm &=&\Omega^{(\pm)}(E_{\alpha}) \phial ,   \\ 
\Omega^{(\pm)}(E_{\alpha}) &=& 1+\gpmal V {\pm i \epsilon
             \over E_{\alpha} -H_0 \pm i\epsilon} , \nonumber  
\ea 
where $\Omega^{(+)}(E_{\alpha})$ and
 $\Omega^{(-)}(E_{\alpha})$ are called wave operators.
Using the adjointness of the Hamiltonian we see that 
\be
\phial=\eta \Omega^{(\pm)}(E_{\alpha})^{\dagger}  \eta \psialpm . 
\ee
From these two relations we can see that
\begin {equation}
\eta \Omega^{(\pm)}(E_{\alpha})^{\dagger}  \eta  
\Omega^{(\pm)}(E_{\alpha})=  1+P_{c} ,
\end {equation}
where $P_{c}$ is the complex energy projector for the free Hamiltonian.
Let us now consider the sum 
\ba
\sum_{\alpha,E_{\alpha}\in \Re}\!\!\!\!&&S_{\beta\alpha}S_{\gamma\alpha}^{*}
    = \sum_{\alpha}\langle\psi_{\beta}^{(-)} |
     \eta \psialp \langle \psi_{\alpha}^{(+)} |
     \eta | \psi_{\gamma}^{(-)}\rangle 
\nonumber \\ 
\!\!\!\!\!    &=& \langle\psi_{\beta}^{(-)} | \eta | \psi_{\gamma}^{(-)}\rangle
\\ 
\!\!\!\!\!    &=&\langle \phi_{\beta} | \Omega^{(-)}(E_{\beta})^{\dagger} \eta
        \Omega^{(-)}(E_{\beta}) | \phi_{\gamma} \rangle 
\nonumber \\ 
 \!\!\!\!\!   &=& \langle \phi_{\beta} |\eta |\phi_{\gamma}\rangle 
    = \delta_{\beta\gamma}  ,
\nonumber 
\ea 
where, in the second step, we have inserted 
the complete set of the full Hamiltonian.
This establishes the unitarity of the $S$-matrix. 
The above proof looks very formal. 
To clearly understand the role of the ghost states in
the theory let us calculate some scattering processes
in the higher derivative $O(N)$ model. 

\subsection{Large-$N$ Scattering Amplitude of $O(N)$ Model}

The basic formula is the perturbation expansion of the $S$-matrix given
by Equation~(\ref{eq:ch3.rmat}) . 
We will consider the large-$N$ limit of the geometric
resummation of the $s$-channel bubble diagram. We will show how the
modified Feynman Rules arise naturally from this calculation.

\begin{figure}[htb]
\vspace{10mm}
\centerline{ \epsfysize=3.0cm 
             \epsfxsize=5.0cm \epsfbox{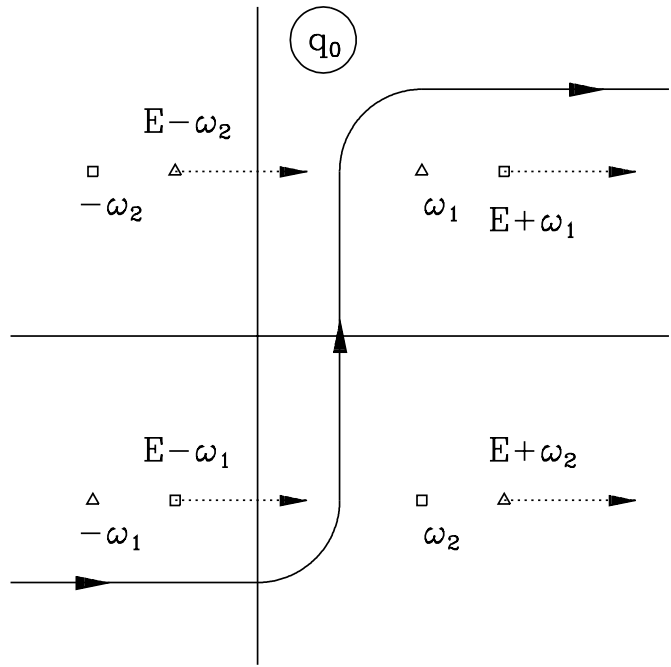}}
\caption{ The complex contour of the integration variable $q_0$.
The triangles are the poles from one of the ghost antighost
contributions. The squares are the poles when the roles of $\omega_1$
and $\omega_2$ are interchanged. As the center of mass energy
$E$ increases, the movement of the poles are also shown by the
arrows. Pinching occurs when $E>(\omega_1+\omega_2)$. }
\label{fig:ch3.cont}
\end{figure}

\begin{figure}[htb]
\vspace{-15mm}
\centerline{ \epsfysize=3.0cm 
             \epsfxsize=5.0cm \epsfbox{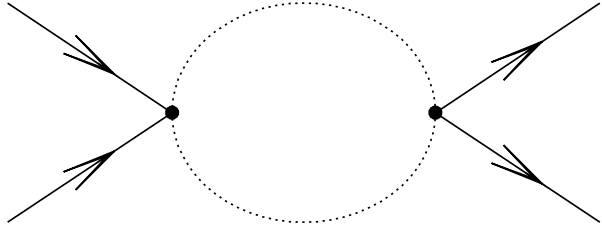}}
\vspace{-15mm}
\caption{ The s-channel one loop amplitude of Goldstone
      Goldstone scattering. The solid lines represent
      incoming and outgoing Goldstone particles. The 
      dashed line can be Goldstone particle, Goldstone
      ghost or antighost particle.}
\label{fig:ch3.oneloop}
\end{figure}

First, we  calculate the $R$-matrix elements to second order 
in bare perturbation
theory of the higher derivative $O(N)$ theory in the broken phase. 
The final large $N$ scattering amplitude can then be formed 
from the geometric resummation of the bubbles and the
large $N$ Higgs propagator.
The calculation in the symmetric phase is quite similar. 
We will parametrize the $R$-matrix elements as 
\be
R_{\alpha\beta}=-(2\pi)^3 \delta^3({\sum \bf p})
                \prod_{ext} \sqrt{ c \over 2\omega V} \cal M_{\alpha
                  \beta} ,
\ee
where the amplitude $\fmab$ is the Feynman amplitude.
The lowest order is trivial namely
\be
\fmab^{(1)} = -8 \lambda_{0} .
\ee 
To the second order, we have to consider the intermediate states
contributions and, in the large-$N$ limit, this reduces to only
$s$-channel scattering. This leads us to the study of the
following one loop contribution, 
\be
\fmab^{(2)} = {96 N^2 \lambda_{0}^2 } 
   \int {d^3 {\bf q} \over {(2\pi)^3}}
   \left( {c_i c_j \over 2 \omega_{i,{\bf q}} 2 \omega_{j,{\bf p-q}}}
      \right) 
{ 2 (  \omega_{i,{\bf q}} + \omega_{j,{\bf p-q}})
    \over
    E^2-( \omega_{i,{\bf q}} + \omega_{j,{\bf p-q}})^2 +i \epsilon } ,
\ee 
in which two types of intermediate states are included, one has
energy $\omega_{i,\bf q}+\omega_{j,\bf p-q}$, the other has energy
$2E+\omega_{i,\bf q}+\omega_{j,\bf p-q}$. 
This form can be brought into the usual loop integral form by using the
identity 
\be
\int_{{\cal C}} {dq_{0} \over 2\pi}
 {1 \over (q_0-E)^2-\omega_1^2 }
      {1 \over q_0^2-\omega_2^2 } = 
{- i \over 2\omega_1 2\omega_2}
 {2(\omega_1+\omega_2) \over E^2-(\omega_1+\omega_2)^2+i\epsilon} ,
\label{eq:ch3.cont}
\ee
where the complex contour is a contour that separates $E-\omega_1$ and 
$-\omega_2$ from $E+\omega_1$ and $\omega_2$ as
shown in Figure~(\ref{fig:ch3.cont}).
However, this type of contour could have some pinching problem
\cite{lee3,polk3}.
The problem only occurs for the 
ghost and antighost pair contribution in
the above equation, i.e. $\omega_1=\omega^{*}_2$.
In order to see how the potential pinching problem occurs,
we have shown 
the movement of the poles in the complex
$q_0$ plane as the center of mass energy increases 
in Figure~(\ref{fig:ch3.cont}) 
The appropriate contour before the pinching is also shown.
It is easy to see that if the center of mass energy $E$ is
less than the so called ghost antighost threshold 
$\omega_1+\omega_2=2 Re\omega_1$, there is no pinching 
and the contour is well defined. As the center of mass energy
increases, four of the eight poles move to the right and 
two of them pinch with $\omega_1$ and $\omega_2$.
One then has to specify how to deform the contour in
the case of pinching. 
\begin{figure}[htb]
\vspace{10mm}
\centerline{ \epsfysize=3.0cm 
             \epsfxsize=5.0cm \epsfbox{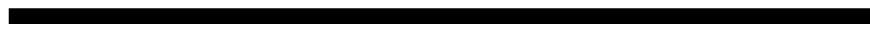}}
\caption{ The complex contour of the integration variable $q_0$
as discussed by Cutkosky et al. 
The pinching is avoided by splitting the ghost and antighost
masses by a small imaginary amount. }
\label{fig:ch3.polk}
\end{figure} 
This type of contour deformation
in the presence of possible pinching was also discussed before by
Cutkosky et al. \cite{polk3}. 
They started directly from the integral representation
and tried to define a contour when the ghost and antighost
masses are not exactly complex conjugate of each other, namely
$M_1-M^{*}_2=i \Delta$, where $\Delta$ is some small parameter.
Then, for every nonvanishing $\Delta$, they were able to find
a suitable contour. The final result is defined to be the limit
where $\Delta \rightarrow 0$.
The corresponding contour is shown in Figure~(\ref{fig:ch3.polk}).
\begin{figure}[htb]
\vspace{10mm}
\centerline{ \epsfysize=3.0cm 
             \epsfxsize=5.0cm \epsfbox{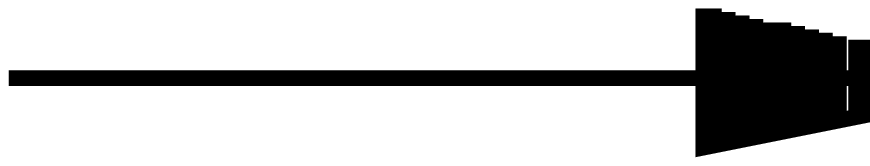}}
\caption{ Our complex contour of the integration variable $q_0$
in the presence of possible pinching.
The pinching is avoided by using the $+i \epsilon$ prescription
which is derived naturally from the Lippmann-Schwinger
formalism of the Hamiltonian. }
\label{fig:ch3.mine}
\end{figure} 
This prescription has a disadvantage that it is 
not justified with any theoretical consideration.
In fact, the Cutkosky prescription is only one of the many ways
to analyticly continue the integral~(\ref{eq:ch3.cont}).
To specify the ``right'' analytic continuation, one would need
some input from the Hamiltonian description of the theory. 
We have started from the Hamiltonian picture of
the theory, so the Hamiltonian should tell us how to
define our contour. Note that pinching only occurs when
the $i \epsilon$ prescription is not applied to the integral.
With the
$i \epsilon$,  however, the pinching is avoided for real value of $s$  and 
we can always find a suitable contour. This contour is shown
in Figure~(\ref{fig:ch3.mine}). It is clear that our contour differs
from the one discussed by Cutkosky et al. and therefore, our final
result is different from theirs. 
Now the Feynman amplitude can be expressed as 
\be
\fmab^{(2)}  = {-i}{96 N \lambda_{0}^2 }\sum_{i,j}  
   \int_{{\cal C}_{ij}} {d^4 {\bf q} \over {(2\pi)^4}}\;\;{c_j \over
q_0^2-\omega_{j,\bf q}^2 +i \epsilon}\;\;
    {c_i \over (q_0-E)^2-\omega_{i,\bf p-q}^2 +i \epsilon} .
\ee 
This amplitude represents the Feynman diagram as shown 
in Figure~(\ref{fig:ch3.oneloop}).
We can shift the integration variable $q_0$ and 
after a Wick rotation we can perform the $q$ integral.  
The integral
itself is finite due to the modification of the propagator at large
momentum and we are left with an integral with Feynman parameter only
\ba 
\fmab^{(2)} &=& {96 N \lambda_0^2 } B(s) , \nonumber\\  
B(s)&=& {-1 \over 16\pi^2}\sum_{i,j} c_i c_j \int_{0}^{1}dx 
\log[xm_i^2+(1-x)m_j^2 -x(1-x)s] . 
\ea 

It is very interesting to study the imaginary part of the bubble
integral. The imaginary part comes only from the angular part of the
argument in the logarithm. The function being quadratic in $x$ has two
roots in the complex $x$ plane which are given by 
\ba
F(s) &\equiv& xm_1^2 +(1-x)m_2^2-sx(1-x) 
      =  s(x-x_1)(x-x_2) , \nonumber \\ 
x_{1,2}&=&{1 \over 2s}(s-m_1^2+m_2^2 
       \pm \sqrt{[s-(m_1+m_2)^2][s-(m_1-m_2)^2]}  
       ) ,
\ea 
where $m_1$ and $m_2$ can take values of three different
masses in the theory. The most important combination is when both are
Goldstone particles. Then, for $s < 0$, we have two conjugate roots whose
real part is exactly $1/2$, therefore, just by symmetry, there is no
imaginary part contribution from this term. This corresponds to the case
that the center of mass energy is less than the threshold. Due to the
massless Goldstone, the lowest threshold is at zero energy.  However, when
$s > 0$ two roots are real and the imaginary part develops, 
we have 
\be
\Im B(s+i\epsilon)= {1\over16\pi} .
\ee
In fact one can show, just from the symmetric form
of the integral, that this is the only imaginary part
contribution to the diagram! 
For example the imaginary part from the
mass pair $\cm$ and $\cm$ exactly cancels the imaginary part from the
mass pair $\cmb$ and $\cmb$ and so on. In the large-$N$ limit
the scattering amplitude is obtained by summing the geometric
chain of the bubbles and the tree contribution from the
intermediate Higgs state
\be
-{N \over 32\pi}A_{00}^{-1}(s)={v^2 \over {s+s^3/M^4}}+
   {1 \over 8\lambda_0}+{N \over 2} B(s) .
\ee  
Taking the imaginary part of this equation, we get the Optical
Theorem in the large-$N$ limit
\be
\Im A_{00}(s)= |A_{00}(s)|^2 .
\ee 
The bubble integral can be exactly worked out. It is a function with
rather complicated analytic structure. 
The detailed discussion is listed in the appendix.

The scattering amplitude
is obtained by taking $s=|s|+i\epsilon$ on the physical sheet.
It is interesting to study the Goldstone-Goldstone scattering cross
section in the large-$N$ limit. 
\begin{figure}[htb]
\vspace{10mm}
\centerline{ \epsfysize=3.0cm 
             \epsfxsize=5.0cm \epsfbox{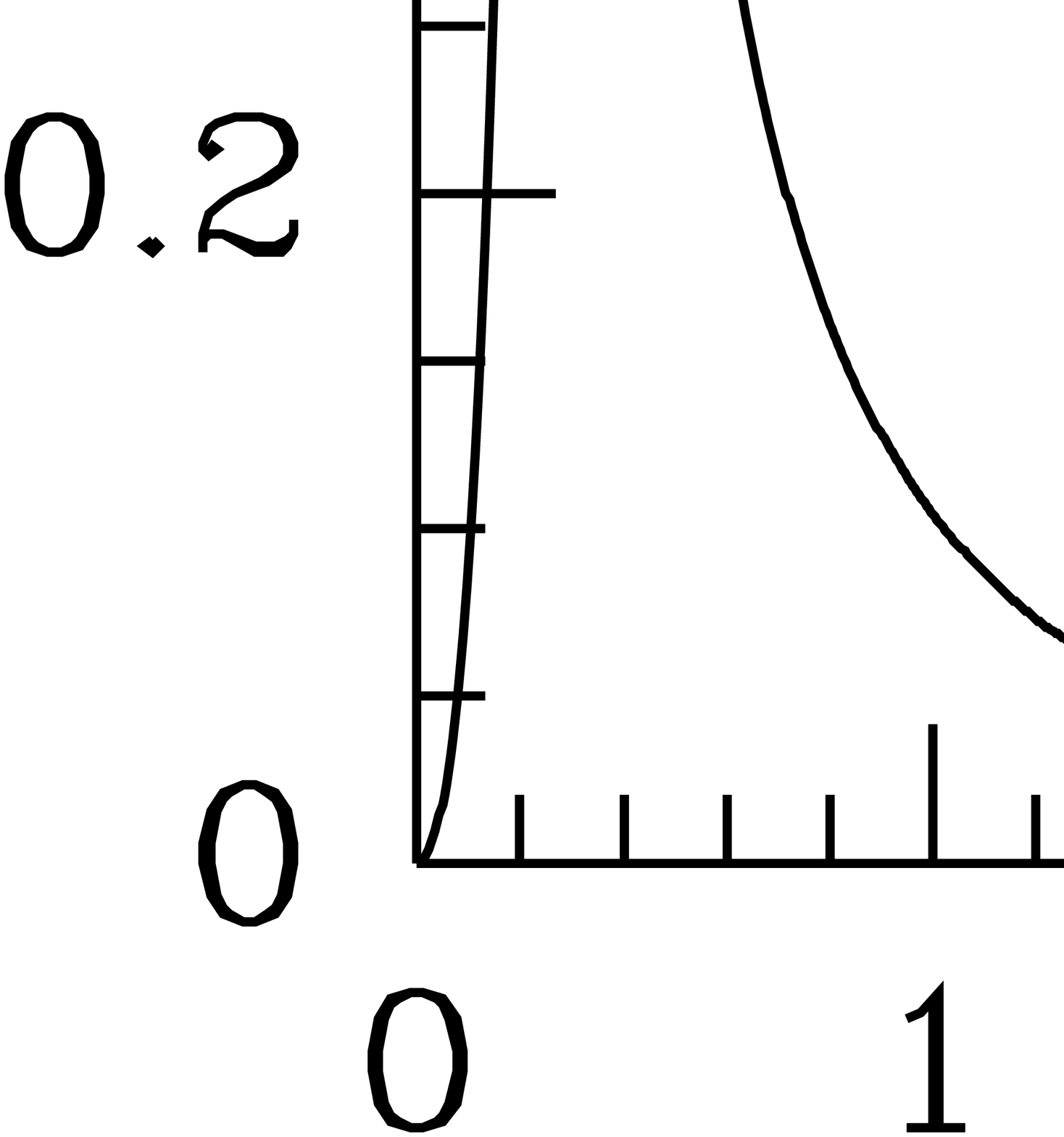}}
\vspace{25mm}
\caption{ The Goldstone Goldstone scattering cross section and phase
shift is plotted against the center of mass energy in large-$N$
expansion for the Pauli-Villars higher derivative
$O(N)$ theory. The input vev value is $v=0.07$ in $M$ unit.
 The peak corresponds to the Higgs resonance, which is at
$m_H=0.28$ in $M$ unit. The scattering cross section is completely
smooth across the so-called ghost pole locations. }
\label{fig:ch3.scatter}
\end{figure} 
In Figure~(\ref{fig:ch3.scatter}) we plotted the cross section
as a function of $\sqrt{s}$ in Goldstone ghost mass unit. Here the
Goldstone ghost pair has a complex phase of $\pi/4$ and the peak
corresponds to the Higgs pole on the second sheet. It is amazing 
that the ghost pair is so well-hidden in the tail of the cross section
that it would be very difficult  
for experimentalists to determine that  there is a ghost pair
hidden somewhere. Also plotted in Figure~(\ref{fig:ch3.scatter})
 is the scattering phase shift
as a function of center of mass energy. We see  that the
phase shift starts out increasing with $\sqrt{s}$ , and at the Higgs pole
it has a sharp rise. If the Higgs particle were infinitely 
narrow then  the rise would be exactly $\pi$. Due to the finiteness of
its width and the Goldstone background, the cross section differs from the
description of Breit-Wigner shape. What is
``unusual'' about this theory is that the phase shift decreases as the
energy gets through the real part of the ghost mass. This is an 
indication of possible acausal behavior in the
scattering, because the sign of $d\delta(s)/ds$ 
determines the relative phase of the scattered wave to the incident
wave. Although for the scattering by a repulsive
potential in the usual theory, this quantity can also be
negative, it would become acausal if this quantity becomes too large,  
 that is, a sharp drop of $\delta(s)$ with respect to $s$. 
In the ordinary theory, this can never happen. 
In the higher derivative theory with the ghost
pair, it could happen if the ghost pair is sufficiently close to the
real axis. It had been argued long ago
by T.~D.~Lee \cite{lee3} that, even in this case, such acausal
behavior would only  occur at microscopic scale
typical of the Compton wave length of the ghost, and it will not lead to
macroscopic disasters. 
In fact, this violation of microscopic causality
is barely visible experimentally.

\section{ Large $N$ Expansion of the Higher Derivative $O(N)$ Model}

The higher derivative $O(N)$ model can be studied in the
large $N$ limit. Many aspects of the theory can be illustrated
in the large $N$ expansion \cite{hhiggs3,dallas3}. 
The general formalism for the large
$N$ expansion has been previously studied \cite{neub3,einh3}. 
Let us  briefly
review the main ideas and focus on its application to the higher
derivative $O(N)$ model, and also emphasize the comparison of the
higher derivative $O(N)$ model with the conventional $O(N)$ model within
the large $N$ approximation.
The Hamiltonian picture that originates from the quantization was
discussed in the last section, so we will
only consider the Euclidean version of the large $N$
approximation here. 
     
\subsection{ General Formalism }

Consider the partition function of the theory as expressed by the
following Euclidean path integral
\ba
{\cal Z}&=& \int {\cal D} \phi e^{-S[\phi]} ,
\nonumber \\
S[\phi]&=& \int d^4 x {1\over 2}\phi^{a} g(-\partial^2) \phi^a
+{\mu^2_0 \over 2}\phi^a\phi^a + {\lambda_0 \over N}(\phi^a\phi^a)^2 ,
\ea 
where the field $\phi^a(x)$ is an $O(N)$ field and the function
$g(-\partial^2)$ is a polynomial function of the operator
$(-\partial^2)$ of the form
\be
g(-\partial^2)=(-\partial^2)+c_4(\partial^4)+c_6(-\partial^6)+\cdots .
\ee
For example, our choice of the Pauli-Villars theory corresponds to
the form of $g(p^2)=p^2+(1/M^4)p^6$.
To perform the large $N$ expansion, it is convenient to introduce the
auxiliary fields $\chi$ such that the path integral is rewritten in the
following form
\be
{\cal Z}= \int {\cal D}\phi{\cal D}\chi
\exp ( -\int d^4 x {1 \over 2}\phi^a (g(-\partial^2) + \mu^2_0 +i \chi)
 \phi^a +{ N \chi^2 \over 4 \lambda_0} ) .
\ee
The effective potential can then be worked out in a standard fashion
\be
U(\bar{\phi}) ={1 \over2 } \bar{\phi}^2 \bar{\chi} -{1 \over 16 \lambda_0}
\bar{\chi}^2 +{\mu^2_0 \over 8 \lambda_0} \bar{\chi} +{N \over 2} 
\int {d^4 k \over (2 \pi)^4} \log(g(k^2)+\bar{\chi})  ,
\ee
where the variable $\bar{\chi}$ is a function of $\bar{\phi}$
determined from the gap equation
\be
\bar{\chi} = \mu^2_0 + {4 \lambda_0 } \bar{\phi}^2
+{ 4 \lambda_0 N } \int {d^4 k \over (2\pi)^4}
{1 \over g(k^2)+ \bar{\chi} } .
\ee
The vacuum can be found from the derivative of the above effective
potential. Due to the gap equation, the derivative has the following
form
\be
U'(\bar{\phi})=\bar{\phi} \bar{\chi} .
\ee
Therefore there could be two phases of the theory. One with
$\bar{\phi} = 0 , \bar{\chi} \neq 0 $ , which is the symmetric
phase; the other one has
$\bar{\phi} \neq 0 , \bar{\chi} = 0 $ , which is the broken phase.
In the broken phase, the vacuum expectation value is obtained via 
\be
0=\mu^2_0+{4 \lambda_0 } v^2 +{4 \lambda_0 N }
\int {d^4 k \over (2\pi)^4} {1 \over g(k^2)} .
\ee
Note that unlike the conventional theory, the integral in the above
equation is finite as long as we have take the highest momentum power
in the propagator to be greater or equal to $6$. 
The critical phase transition line is obtained by setting $v$ to
zero in the above equation, i.e. 
\be
\label{eq:ch3.crline}
0=\mu^2_0+{4 \lambda_0 N }
\int {d^4 k \over (2\pi)^4} {1 \over g(k^2)} .
\ee
As we will see in Chapter~(\ref{ch:BOA}), the large $N$ prediction
of the critical line is in very good agreement with the simulation.

We can work out the propagator of the fields in the large
$N$ approximation. If we are in the symmetric phase, the leading order
correction to the propagator  is just from the mass renormalization.
Therefore, we have 
\ba
\label{eq:ch3.massgap}
< \phi^a(p)\phi^b(p) > &=& { \delta^{ab} \over p^2 + m^2 } ,
\nonumber \\
m^2 &=& \mu^2_0 + 4 \lambda_0 N \int {d^4 k \over (2\pi)^4}
     {1 \over g(k^2) +m^2 } .
\ea
In the broken phase, the Goldstone
propagator remains unchanged to the leading order but the longitudinal
Higgs propagator is modified by the bubble summation of the Goldstone
intermediate states. Thus, we have
\ba
\label{eq:ch3.bubble}
\Gamma_{\sigma\sigma}(p^2)&=& g(p^2) + 
{ 8 \lambda_0v^2 \over 1 + {4 \lambda_0 N } B(p^2) } ,
\nonumber \\
B(p^2)&=& \int { d^4 k \over (2\pi)^4} {1 \over g((p-k)^2)g(k^2) } .
\ea
The scattering amplitude can be worked out in both the symmetric and
broken phase. In the symmetric phase,
\ba
- { N \over 32 \pi} A^{-1}_{00}(p^2) &=& {1 \over 24 \lambda_0}
+ {N+8 \over 6} I(p^2) ,
\nonumber \\
I(p^2)&=&\int {d^4 k \over (2\pi)^4}
{ 1 \over (g((k-p)^2)+m^2)(g(k^2)+m^2) } ,
\ea
where $m^2$ is related to the bare mass parameter
$\mu^2_0$ according to  Equation~(\ref{eq:ch3.massgap}).
If we define the scattering amplitude at $p^2=0$ to be
$-{3N/4\pi}\lambda_R$, we can express the above equation
in terms of the renormalized coupling constant $\lambda_R$
\be
\label{eq:ch3.lambdar}
- { N \over 32 \pi} A^{-1}_{00}(p^2) = {1 \over 24 \lambda_R}
+ {N+8 \over 6} [I(p^2)-I(0)] .
\ee
In the broken phase, we have
\be
- { N \over 32 \pi} A^{-1}_{00}(p^2) = {1 \over 8\lambda_0}
+ {N \over 2} \int {d^4 k \over (2\pi)^4}
{ 1 \over g((k-p)^2)g(k^2) }
+{v^2 \over g(p^2)}  .
\ee
Note that although  we are dealing with the Euclidean 
scattering amplitude here in
large $N$, it should be understood as the amplitude arising from the
Hamiltonian formalism described in the previous section.
As long as the correct complex contour integration is implemented,
and the analytic properties of these amplitudes are understood,
the Euclidean amplitude will also give us the correct physical picture.
 
We can also modify the above formalism to the theory
on the lattice in a finite volume. All we have to do is to change the 
integral into finite lattice summations.

Let us now show some examples of the application of the large $N$
results and see what we can learn from it.

\subsection{ Renormalized Coupling Constant in the Symmetric Phase}

In the first example, we compare the higher derivative $O(N)$
model on the lattice and the conventional $O(N)$ model on the lattice
in the symmetric phase. As described in Equation~(\ref{eq:ch3.lambdar}), the
renormalized coupling constant $\lambda_R$ can be defined as
\ba
\lambda_R&=&{ \lambda_0 \over 1+ {4 \lambda_0 }(N+8)I(0) } ,
\nonumber \\
I(0)&=& {1 \over V} \sum_{k} { 1 \over (g(k^2)+m^2)^2 } ,
\ea 
where everything is measured in lattice unit. 
In this formula, the factor $(N+8)$ is the exact
group theory factor. However, in the naive large $N$ approximation,
 we should use $N$ instead of $(N+8)$. When we apply this to
the $O(4)$ model, this makes a factor of $3$ difference.
We therefore have to conclude that the leading order large
$N$ results are ambiguous when applied to $N=4$.
As we will see in the next example,  similar situation
occurs for the broken phase.
We can calculate this 
renormalized coupling constant at 
the same correlation length $\xi \equiv 1/m$ in the conventional $O(N)$ model
and in the Pauli-Villars theory, with some fixed
value of $M$, for every value of the bare coupling constant
\begin{figure}[htb]
\vspace{15mm}
\centerline{ \epsfysize=3.0cm
             \epsfxsize=5.0cm 
             \epsfbox{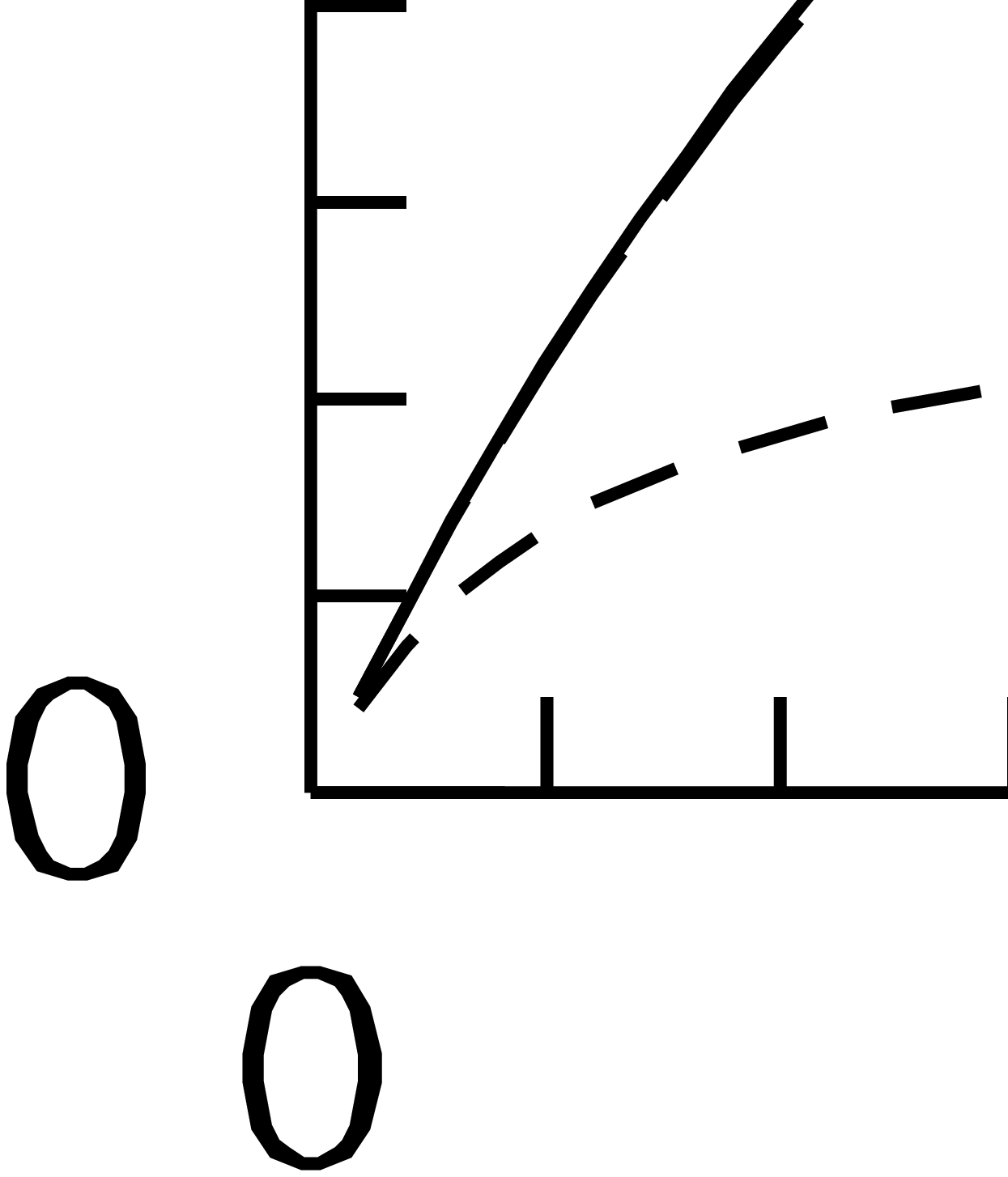} }
\vspace{17mm}
\caption{ The comparison of the large $N$ renormalized 
coupling constant in the symmetric
phase is shown for three cases: continuum Pauli-Villars, lattice
Pauli-Villars and the conventional $O(N)$ model. For this choice of the
correlation length, the lattice effects are small and the Pauli-Villars
theory shows much stronger interaction when compared with the
conventional $O(N)$ model. We have modified the naive large $N$
formula  so that  the right group theory factor is substituted, 
 i.e. $N+8=12$. }
\label{fig:ch3.symmN}
\end{figure}
$\lambda_0$. The magnitude of this quantity reflects the strength of
the interaction in the symmetric phase. 
In Figure~(\ref{fig:ch3.symmN}) , the comparison
between the two theories is shown for $m=0.3,M=1.0$ for every given
$\lambda_0$. The lattice summation is calculated on a $32^4$ lattice
with the naive lattice discretization of the Laplacian.
The continuum Pauli-Villars result is also shown in the figure.
For this choice of the correlation length, the lattice effects are
rather small in  both theories. It is clear that
the Pauli-Villars theory has stronger interaction, 
about a factor of 4, for the same 
correlation length when compared with the conventional $O(N)$ model.
As mentioned above, the large $N$ expansion has its own ambiguities, 
so we do not anticipate this large $N$ result to give us a precise
quantitative description of the theory. However,  we do expect that the 
increase of the coupling constant in the Pauli-Villars theory
{\em{relative}} to the conventional theory should also be present 
in the full theory. Similar
results in the broken phase also support this picture, 
as we will see in the next subsection.
 
\subsection{Higgs Mass and Width for Conventional $O(N)$ Model } 

Now, we will examine the situation in the broken phase of the theory.
We will first briefly review the large $N$ result for the conventional
$O(N)$ model with a hypercubic lattice regulator \cite{neub3,einh3} in
the broken phase. The large $N$ Goldstone propagator will remain the
free propagator in the leading order of $1/N$ expansion, but the 
large $N$ Higgs propagator will be
\ba
\label{eq:ch3.bublat}
\Gamma_{\sigma\sigma}(p^2)&=& \hat{p}^2 + 
{ 8 \lambda_0v^2 \over 1 + {4 \lambda_0 N } B(p^2) } ,
\nonumber \\
B(p^2)&=& \int { d^4 k \over (2\pi)^4} {1 \over \widehat{(p-k)}^2 \hat{k}^2 }
.
\ea
In general, the lattice bubble is  a very 
complicated function of $p$. However in the
limit of small $p^2$, which is the regime of
physical interests, it has been worked out and has the following
asymptotic expression \cite{lusc3}
\be
B(p^2)={1 \over 16\pi^2} ( -\log(p^2) + c_{latt} + {\cal O}(p^2) ) ,
\ee
where the constant $c_{latt}=5.79200957$ for the hypercubic lattice.
We can now  use this relation to find the complex pole to the Higgs
propagator. Since the higher order terms  are 
neglected in the bubble, it is sufficient to keep
only the leading term in the lattice momentum.
Setting $(-p^2)=s=(m_H-i\Gamma_H /2)^2=r^2e^{-2i\theta}$, we have (taking the
second Rieman sheet value for the logarithm)
\ba
r&=&\sqrt{ {32\pi^2 v^2 \over N}{\sin 2\theta \over \pi+2\theta} } ,
\nonumber \\
\cos 2\theta &=& {\sin 2\theta \over \pi+2\theta} 
\left( {4\pi^2 \over \lambda_0 N} + c_{latt}      
-\log{32\pi^2 v^2 \over N}
-\log{\sin 2\theta \over \pi+2\theta}  \right) .
\ea
The phase $\theta$ is first determined from the second equation 
and then substituted into the first one to get the real and
imaginary part of the Higgs pole.
\begin{figure}[htb]
\vspace{13mm}
\centerline{ \epsfysize=3.0cm
             \epsfxsize=5.0cm
             \epsfbox{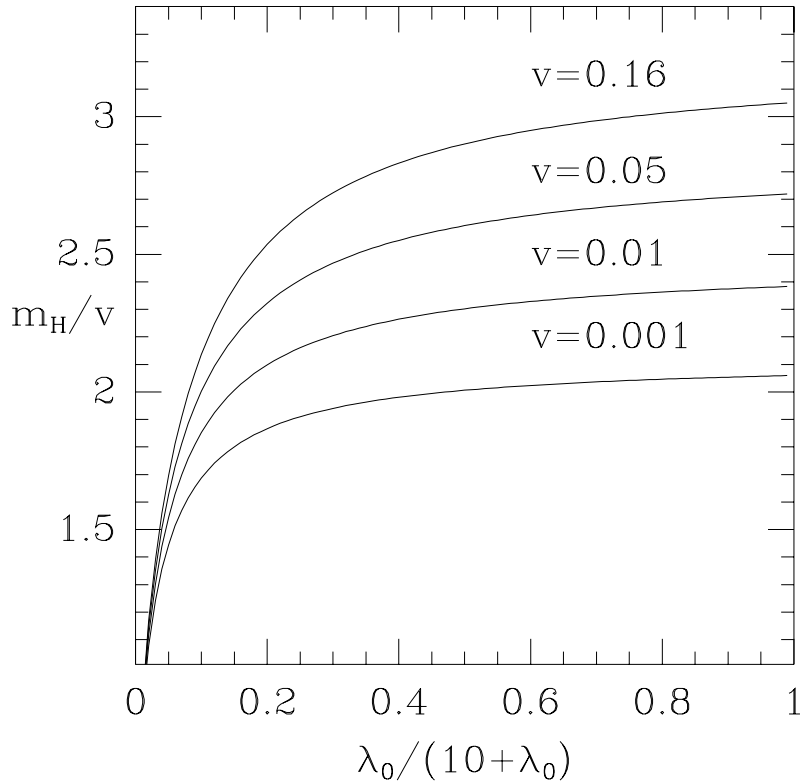} }
\vspace{17mm}
\caption{ The large $N$ result of the Higgs mass over vev ratio 
$m_H/v$ as a function of the bare coupling constant for the conventional
$O(N)$ model with a hypercubic lattice regulator. Four curves
correspond to different $v$ values (in lattice units) as indicated. 
$N$ has been set to $4$ in the calculation. } 
\label{fig:ch3.bNlat}
\end{figure} 
The result is summarized in Figure~(\ref{fig:ch3.bNlat}) . For the Higgs
correlation length of about $2$, the ratio $m_H /v $ is only about
$3$. The correlation length $2$ is chosen because if the Higgs is
too heavy in lattice units, the lattice effects would become 
significant and the theory  would no longer
describe continuum physics \cite{lusc3}.

Another feature that we can study is the width of the Higgs particle.
There has been quite a lot of confusion even with the conventional
$O(N)$ model. The large $N$ width of the $O(N)$ model was first 
carefully studied by Einhorn using a sharp momentum cutoff \cite{einh3}.
He found that the large $N$ formula, if $N=4$ is substituted in,  
gives too large a width (40 percent larger) when compared
with the perturbation theory of the $O(N=4)$ model.
Therefore, the large $N$ results seem to  indicate
that the theory is more strongly interacting than the perturbative
predictions. Some authors interpret this finding 
as genuine nonperturbative effects of the model \cite{neub3}.
However, we do not think this is true for the following
two reasons. First of all, the large $N$ result of the width does not
agree with perturbation theory, even for very weak couplings. In this 
regime, the next to leading term of the width has been calculated in
perturbation theory. The correction is very small, typically of the 
order of one percent. Therefore, it is unlikely that even higher order
terms will change this perturbative result significantly. Secondly, the
perturbative result has been proven to be correct by extensive
nonperturbative Monte Carlo simulation studies. No mysterious 
nonperturbative effects as predicted by large $N$ have been found.

We believe this discrepancy is because of the ambiguity
within the large $N$ framework. The large $N$
result can only be accurate to about 20 to 30 percent because of the
large subleading $1/N$ terms at $N=4$. After all, $N=4$ is too
far from $N=\infty$. This has been previously pointed out by
Kuti, et. al. \cite{kutiN3}. 
More importantly, we have over
estimated the decay channel of the Higgs particle  
in the naive large $N$ formula.
At $N=4$, the Higgs particle can decay into $3$ colors of the Goldstone
pairs while the large $N$ formula counts $4$. When we take this into
account and substitute $(N-1)$ for $N$ in the large $N$ formula, 
we expect compatible results with the perturbation theory.
\begin{figure}[htb]
\vspace{13mm}
\centerline{ \epsfysize=3.0cm
             \epsfxsize=5.0cm
             \epsfbox{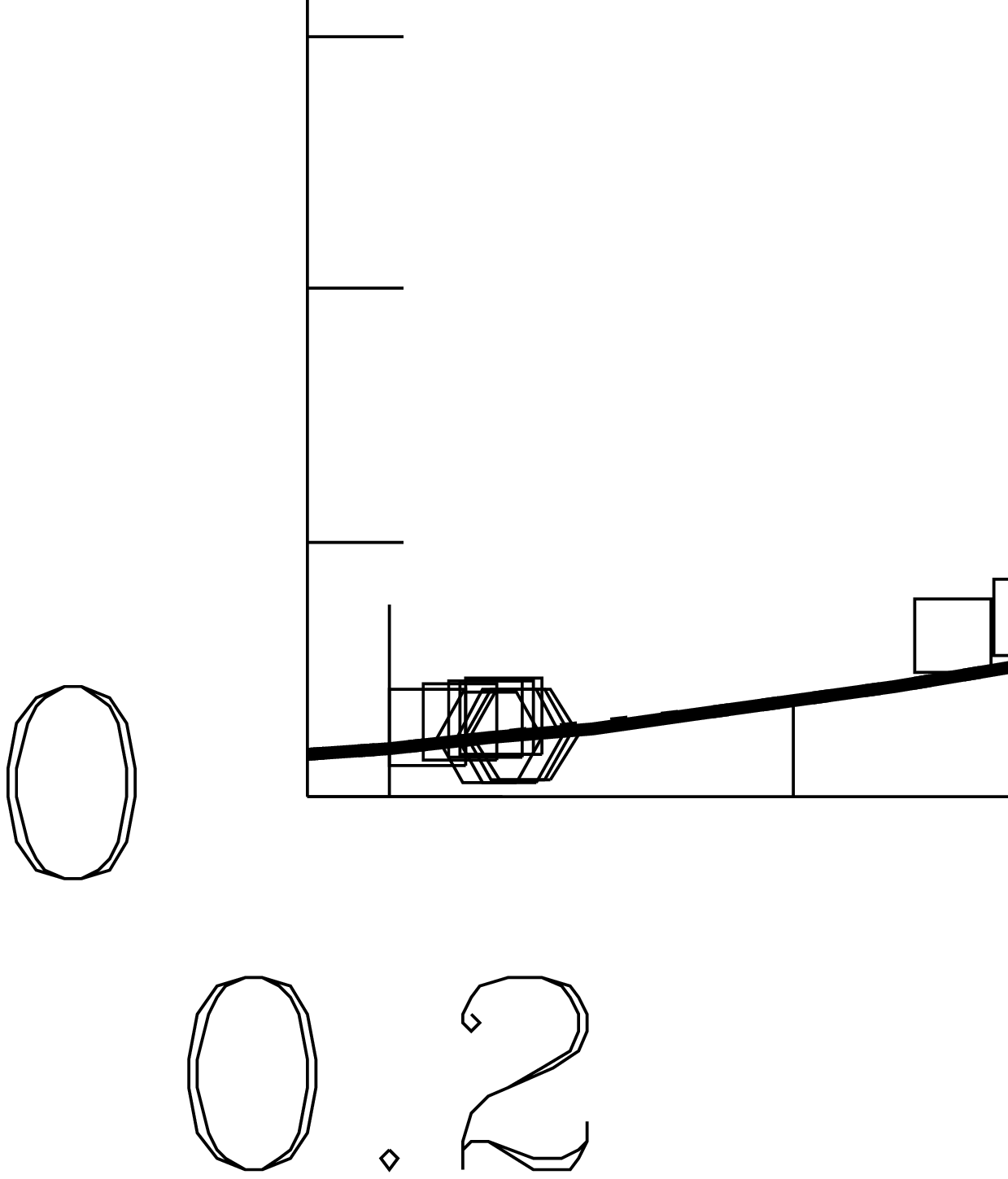} }
\vspace{17mm}
\caption{ The large $N$ results for the width of the Higgs particle as
a function of the Higgs mass is shown. The open squares are the naive
large $N$ prediction for $O(4)$ model. The open hexagons are the large
$N$ results after the number of decay channels has been corrected. The solid
line is the leading order perturbation result and the dashed line is
the perturbation result up to the second order. The corrected large 
$N$ width agrees with the perturbative prediction very well in the
weakly interacting regime as it should. The naive large $N$ result
overshoots by about $30$ to $40$ percent. } 
\label{fig:ch3.latwid}
\end{figure}
In Figure~(\ref{fig:ch3.latwid}), 
we have plotted the two large $N$ results of the width 
as a function of the Higgs mass and compared them with the
perturbative results. As expected, the corrected large $N$ width
agrees reasonably well with the perturbative predictions, especially 
in the weakly interacting region or small Higgs mass. The naive large
$N$ result, however, overshoots by about $30$ to  $40$ percent simply
because it 
fails to identify the correct number of decay channels of the Higgs
particle. From this calculation, we conclude that the large $N$
approximation has its own ambiguities when it is applied to 
$N$ values that are not very large. Therefore, one must modify the naive 
large $N$ formula in order to get meaningful quantitative results. 

\subsection{ Higgs Mass and Width for Higher Derivative $O(N)$ Model } 

In higher derivative theory, things are getting more
complex because of the ghost pair.
One could try to evaluate the continuum bubble integral in 
Equation~(\ref{eq:ch3.bubble}) and solve for the complex pole
of the Higgs propagator. Note that this integral is
finite and no regulator has to be introduced.
\begin{figure}[htb]
\vspace{10mm}
\centerline{ \epsfysize=3.0cm
             \epsfxsize=5.0cm
             \epsfbox{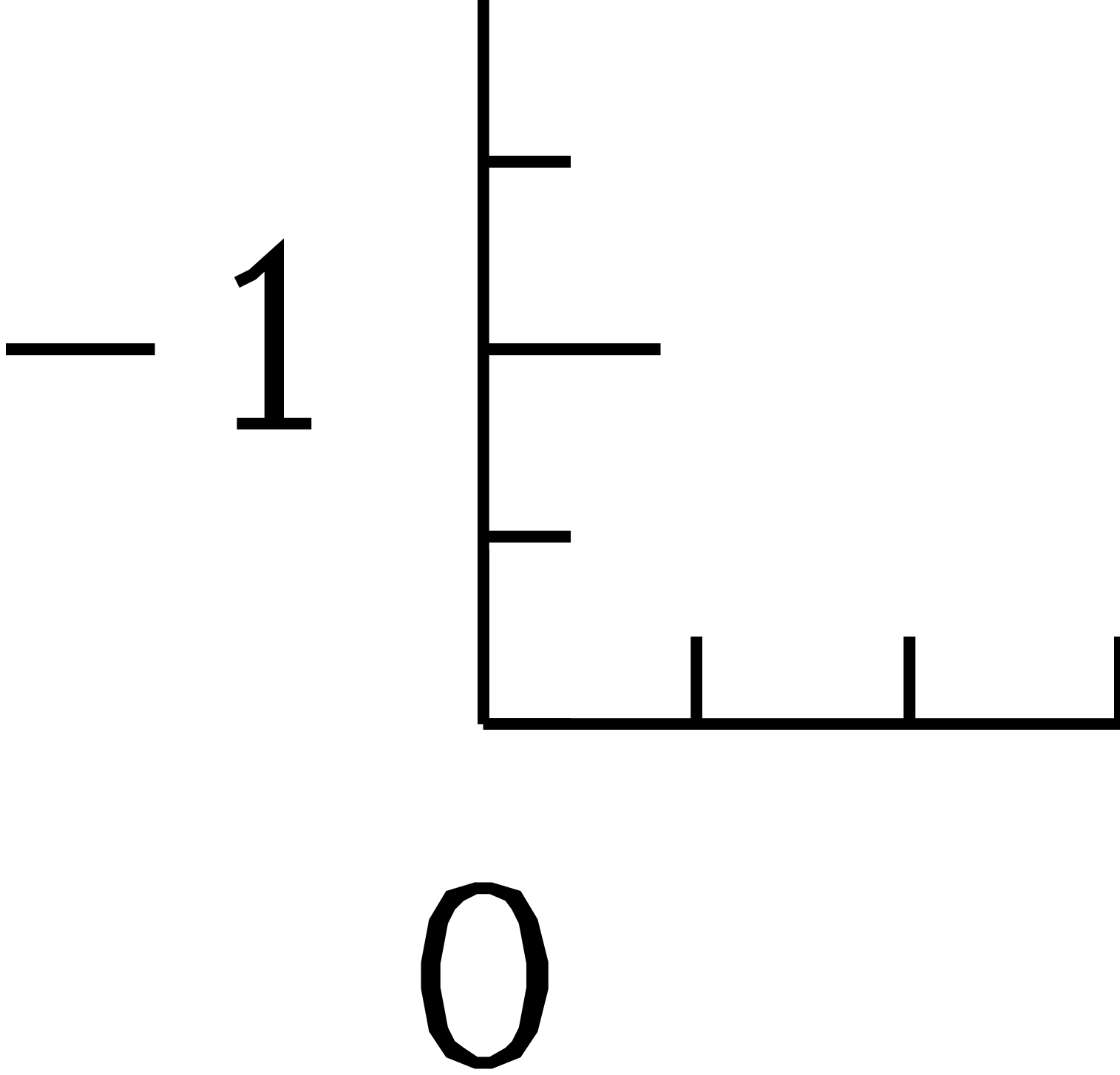} }
\vspace{15mm}
\caption{ The complex poles of the large $N$ Higgs propagator is shown
on the first and the second Riemann sheets. The bare coupling constant
is set to infinity in this figure. The open hexagonal points represent 
the ghost pair poles on the first Riemann sheet. The filled hexagonal
points are the 'image' of the ghost on the second Riemann sheet. The
filled circles are the Higgs poles on the second sheet. The size of the
points reflects the different $v$ values. }
\label{fig:ch3.polesn}
\end{figure}
The precise form of the bubble integral
is very complicated and is listed in the Appendix of this chapter.
The analytic structure ( Riemann sheets and cuts ) are also
quite complex, as described in the Appendix.
The resulting function is then substituted into the full large 
$N$ Higgs propagator to solve for the poles. The complex pole structure of
the function is also very complicated due to the existence of the
ghost states. 
The poles
are numerically searched for a given parameter $v$ in $M$ unit 
and a fixed value of the bare coupling constant $\lambda_0$.
The result is shown in Figure~(\ref{fig:ch3.polesn}).
One has to be careful with the Riemann sheet 
structure of the function in order to get the right result. 
The poles are characterized by their positions on the Riemann sheets. On
the first Riemann sheet, due to the ghost states, one finds a conjugate
pair of poles represented by the open hexagonal points in the figure.
They are moving towards the higher energy values as the interaction is
turned on. These complex conjugate ghost pairs have 
``shadow images'' on the second Riemann
sheet which are represented by the filled hexagonal points. Because
of the interaction with the Higgs pole on the second sheet, these poles
are not moving symmetrically. The conventional Higgs poles are
on the second Riemann sheet, represented by the filled circles. As the
vacuum expectation value
is increases  in $M$ unit, the Higgs pole is moving
towards the higher energy range. When the Higgs pole is at very low
energy and far away from the ghost poles, the effects of the ghost
states can be viewed as an effective cutoff to the conventional theory.
This can also be seen in Equation~(\ref{eq:ch3.bubble}). When $p^2$ is
small, the bubble integral becomes very simple and can be very well
approximated by
\be
B(p^2) \sim {1 \over 16\pi^2}( -\log(p^2) + 1/2) .
\ee
But, if the Higgs pole is getting closer to the energy scale of 
the ghost poles, the higher
derivative theory feature has great importance and viewing the ghosts as
the effective cutoff to the conventional theory  becomes meaningless. 

\begin{figure}[htb]
\vspace{13mm}
\centerline{ \epsfysize=3.0cm
             \epsfxsize=5.0cm
             \epsfbox{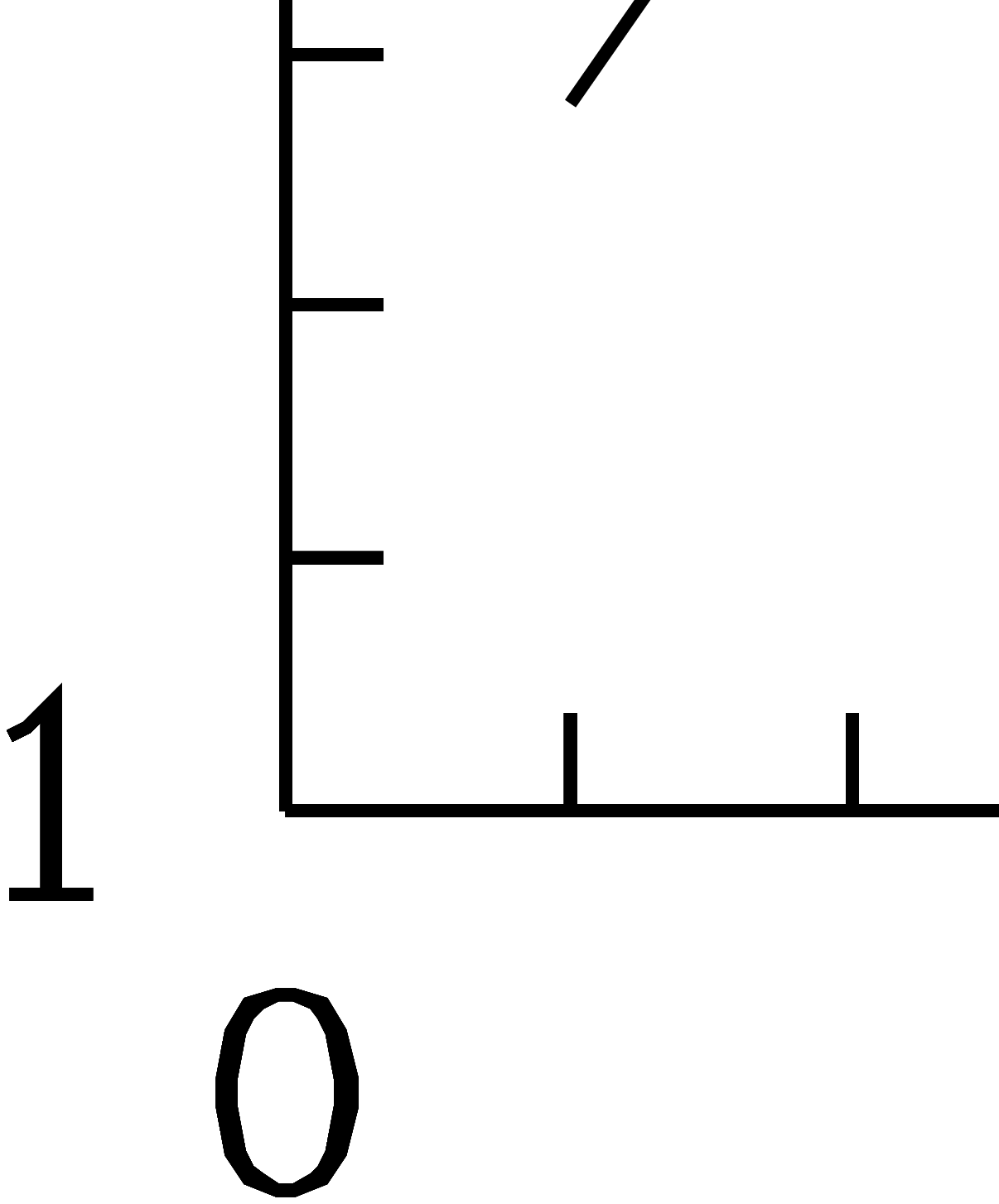} }
\vspace{15mm}
\caption{ The large $N$ result of the ratio $m_H/v$ as a function of
the bare coupling constant for various values of the vacuum expectation
value (measured in $M$ units) for the higher
derivative $O(N)$ theory. The maximum ratio saturates to about
$4$ at infinite bare coupling constant. } 
\label{fig:ch3.bn}
\end{figure}

Identifying the real part of the Higgs pole with the mass parameter
and the imaginary part with the half width, we can plot the ratio
$m_H/v$ as a function of the bare coupling constant, which is shown
in Figure~(\ref{fig:ch3.bn}). 
In this figure, we have selected $4$ different vev 
values and the ratio saturates to about $4$ when the bare coupling
constant is brought to infinity. When we set the physical value of the 
vacuum expectation value to $250$ GeV, this implies a Higgs particle
with the mass $m_H=1$ TeV. 
This should be compared with the result of the conventional $O(N)$ model
discussed earlier in Figure~(\ref{fig:ch3.bNlat}).
Although the absolute values of the 
Higgs mass may be somewhat ambiguous due to the large $N$
approximation,
this result indicates there is a $30$ percent {\em{relative}} increase in
the Higgs mass over vev ratio when the Pauli-Villars theory is
compared with the conventional $O(N)$ model on the hypercubic lattice.
So we would expect the full Pauli-Villars theory should also
generate a larger $m_H/v$ ratio compared with the conventional
theory. Recall that the Higgs mass bound for the conventional
theory is about $750$ GeV (which is a ratio of $3$), we 
expect the Pauli-Villars theory could have a heavy Higgs particle
in the TeV range. In fact, this hint from the large
$N$ expansion initiated our nonperturbative study of the
Pauli-Villars theory \cite{hhiggs3}. 
As we will see in the coming chapters, this
scenario of strongly interacting Higgs sector in the Pauli-Villars
theory is confirmed by our nonperturbative simulation results. 

We can plot the width of the Higgs particle  as a function
of the Higgs mass, just as we did for the conventional theory.
\begin{figure}[htb]
\vspace{10mm}
\centerline{ \epsfysize=3.0cm
             \epsfxsize=5.0cm
             \epsfbox{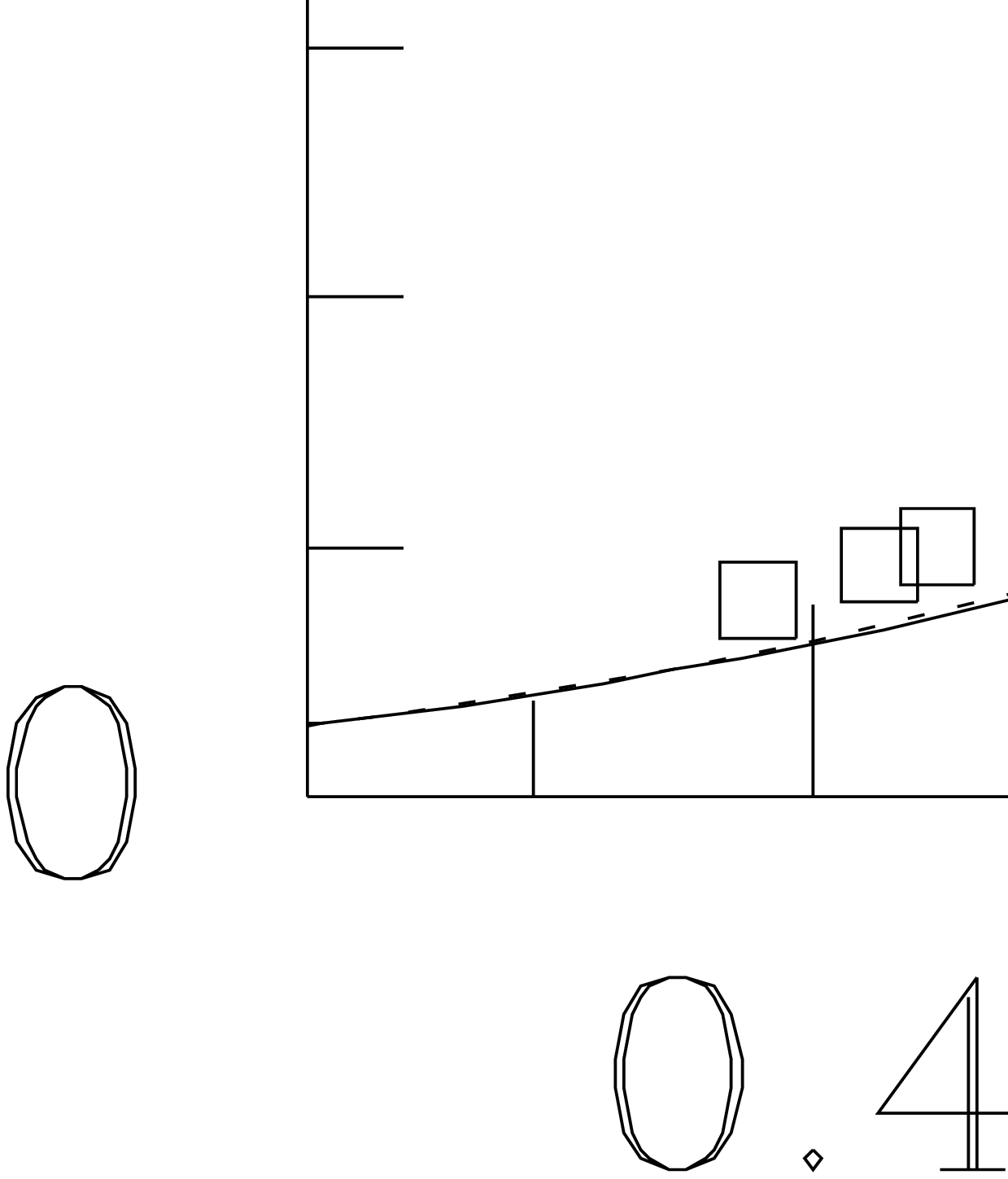} }
\vspace{17mm}
\caption{ The large $N$ result for the width of the Higgs particle as
a function of the Higgs mass is shown in the Pauli-Villars
higher derivative $O(N)$ theory. The open squares are the naive
large $N$ prediction at $N=4$. The open hexagons are the large
$N$ results after the number of decay channels has been corrected. The solid
line is the leading order perturbation result and the dashed line is
the perturbation result up to the second order. The corrected large 
$N$ width agrees with the perturbative prediction very well in the
weakly interacting regime as it should. The naive large $N$ result
overshoots by about $30$ to $40$ percent. } 
\label{fig:ch3.pvwid}
\end{figure}
In Figure~(\ref{fig:ch3.pvwid}), 
the similar plot for the higher derivative
Pauli-Villars
theory is shown. One has to again 
modify the naive large $N$ results for
the right decay channels. At very low energy, this result agrees with
the perturbative result, which means that the theory can be viewed as
a Pauli-Villars regulated conventional theory. However, when the
Higgs mass is getting heavier, it deviates quite rapidly from the perturbative
result, and in this range, the higher derivative theory
does not resemble a regulated conventional theory.

\section{ Perturbation Theory of the Higher Derivative $O(N)$ Model}

The renormalized perturbation theory of the higher derivative
scalar $O(N)$ model can be established in the usual way, except we 
pay special attention to the role 
of the ghost pair. In the low energy regime, we expect to
recover the conventional theory. When the energy scale is increased,
one should see the ghost pair begins to play a more important role.
To incorporate this energy dependence, a mass dependent renormalization
scheme is needed.  We now illustrate this briefly by considering the model
in the broken phase. 

\subsection{ Lagrangian and Renormalization Conditions }

Let us consider the following Euclidean Lagrangian,
\ba
{\cal L}_E &=& {1 \over 2} \phi^a (-\Box
          - {\Box^3 \over M^4} ) \phi^a 
          -{1 \over 2} \mu^2 \phi^a\phi^a 
           +\lambda  (\phi^a \phi^a)^2  ,
\nonumber \\
\delta{\cal L}_E &=& {\delta Z_1 \over 2} \phi^a (-\Box)\phi^a 
      +{\delta Z_3 \over 2} \phi^a (-{\Box^3 \over M^4})\phi^a 
          -{\delta \mu^2 \over 2} \phi^a\phi^a 
           +\delta \lambda  (\phi^a \phi^a)^2  .
\ea
We can define the bare fields and bare parameters according to 
\ba
\phi^a_0&=& Z^{1/2}_1 \phi^a,\;\;\; Z_1=1+\delta Z_1,
\nonumber \\
M^{-4}_0&=&{Z_3 \over Z_1} M^{-4},\;\;\; Z_3=1+\delta Z_3,
\nonumber \\
\mu^2_0&=& (\mu^2 +\delta \mu^2)/Z_1 ,
\nonumber \\
\lambda_0&=& (\lambda +\delta \lambda)/Z^2_1 .
\ea 
Then, the total Lagrangian can be written as
\be
{\cal L}_{E0} = {1 \over 2} \phi^a_0 (-\Box
          - {\Box^3 \over M^4_0} ) \phi^a_0 
          -{1 \over 2} \mu^2_0 \phi^a_0\phi^a_0 
           +\lambda_0  (\phi^a_0 \phi^a_0)^2  .
\ee
In the broken phase, it is convenient to separate the Higgs field and
the Goldstone fields as
\be
\phi^a=\left( \begin{array}{cccccc}
              \pi^1(x) \\
                .      \\
                .      \\
                .      \\
              \pi^{N-1}(x) \\
              v+\sigma(x)
              \end{array} \right)
\ee
where $v=\mu^2/4\lambda$ is the renormalized vev. We can then 
write down various propagators to one loop order and  impose the 
following mass dependent renormalization conditions,
\ba
\delta v&=& 0,   
\nonumber \\
{ d \over dp^2} \Gamma^{\pi\pi}(p^2) |_{p^2=0} &=& 1,
\nonumber \\
({ d \over dp^2})^3 \Gamma^{\pi\pi}(p^2) |_{p^2=M^2} &=& M^{-4},
\nonumber \\
\Gamma^{\sigma\sigma}(\kappa^2) &=& 
Z_1\kappa^2+Z_3\kappa^6/M^4+ m^2(\kappa).
\ea 
Notice that the above renormalization conditions uniquely determines the
four renormalized parameters. The arbitrary scale $\kappa$ is introduced
to avoid the infrared divergences. All the renormalized parameters will
depend on this running scale through the above definitions. It is easy 
to fix the counter terms according to the above equations,
\ba 
\delta Z_1&=& {\lambda m^2} B'_{\sigma\pi}(0),
\nonumber \\
M^{-4}\delta Z_3 &=& {4 \lambda m^2 \over 3}B'''_{\sigma\pi}(M^2),
\nonumber \\
{\delta \lambda \over \lambda^2} &=&
 36 B_{\sigma\sigma}(\kappa^2)+4(N-1) B_{\pi\pi}(\kappa^2), 
\nonumber \\
{\delta m^2 \over m^2}&=&{\delta \lambda \over \lambda}
 +{T^{\sigma} \over m^2}+{N-1 \over 3}{T^{\pi} \over m^2},
\ea
where the bubble integrals $B_{\sigma\sigma}$,$B_{\sigma\pi}$,
,$B_{\pi\pi}$ and the tadpoles $T^{\sigma}$,$T^{\pi}$ are listed
below:
\ba
B_{\sigma\sigma}(p^2)&=&\int {d^4 k \over (2\pi)^4} 
{1 \over (k^2+m^2+k^6/M^4)((k-p)^2+m^2+(k-p)^6/M^4) },
\nonumber \\
B_{\sigma\pi}(p^2)&=&\int {d^4 k \over (2\pi)^4} 
{1 \over (k^2+m^2+k^4/M^4)((k-p)^2+(k-p)^6/M^4) },
\nonumber \\
B_{\pi\pi}(p^2)&=&\int {d^4 k \over (2\pi)^4} 
{1 \over (k^2+k^4/M^4)((k-p)^2+(k-p)^6/M^4) },
\nonumber \\
T^{\sigma}&=&\int {d^4 k \over (2\pi)^4} 
{1 \over k^2+k^4/M^4+m^2 },
\nonumber \\
T^{\pi}&=&\int {d^4 k \over (2\pi)^4} 
{1 \over k^2+k^4/M^4}.
\ea

\subsection{ One-loop Mass-dependent Beta-functions }

Now we can  work out the one loop mass dependent $\beta$-function of
the theory, which is obtained by noticing that the bare coupling
constant $\lambda_0$ does not depend on the renormalization scale
$\kappa$. The result can be written in the following form,
\ba
{1 \over \lambda}\beta_{\lambda}&=& {9\lambda \over 2\pi^2}
\left( \sum^2_{i,j=0} a_i a_j \int^1_0 dx
 {x(1-x)\kappa^2 \over x(1-x)\kappa^2+x\lambda_i+(1-x)\lambda_j} \right.
\nonumber \\
&+& \left. {N-1 \over 9} \sum^2_{i,j=0} b_i b_j \int^1_0 dx
 {x(1-x)\kappa^2 \over x(1-x)\kappa^2+x\xi_i+(1-x)\xi_j} \right),
\ea
where $\hat{\kappa}^2=\kappa^2/M^2$ and  the 
mass parameters $\lambda_i$, $\xi_i$ and their
corresponding residues are determined from the following decomposition:
\ba
{1 \over k^2 +k^6+(m/M)^2} & \equiv &
{1 \over k^2+\lambda_0}+{1 \over k^2+\lambda_1}+{1 \over
k^2+\lambda_2} ,
\nonumber \\
{1 \over k^2 +k^6} & \equiv &
{1 \over k^2+\xi_0}+{1 \over k^2+\xi_1}+{1 \over
k^2+\xi_2} .
\ea
The important feature of these coeffecients is 
\be
\sum^2_{i=0}a_i=\sum^2_{i=0}b_i=0 .
\label{eq:ch3.sum0}
\ee
Now it is easy to see how the effective coupling constant evolve with the
energy scale $\kappa$. When $\kappa/M$ is very small, the ghost pair 
contributions to the beta-function is negligible. The summation in the
beta-function reduces to only the $i=j=0$ contribution, which is the
conventional, well-known beta-function of the $O(N)$ model in the broken
phase. As $\kappa/M$ increases, the ghost contributions become
increasingly important. When the energy scale is well above the ghost scale,
the integral in the beta-function reduces to $1$ and the quantity in the
bracket vanishes due to Equation~(\ref{eq:ch3.sum0}). This means that, at
high energies, beta-function of the theory vanishes. Therefore, as
the energy scale increases,  the
running coupling constant $\lambda(\kappa)$ also increases. However, at
the scale of the ghost pair or higher, 
the coupling constant gradually flattens out to some finite number.
This is a very different feature when compared with the conventional
$O(N)$ model. In the conventional $O(N)$ model, the running coupling
constant keeps increasing and  becomes divergent at the so-called
Landau ghost energy scale. In our higher derivative theory, we have
replaced the Landau ghost with real ghost pair and the running coupling
constant will remain finite for all energies.

\section{Appendix}

In this appendix we list the explicit form of the bubble integral 
and discuss some analytic properties of such.

\begin{figure}[htb]
\vspace{10mm}
\centerline{ \epsfysize=3.0cm 
             \epsfxsize=5.0cm \epsfbox{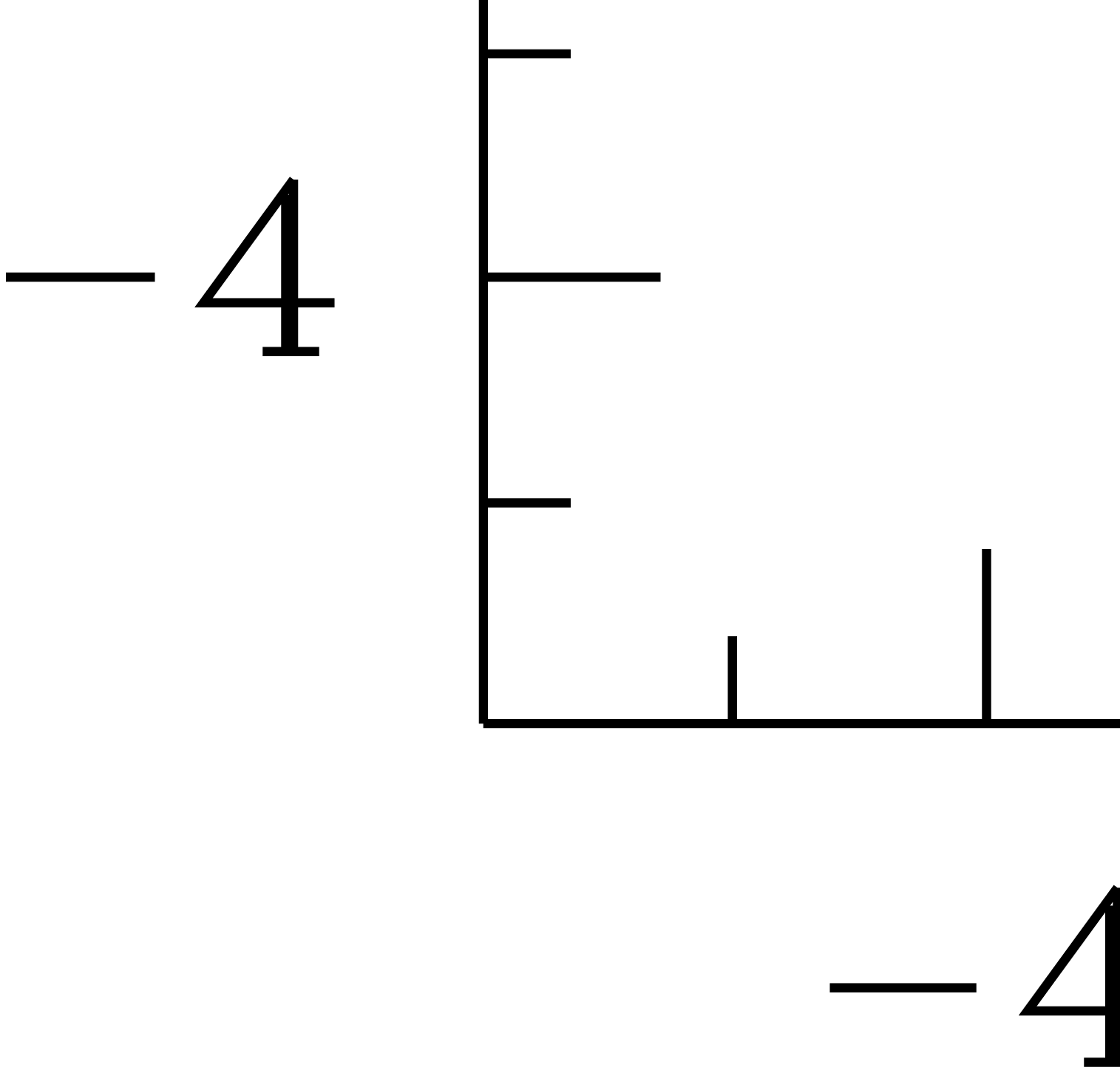}}
\vspace{15mm}
\caption{ The Rieman sheet structure of the function $B(s)$ is shown.
The filled squares represent the starting points of different cuts.
The hyperbola cut is due to the ghost-antighost contribution $f(s)$.}
\label{fig:ch3.sheet}
\end{figure}

The function is given by the parametric integral representation 
\be 
B(s)= {-1 \over 16\pi^2}\sum_{i,j} c_i c_j \int_{0}^{1}dx 
\log[xm_i^2+(1-x)m_j^2 -x(1-x)s] ,
\ee 
where the sum over $i$ and $j$ runs from $0$ to $2$ with the following
values of $c_i$ and $m_i^2$
\ba
m_{0}^2=0 , && c_{0}=1 ,  \\
m_{1}^2=e^{+2i\Theta} , 
&& c_{1}={-ie^{-2i\Theta} \over 2\sin2\Theta} , \nonumber\\
m_{2}^2=e^{-2i\Theta} ,  
&& c_{2}={+ie^{+2i\Theta} \over 2\sin2\Theta} . \nonumber
\ea
The integral can be worked out explicitly  with the result 
\ba 
B(s)&=&\!\!\!-{1 \over 16\pi^2}\;\left\{\;c_0^2
               \;\log(-s) \right.\nonumber\\
\!\!\!\!\!\!\!\!\!\!\!\!&+&\!\!\!c_1^2\left[
               +2i\Theta-{\sqrt{(1-s/4\cm^2)(-s/4\cm^2)}
               \over 2(s/4\cm^2) } 
 \log(
               { {\sqrt{1-s/4\cm^2} + \sqrt{-s/4\cm^2} }
               \over   
                {\sqrt{1-s/4\cm^2} - \sqrt{-s/4\cm^2} } })^2 \right] 
               \nonumber \\
\!\!\!\!\!\!\!\!\!\!\!\!&+&\!\!\!c_2^2 \left[
               -2i\Theta-{ \sqrt{(1-s/4\cmb^2)(-s/4\cmb^2)}
               \over 2(s/4\cmb^2) } 
\log(
               { {\sqrt{1-s/4\cmb^2} + \sqrt{-s/4\cmb^2} }
               \over   
               {\sqrt{1-s/4\cmb^2} - \sqrt{-s/4\cmb^2} } })^2 \right]
                  \nonumber \\
\!\!\!\!\!\!\!\!\!\!\!\!&+&\!\!\!2c_0c_1\;[+2i\Theta +(1-{\cm^2 \over s}) 
                       \log(1-{s \over \cm^2})\;]  \\
\!\!\!\!\!\!\!\!\!\!\!\!&+&\!\!\!
 \left. 2c_0c_2\;[-2i\Theta +(1-{\cmb^2 \over s}) 
      \log(1-{s \over \cmb^2})\;]   
+2c_1c_2 f(s) \right\} , \nonumber
\ea 
where the last term is the ghost-antighost contribution and 
the function $f(s)$ is given by
\ba
f(s)\!\!&=&{i\sin2\Theta \over s}
                 \left(\log({s-2\cos2\Theta+\Delta(s) 
                        \over -2\cmb^2}) 
            +\log({-2\cm^2 \over
                       s-2\cos2\Theta+\Delta(s)}) \right) \nonumber\\ 
\!\!\!\!\!\!&-&\!\!\!{ \Delta(s) \over 2s}
                \left( \log({s-2\cos2\Theta-\Delta(s) 
                       \over -2\cmb^2}) 
             -\log({s-2\cos2\Theta-\Delta(s) 
                  \over -2\cmb^2}) \right)\; , \nonumber\\ 
\Delta(s)&=&\sqrt{(s-4\cos^2\Theta)(s+4\sin^2\Theta)} . 
\ea 
The logarithm 
functions in the above equations take the complex angle between $\pi$
and $-\pi$. 
The function $f(s)$ was worked out long ago by Lee and
Wick but our results is different from theirs \cite{lee3}. 
Their results correspond to
combining the two logarithms in the above equation, 
which is not always legitimate 
because of the restricted
range of the complex phase of the arguments under the logarithms. The
Rieman sheet structure of this function is highly nontrivial as shown 
in Figure~(\ref{fig:ch3.sheet}).
Our function agrees with Lee's function when
$Re(s)<2\cos2\Theta$. The function $f(s)$ has a cut which is
a hyperbola whose center is at $(2\cos2\Theta,0)$  in the complex 
$s$ plane. The function $f(s)$  
has a finite jump anywhere across the cut
except  at $s=4\cos2\Theta$ where it is continuous. 
This function is  analytic everywhere else away from the cut.
The so-called  ``ghost-antighost threshold'' is not a real one, and no
imaginary part contribution will arise when the center of mass energy
steps through $4\cos^2\Theta$. This is  necessary for the unitarity 
to hold.                     
Other parts in the $B(s)$ function have cuts starting at the ghost pole
location.
In the small $s$ and large $s$ region the function $B(s)$ simplifies to
\ba
B(s)\!\!\!\!\!\!&&\stackrel{s \rightarrow 0}{\sim} 
-{ 1 \over 16 \pi^2} \left(\log(s) -i\pi +1/2 +O(s) \right) ,  
\nonumber \\
B(s)\!\!\!\!\!\!&&\stackrel{s \rightarrow \infty}{\sim} 
-{ 1 \over 16 \pi^2} \left( -i\pi +O(1/s) \right)  .
\ea 
This concludes our discussion of the analytic properties of the
function.

\vfill\eject

%% file: c4.tex
\chapter{Higher Derivative Field Theories on the Lattice}  
\label{ch:BOA}

\section{The Naive Lattice Action and Phase Diagram}

\begin{figure}[htb]
\vspace{5mm}
\centerline{ \epsfysize=3.0cm 
            \epsfxsize=5.0cm \epsfbox{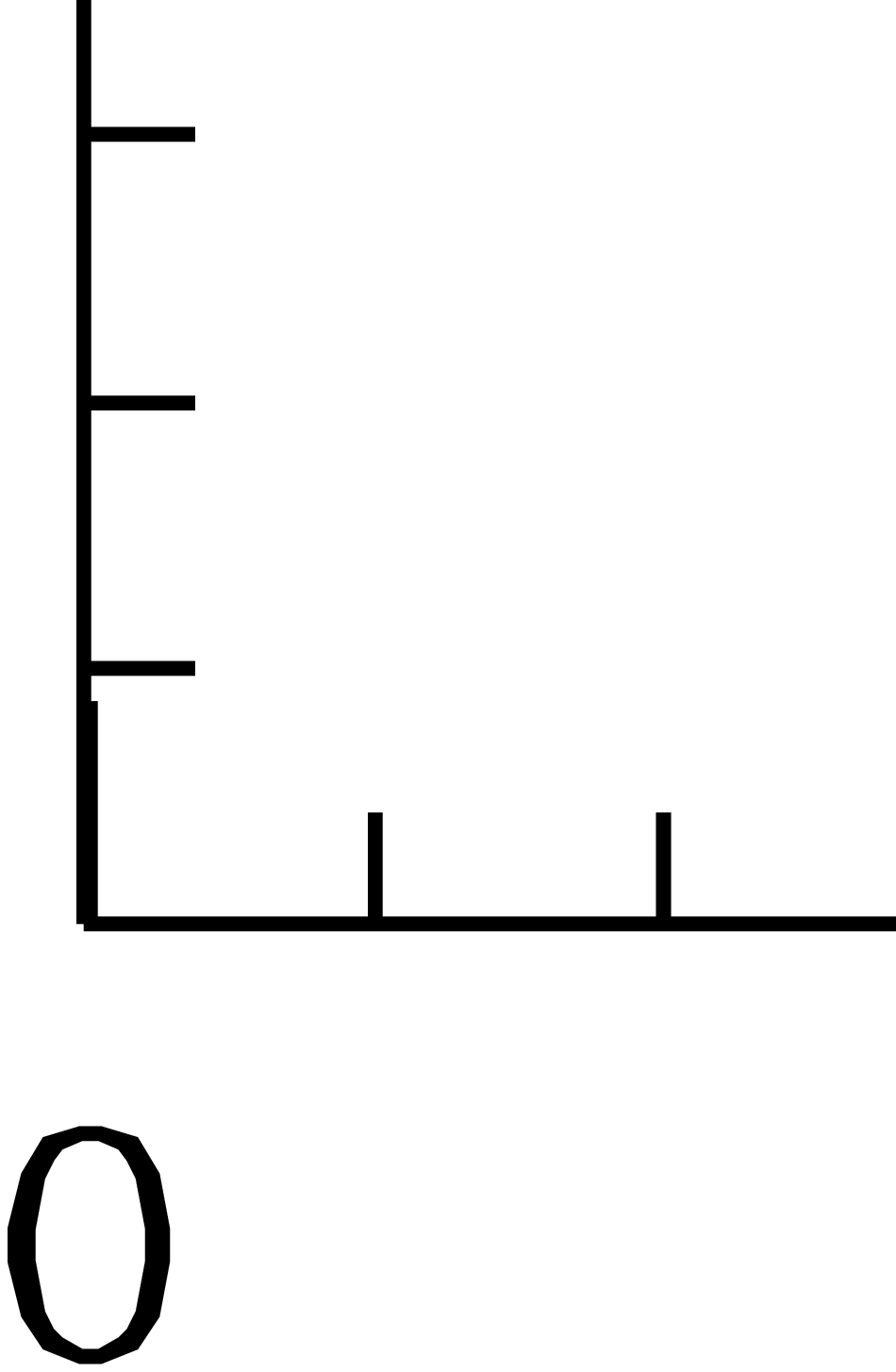}}
\vspace{17mm}
\caption{ The phase diagram of the lattice model at infinite bare
coupling. Data points are obtained from 
Monte Carlo simulations. 
The dotted line is calculated in the large-N expansion. 
The solid line displays a fixed $M_R/m_H$ ratio towards the continuum
limit of the higher derivative theory.}
\label{fig:ch4.phase}
\end{figure}

The need of a lattice for the higher derivative scalar field theory
presented in the previous chapter is not 
for the purpose of regularization, but rather,
to make the degree of freedom finite so that a nonperturbative study
of the model can be performed in computer 
simulations. The lattice spacing $a$   
introduces a new short distance energy scale with the associated
momentum cutoff $\Lambda=\pi/a$. In order to recover the higher
derivative field theory in the continuum, we would have to work
towards the $\Lambda/M \rightarrow \infty$  limit with a fixed ratio
of $M/m_H$. The lattice action we choose to study \cite{hhiggs4,dallas4} is  
\be
{\cal L}_E=-\kappa \phi(x)(-\Box-{\Box^3 \over M^4})\phi(x)
          +(1-8\kappa)\phi(x)^2 -\lambda(\phi(x)^2-1)^2 ,
\ee
where the $\Box$ is the lattice Laplace operator. The phase structure
of this lattice model is quite similar to the conventional $O(N)$ scalar
field theory. It has two phases as shown in
Figure~(\ref{fig:ch4.phase}). The $O(N)$ symmetric phase 
is separated from the broken phase with residual $O(N-1)$ symmetry
by a second order phase transition line for every value of the
lattice coupling constant $\lambda$ in the $(\kappa,M)$ plane.
Near the critical line, we expect to recover the continuum theory
without the lattice artifacts. However, the critical behavior of 
our model is more complicated than the conventional $O(N)$ model. It can
represent different universal continuum theories along different 
paths towards the critical line. Tuning the value of $\kappa$ towards
the critical line for any fixed value of $M$ corresponds to the
trivial field theory in the continuum. In this limit, the operator
$\phi\Box^3\phi$ becomes irrelevant in the critical region. 
However, if we tune the value of $\kappa$ towards the critical line
in such a way that the ratio $M_R/m_H$ remains fixed, we will recover
the continuum higher derivative field theory with the corresponding
ratio of the ghost mass parameter and the Higgs mass. In this limit
the operator $\phi\Box^3\phi$ cannot be viewed as an irrelevant
operator \cite{neub4} in the Lagrangian. Thus, 
it becomes clear, from the
discussion above,  that if we want to study the higher derivative 
field theory, we have to work towards the second limit.  

In the practical application, however, this limit is not very
easy to arrange. One reason is that if we want our results
to represent the continuum results, we have to keep the ghost mass
parameter $M$ reasonably small in lattice units 
in order to get rid of the lattice 
effects associated with it. On the other hand, we need to put the
Higgs mass below the ghost mass parameter. Therefore, we are 
very restricted in the parameter space. On the one hand, making
the Higgs mass smaller will lead to huge finite size effects  for
the practical lattice sizes; on the other hand, making the Higgs mass 
larger will push up the ghost mass and will result in large lattice
effects. So we have a rather narrow range in the Pauli-Villars
correlation length $M/m_H$. Typical values we took in the beginning
of our simulation were: $M=0.8\sim1.0$, $m=0.3\sim0.4$.  
This, of course, was unsatisfactory because the ratio 
$M/m_H=2\sim 3$ is too narrow of a range. If we
view this theory as a Pauli-villars regulated theory, for example, we
would hope to see the conventional scaling behavior in the
large $M/m_H$ limit. It turns out that the scaling form 
may apply only for rather large $M/m_H$ values  which is
impossible for us to investigate using this naive lattice action.
Also, due to this restricted range, it was also impossible for
us to study the scattering phase shift profile of the model. 
This type of analysis offers us a very good way of extracting
the mass value for an unstable particle in the finite box ( see
Chapter~(\ref{ch:LUSC}) for full discussion ).
This restriction in the parameters is purely due to the
introduction of the underlying lattice structure. When 
we were able to eliminate most of the lattice effects, we were then 
able to enlarge our parameter space quite substantially.
Therefore, the need for an improved lattice action
becomes quite obvious.

\section{ The Improved Lattice Action }
 
Improving the lattice action so that it has better
Euclidean invariance was studied long  ago \cite{syman4}. Our choice 
of the improvement corresponds to modifying the lattice 
Laplacian so that it resembles the continuum 
Laplacian. Therefore, we take
\be
p_{I}^2= \hat{p}^2 + {a_1 } \sum_{\mu} \hat{p}_{\mu}^4 
         + {a_2 } \sum_{\mu} \hat{p}_{\mu}^6 
         + {a_3 } \sum_{\mu} \hat{p}_{\mu}^8 
         + {a_4 } \sum_{\mu} \hat{p}_{\mu}^{10} 
         + {a_5 } \sum_{\mu} \hat{p}_{\mu}^{12} 
         + {a_6 } \sum_{\mu} \hat{p}_{\mu}^{14} , 
\ee
where the coefficients are given by the following table
\begin{table}[htb]
\begin{tabular}{|cccccccccc|}
\hline
           $a_1$
          &$a_2$
          &$a_3$
          &$a_4$
          &$a_5$
          &$a_6$
          &$a_7$
          &$a_8$
          &$a_9$
          &$a_{10}$ \\
\hline
          ${1\over 12}$
         &${1\over 90}$
         &${1\over 560}$
         &${1\over 3150}$
         &${1\over 16632}$
         &${1\over 84084}$
         &${1\over 411840}$
         &${1\over 1969110}$
         &${1\over 9237800}$
         &${1\over 42678636}$\\
\hline
\end{tabular}
\end{table}

In fact, we calculated the renormalized coupling constant in the
large $N$ limit and we found that this improved action significantly
decreased the lattice effects. With this improved lattice action, even at
$M=2.0$, there was negligible lattice effects on the large $N$  
results. The phase diagram of the improved action is similar to 
the naive action.   

The improved action offers us another power: the possibility
of performing a phase shift simulation on the higher derivative
$O(N)$ model. This is the subject in Chapter~(\ref{ch:LUSC}).
As we will demonstrate, without the improved action, we are in the
parameter range that is impossible for this type of simulation because 
we would need unrealisticly large lattices to extract the
phase shift. With the improved action, this type of calculation
becomes possible.

\section{ The Rotator States and Born Oppenheimer Approximation }

Studying the higher derivative $O(N)$ model in the broken phase and 
the corresponding Higgs mass problem requires a better understanding
of the symmetry breaking mechanism in the finite volume. In fact this
is already an important issue in the conventional $O(N)$ theory 
without the higher derivative terms added.
The symmetry breaking mechanism has been understood very well
in the infinite volume limit. However, it has not been answered 
satisfactorily in the $O(N)$ model in a finite volume.

There are several complications. First of all, the notion of
symmetry breaking in the infinite volume cannot be applied to
a system in a finite volume. Strictly speaking, in a finite volume, 
the symmetry is never broken. 
Secondly, it turns out that the dynamics of the zeromode 
are crucial for the understanding, and  the zeromode 
is coupled to other modes in a complicated
way.  For the one component $\phi^4$ theory, Hartree type of
approximation will give us a very good description of the
symmetry breaking. For the $O(N)$ model, extra care must 
be paid to the motion of the zeromode and new approximation
schemes are needed for the understanding of the problem. 
This section consists of several parts. 
In the first part, we  will review what is known to the
symmetry breaking in a ordinary one-component $\phi^4$
theory in the broken phase. It turns out that this is
a very instructive model to study. In the second part,  
the conventional $O(N)$ model is studied in the broken phase.
Here we introduce the Born-Oppenheimer Approximation (or Adiabatic 
Approximation) and fully investigate 
the dynamics of the zeromode. 
In the third part, we consider some important 
applications of the Born-Oppenheimer Approximation.
The machinery is applied to the ground state 
and  higher energy excited states.
The rotator correction to the energy of these states
is calculated. This  will serve as a theoretical guide line to
the analysis of the simulation results in Chapter~(\ref{ch:SIMU}).
Then, 
the higher derivative $O(N)$ theory is presented in the
next section. 

\subsection{ Symmetry Breaking of the One-component $\phi^4$ Model }
 
Consider the one component $\phi^4$ theory in a cubic box. The 
Hamiltonian of the theory is given by
\be
{\cal H} = {1 \over 2} \pi^2 +{1 \over 2} (\nabla \phi)^2 -
           {1 \over 2} \mu_0^2 \phi^2 +{\lambda_0 } \phi^4 .
\ee
This Hamiltonian is obviously invariant under the 
change $\phi \rightarrow -\phi$. That is to say
that one can construct a parity operator $P$, which
flips the sign of the $\phi$ field and it 
commutes with the Hamiltonian. Therefore, 
all the eigenstates of the Hamiltonian can be
chosen to have a definite parity.
 
We can build up two approximate ground states of the 
Hamiltonian $|\pm\rangle$, which are Gaussian wavefunctions centered at
$\pm v$ respectively. 
This picture  is very well illustrated
by the so-called ``Hartree approximation''. We start with a trial
wave functional which is a Gaussian
\be
\Psi(\phi) = N \exp \left( -{1 \over 2}
(\phi(x)-v)G^{-1}(x,y)(\phi(y)-v) \right)
\ee
where $v$,$G^{-1}(x,y)$ are variational parameters and the summation
over $x$ and $y$ is implied. The approximate ground state of the
system can be found by using the minimization condition of the
energy. This condition in the broken phase will give us two
solutions for the parameter $v$, namely, 
$v=\pm \sqrt{\mu^2_0/4\lambda_0}$ and the propagator $G^{-1}(x,y)$ 
is given by
\be
G(x,y)={1 \over 2} \int {d^3k \over (2\pi)^3}
 {e^{i \bk \cdot \bx} \over \sqrt{\bk^2 + m^2_R} }
\ee
If we denote these two states as $| \pm \rangle$ then we see 
they satisfy the following properties
\be
P |\pm \rangle =|\mp \rangle, \;\; \langle +|-\rangle=e^{-v^2m_RL^3} .
\ee
Note that the 
states $| \pm \rangle$ are not orthogonal to
each other in the finite volume. 
The true ground state and the first excited state are given by the
symmetric and antisymmetric linear combination of these two
states 
\be 
|0\rangle = {1 \over \sqrt{2}} (|+\rangle + |-\rangle ) ,
\;\;\;\;
|1\rangle = {1 \over \sqrt{2}} (|+\rangle - |-\rangle ) .
\ee
The true ground state is a parity even
state, while the first excited state is a parity odd state. 
 The energy difference between the two is exponentially
small when the volume is large.
This means that if the
system were started at one of the minimum, after a long enough 
time, there is a finite
probability of finding the system
tunneled  to the other minimum. The typical time
scale for this is $1/\Delta E$, where $\Delta E$ is the 
energy difference between the ground and the first excited state. 
    
If we use a infinite volume, the state 
$|+\rangle$ would be exactly orthogonal to the state
$|- \rangle$; then  the system starting from one
particular minimum of the potential as the true vacuum 
will stay there, without knowing
the other one and the $\phi \rightarrow -\phi$ 
symmetry is broken. However, in a
finite but large volume, the system 
will stay around one minimum for such  a 
long enough time that we may say 
the symmetry is ``almost broken''. 
 
In the one component model, the symmetry is a discrete symmetry
and the Hartree approximation gives us a very good understanding
of the symmetry breaking mechanism in the finite volume.
Nonperturbative works have also been done to measure the 
energy gap between the ground state and the first excited
state, which is related to the surface tension of the 
system.

\subsection{$O(N)$ Model: General Setup}

The situation is much more complicated when we try to do
a similar analysis for the $O(N)$ model. The main reason 
is that the symmetry is a continuous symmetry, therefore,  
the dynamics of the zeromode is much more complicated.

We could first try out the Hartree approximation, but it 
will not give us the right energy spectrum of the theory. 
This is because the Hartree approximation
treats every mode of the system equally. In
the one component model this is valid, but it is not valid for the 
$O(N)$ model. In the $O(N)$ model there exists one special mode, 
that is,  the direction of the zero Fourier mode which 
can be characterized by an $O(N)$ unit vector. In a large but
finite volume, this mode 
is a slow varying mode when compared with the other 
modes. It is the counter part of the parity operator
in the one component model. The only difference is that, 
in the one component 
model, the parity only takes discrete values and is not
dynamical. In the O(N) model, however, this unit vector lives on
a $(N-1)$-sphere and has its own dynamics.
Therefore, we  expect that the Born-Oppenheimer
Approximation (BOA), also known as the
Adiabatic Approximation,  will give  us a very good description 
of the zeromode dynamics.

The Born-Oppenheimer Approximation was first introduced
in the study of diatomic molecules. In the molecular
problem, there are two types of degrees of freedom. The motion
of the electron is called ``fast'', and the motion of
the nucleus is called ``slow''. Therefore, when solving the
energy eigenvalues of the system, one should first nail down
the slow variable, namely the configuration of the
nucleus, and solve the fast variable spectrum. In this step, 
the configuration of the slow variable is treated as an
external field. The eigenvalues and eigenstates that 
come out will, in general, depend on the prescribed 
configuration of the slow variable. These eigenvalues
are then taken back into the Schr\"odinger equation for
the slow variables as the effective potential, which 
reflects the feedback of the fast variable to the
slow variable. Finally, the Schr\"odinger equation for
the slow variable is solved to get the spectrum of the
molecule.

The spectrum of the molecule has a three fold hierarchy:
electron energy, oscillation energy and rotational energy.
These energy gaps are  characterized by different powers of
a small parameter, which is the ratio $m_e /m_N$, where
$m_e$ is the mass of the electron and $m_N$ is the mass
of the nucleus. Born-Oppenheimer is a very good
approximation for the molecule, since this ratio is so small.
It is not hard to imagine that a Hartree approximation to the
molecular problem would be a poor choice, since it 
treats the electron (fast variable) and the nucleus (slow variable)
equally, while ignoring the enormous difference in the mass of the two.
Similarly, in our application, Hartree is a poor approximation
 for the same reason.
To get the right picture, one has to separate the special
zeromode and use the Born-Oppenheimer type of approximation.

In our model, we will treat the direction of the
zeromode as the only slow variable. We will use the 
same Born-Oppenheimer type of spirit to solve for
the energy levels of our model.  
 
We begin with the Hamiltonian
\be
 H = \sum_{\bx}{1 \over 2} \pi^a\pi^a +{1 \over 2} \nabla \phi^a
           \nabla \phi^a -
           {1 \over 2} \mu_0^2 \phi^a\phi^a 
           +{\lambda_0 } (\phi^a \phi^a)^2 , 
\ee
where for convenience we have discretize the system on a 
cubic lattice. The operator $\pi^a(\bx)$ is just the
derivative operator  $(-i)\partial/\partial \phi^a(\bx)$ in the
field variable diagonal representation.
Two classes of  symmetry operators that commute with this Hamiltonian are 
very important. First, there are  
global $O(N)$ symmetry generators $Q^{ab}$ given by
\be
Q^{ab}=\sum_{\bx} \phi^{a}(\bx)\pi^{b}(\bx)
                 -\phi^{b}(\bx)\pi^{a}(\bx)
\ee
We also have the $3$-momentum operators $P_i$
\be
P_i = \sum_x [\phi^a(\bx+\bei)-\phi^{a}(\bx)]\pi^{a}(\bx) ,
\ee 
where $\bei$ is the unit vector in the $i$ direction. It is
trivial to verify that these operators commute with the full
Hamiltonian hence are symmetries of the theory.

We now  introduce the Fourier modes of the field variable,
\ba 
\phi^a(\bx) &=& \bar{\phi}^a +
            {1 \over \sqrt{V}} \sum_{\bk > 0}  
 \phi^a_{\bk} e^{i\bk \cdot \bx}
           +\phi^{a*}_{\bk} e^{-i\bk \cdot \bx} ,
\nonumber \\
\pi^a(\bx) &=& {(-i) \over V}{\partial \over \partial \bar{\phi}^a}
           +{(-i) \over \sqrt{V}} \sum_{\bk > 0}
 e^{-i\bk \cdot \bx}{\partial \over \partial \phi^a_{\bk}}
    +e^{+i\bk \cdot \bx}{\partial \over \partial \phi^{a*}_{\bk}} ,
\ea
where $V=L^3$ is the $3$-volume of the box. As we mentioned 
above, the zero mode $\bar{\phi}^a$ plays a very important role
in the broken phase. Therefore we have singled out this mode from the
nonzero momentum modes. Let us define:
\ba
\bar{\phi}^a &=& (v+\sigma) n^a , \;\;\; n^a n^a=1 ,
\nonumber \\
P_{L}^{ab}&=&n^a n^b , \;\;\;\; P_{T}^{ab}=\delta^{ab} -n^a n^b ,
\ea
where $v=\mu_0^2/4\lambda_0$ is the 
vev of the theory.
In the Born-Oppenheimer type of 
approach, we will treat the direction of $\bar{\phi}^a$, namely
 $n^a$,  as the only 
slow varying variable and treat the rest as fast variables.
The justification of this will be seen shortly. 
Then the Hamiltonian can be expressed in terms of
these Fourier modes.

We use the radial variables for the mode
$\bar{\phi}^a$. Thus we write the wavefunctional of
the system as $\Psi=\rho^{-(N-1)/2}\psi$ and the
effective Hamiltonian for $\psi$ will contain only the 
second derivative with respect to $\rho$. For the nonzero
Fourier modes, let us introduce the creation and 
annihilation operators as
\ba 
\label{eq:ch4.creatl}
L^a_{\bk} &=& {1 \over \sqrt{2}}P_{L}^{ab}
         (\sqrt{\Omega_{\bk}} \phi^a_{\bk}
        + {1 \over \sqrt{\Omega_{\bk}}}
        {\partial \over \partial\phi^{b*}_{\bk} } ) ,    
\nonumber \\
L^{a\dagger}_{\bk} &=& {1 \over \sqrt{2}}P_{L}^{ab}
         (\sqrt{\Omega_{\bk}} \phi^{a*}_{\bk}
        - {1 \over \sqrt{\Omega_{\bk}}}
        {\partial \over \partial\phi^{b}_{\bk} } )    , 
\nonumber \\
L^{a}_{-\bk} &=& {1 \over \sqrt{2}}P_{L}^{ab}
         (\sqrt{\Omega_{\bk}} \phi^{a*}_{\bk}
        + {1 \over \sqrt{\Omega_{\bk}}}
        {\partial \over \partial\phi^{b}_{\bk} } )     ,
\nonumber \\
L^{a\dagger}_{-\bk} &=& {1 \over \sqrt{2}}P_{L}^{ab}
         (\sqrt{\Omega_{\bk}} \phi^a_{\bk}
        + {1 \over \sqrt{\Omega_{\bk}}}
        {\partial \over \partial\phi^{b*}_{\bk} } )     ,
\ea
where $\Omega_{\bk}=\sqrt{m_0^2+\bk^2}$ is the higgs excitation.
We can define the Higgs creation and annihilation operators as
\be
h_{\bk}=n^a L^a_{\bk}, \;\;\;\;\;h^{\dagger}_{\bk}=n^a L^{a\dagger}_{\bk}, 
\;\;\;\;\;
\sigma= {1 \over \sqrt{2Vm_0}} (h_0 + h^{\dagger}_0) .
\ee
Similarly, we can define the transverse Goldstone creation and 
annihilation operators as  
\ba 
\label{eq:ch4.creatt}
T^a_{\bk} &=& {1 \over \sqrt{2}}P_{T}^{ab}
         (\sqrt{\omega_{\bk}} \phi^a_{\bk}
        + {1 \over \sqrt{\omega_{\bk}}}
        {\partial \over \partial\phi^{b*}_{\bk} } ) ,    
\nonumber \\
T^{a\dagger}_{\bk} &=& {1 \over \sqrt{2}}P_{T}^{ab}
         (\sqrt{\omega_{\bk}} \phi^{a*}_{\bk}
        - {1 \over \sqrt{\omega_{\bk}}}
        {\partial \over \partial\phi^{b}_{\bk} } )    , 
\nonumber \\
T^{a}_{-\bk} &=& {1 \over \sqrt{2}}P_{T}^{ab}
         (\sqrt{\omega_{\bk}} \phi^{a*}_{\bk}
        + {1 \over \sqrt{\omega_{\bk}}}
        {\partial \over \partial\phi^{b}_{\bk} } )     ,
\nonumber \\
T^{a\dagger}_{-\bk} &=& {1 \over \sqrt{2}}P_{T}^{ab}
         (\sqrt{\omega_{\bk}} \phi^a_{\bk}
        + {1 \over \sqrt{\omega_{\bk}}}
        {\partial \over \partial\phi^{b*}_{\bk} } )     ,
\ea
where $\omega_{\bk} = |{\bk}|$ is the Goldstone energy.
In terms of these operators, we can rewrite the effective
Hamiltonian in the following form
\ba
H&=& \sum_{\bk} \Omega_{\bk} h^{\dagger}_{\bk} h_{\bk}
    +\sum_{\bk \neq 0} \omega_{\bk} T^{a\dagger}_{\bk} T^{a}_{\bk}
          +H_{\rm int} +{ L^2 + \Delta_N \over 2 V (v+\sigma)^2} ,
\nonumber \\
H_{\rm int}&=& \sum_{\bx} {4 \lambda_0 v } 
       h(h^2 + \tilde{\phi^a_T}\tilde{\phi^a_T} )        
     + {\lambda_0 }(h^2 + \tilde{\phi^a_T}\tilde{\phi^a_T})^2 ,
\ea
where the fields $h(\bx)$ and $\tilde{\phi^a_T}$ are given by
\ba
h(\bx)&=& \sum_{\bk} {1 \over \sqrt{2V\Omega_{\bk}}}
       ( h_{\bk}e^{i \bk \cdot \bx}
       +h^{\dagger}_{\bk}e^{-i \bk \cdot \bx} ) ,
\nonumber \\
\tilde{\phi^a_T}(\bx)&=& \sum_{\bk \neq 0} {1 \over \sqrt{2V\omega_{\bk}}}
       ( T^{a}_{\bk}e^{i \bk \cdot \bx}
       +T^{a\dagger}_{\bk}e^{-i \bk \cdot \bx} ) ,
\ea 
and the constant $\Delta_N = (N-3)(N-1)/4$. The operator
$L^2 \equiv L^{ab}_0 L^{ab}_0 /2$ is the $O(N)$ Casimir of
the zeromode variable.
The above creation and annihilation operators satisfy the following
commutation relations
\be
[T^a_{\bk}, T^{b\dagger}_{\bp}]=P^{ab}_T \delta_{\bk \bp},
\;\;\;
[L^a_{\bk}, L^{b\dagger}_{\bp}]=P^{ab}_L \delta_{\bk \bp} .
\ee
It is also very convenient to introduce the following decomposition
for the fields. For a given $O(N)$ unit vector $n^a$, we can find
additional $N-1$ unit vectors which, together with $n^a$, form a
complete set in the $O(N)$ space. We therefore define
\ba
\label{eq:ch4.compht}
n^a_0 &\equiv& n^a , \;\; n^a_{\alpha} n^a_{\beta}
=\delta_{\alpha \beta} , \;\; \alpha, \beta=0,1,...N-1 ,
\nonumber \\
T^a_{\bk}&=& n^a_i T_{i \bk}, \;\;
T_{i \bk}= n^a_i T^a_{\bk} ,
\nonumber \\
L^a_{\bk}&=& n^a_0 h_{\bk}, \;\;
h_{\bk}= n^a_0 L^a_{\bk} .
\ea 
It is readily checked that these operators satisfy the standard
commutation relations
\be
[h_{\bk},h^{\dagger}_{\bp}]=\delta_{\bk \bp},
\;\;\;
[T_{i \bk},T^{\dagger}_{j \bp}]=\delta_{ij}\delta_{\bk \bp}.
\ee
Note that due to the leftover $O(N-1)$ symmetry, the determination of
the unit vectors $n^a_i$ is not unique. However, the physical quantities
will not  depend on this ambiguity. Moreover we can calculate the 
commutator of the operator $L^{ab}_0$ with the unit vectors,
\be
[L^{ab}_0, n^c_i] =(-i) n^{[a}_{\alpha} \delta^{b]c}, 
\;\;\;\;\; \alpha=0,1,...N-1 .
\ee 
Now we can set up a basis in our Hilbert space from the eigenstate
of the free Hamiltonian. We will also choose the angular momentum
eigenstate of the $n^a$ variable, namely
\be
|n,\{ n^L_{\bk},n^T_{\bk} \},lm\rangle = |n\rangle \otimes
  \prod_{\bk \neq 0} |n^L_{\bk} \rangle \otimes |n^T_{\bk} \rangle
    \otimes |lm\rangle ,
\ee 
where the state $|lm\rangle$ is the eigenstate of $L^2$ with eigenvalue
$l(l+N-1)$. This is just a symbolic notation
of the state. Strictly speaking, for
$O(N)$ model, we need more quantum numbers to specify 
the state. Note that the oscillator part of the state actually 
depends on the unit vector $n^a$ via the definition of the
longitudinal and transverse projection, although
the eigenvalue does not. Therefore, if we were to
act the operator $L^2$ on the states above, it would not only 
act on the state $|lm\rangle$, but also act on the rest of the components
( except $|n\rangle$, of course, since it is the radial zero 
momentum mode ). As we will see below, the Born-Oppenheimer 
Approximation will first neglect the effect of $L^2$ on the
fast modes and only consider the slow mode part of the
state, i.e. $|lm \rangle$. The next order correction has
to take this into account and the BOA is valid when 
the correction is small.  
 
The global $O(N)$ generator $Q^{ab}$ now can be expressed in terms
of the creation and annihilation operators defined in
Equation~(\ref{eq:ch4.compht}) 
\be
Q^{ab}=L^{ab}_0-i\sum_{\bk \neq 0} 
 T^{\dagger}_{\alpha \bk}T_{\beta \bk} n^{[a}_{\alpha} n^{b]}_{\beta} ,
\ee
where the index $\alpha$ and $\beta$ run from $0$ to $(N-1)$.
The operators $T_{0 \bk}$ and the  functions $f_{\bk}$, $g_{\bk}$ are 
defined by the following:
\ba
T^{\dagger}_{0 \bk}=f_{\bk}h^{\dagger}_{\bk}-g_{\bk}h_{-\bk} ,
&&
T_{0 \bk}=f_{\bk}h_{\bk}-g_{\bk}h^{\dagger}_{-\bk} ,
\nonumber \\
f_{\bk}={1 \over 2}( \sqrt{{\Omega_{\bk} \over \omega_{\bk}}}
           +\sqrt{{\omega_{\bk} \over \Omega_{\bk}}} ) ,
 &&    
g_{\bk}={1 \over 2}( \sqrt{{\Omega_{\bk} \over \omega_{\bk}}}
           -\sqrt{{\omega_{\bk} \over \Omega_{\bk}}} ) .
\ea 
We will need the result of $L^2$ acting on the oscillator
states. Let us consider the object  
$L^2 |0_{\bk \neq 0}\rangle$.
This can be easily obtained by noticing that the state is
annihilated by the global $O(N)$ generators $Q^{ab}$. 
Therefore we would induce that
\be
\label{eq:ch4.lab0}
L^{ab}_0|0_{\bk \neq 0} \rangle =\sum_{\bk \neq 0}  
      i g_{\bk} n^{[a}_{0} n^{b]}_{i} 
h^{\dagger}_{-\bk}T^{\dagger}_{i \bk}
          |0_{\bk \neq 0}\rangle .
\ee
We will also need the commutation relations between
$L^{ab}_0$ and the annihilation operators
\ba
\label{eq:ch5.commu}
[ L^{ab}_0 , L^c_{\bk} ]&=& (-i) ( n^{[a}\delta^{b]c} 
                 h_{\bk} 
         +n^c ( f_{\bk} n^{[a} T^{b]}_{\bk}
          +g_{\bk} n^{[a} T^{b\dagger]}_{-\bk}) ) ,
\nonumber \\
{[ L}^{ab}_0 , T^{c}_{\bk} ]&=& (-i) ( n^{[a} \delta^{b]c} 
   (f_{\bk}h_{\bk}-g_{\bk}h^{\dagger}_{-\bk})
   + n^c n^{[a} T^{b]}_{\bk} ) ,
\ea
or equivalently, in terms of operators $T_{i \bk}$ and $h_{\bk}$, 
we have 
\ba
\label{eq:ch5.commucompo}
{[ L}^{ab}_0 ,h_{\bk} ] &=& (-i) n^{[a}_0 n^{b]}_i 
          (f_{\bk} T_{i \bk} + g_{\bk} T^{\dagger}_{i -\bk}) ,
\nonumber \\
{[ L}^{ab}_0 ,T_{i \bk} ] &=&(-i)n^{[a}_i n^{b]}_j T_{j\bk}+ 
        (-i) n^{[a}_0 n^{b]}_i 
          (f_{\bk} h_{\bk} - g_{\bk} h^{\dagger}_{- \bk} ) .
\ea
The corresponding commutators with the creation operators
can be obtained from  hermitian conjugation of the 
above equations. We now  have all the tools to study the
rotator energy spectrum.

\subsection{$O(N)$ Model: The Ground States}

The full ground state of the free Hamiltonian consists of two
parts. One is the ground state of oscillator states. The other part
is the rotator states.
\be
| G,lm \rangle =| 0_{osc} \rangle \otimes | lm \rangle .
\ee
Note that, before the rotator energy contributions are taken into
account, the degeneracy of the ground state is infinite, since any
rotator state $| lm \rangle$ will belong to the same energy. 
It is very easy to check that all these states are eigenstates of 
the momentum operator with eigenvalue of $0$. They can also be taken
as the eigenstates of the appropriate $O(N)$ operators. To see this, 
notice that when the operator $Q^{ab}$ is applied to the states, it 
generally has two contributions. One is from $Q^{ab}$ acting on the
oscillator state $| 0_{osc} \rangle$, which is zero in this case; the
other is from $Q^{ab}$ acting on the rotator states $|lm \rangle$, which
is equivalent to $ L^{ab}_0 |lm \rangle$. Therefore, we have
\be
Q^{ab} |G,lm \rangle = |0_{osc} \rangle \otimes L^{ab}_0 |lm \rangle .
\ee 
By taking the states $|lm \rangle$ to be the eigenstates of the 
zeromode Casimir, we also make the ground states to have the 
appropriate $O(N)$ charges. Basically, the oscillator ground state
has $O(N)$ charge $0$, and all the $O(N)$ charges comes from the
rotator states.

As mentioned above, the leading order ground states are infinitely
degenerate due to the rotator states. This degeneracy is lifted 
once the first nonvanishing rotator correction is taken into account.
To do this, let us evaluate the matrix element of the rotator 
energy operator $(1/2)L^{ab}_0L^{ab}_0\omega_r$ among the ground states 
\ba
\!\!\!\!L^{ab}_0 |0_{osc}\rangle \otimes |lm \rangle &=&
(L^{ab}_0 |0_{osc}\rangle) \otimes |lm \rangle +
|0_{osc}\rangle \otimes (L^{ab}_0 |lm \rangle )
\nonumber \\
\!\!\!\!\!\! &=&\sum_{\bk \neq 0}  
      i g_{\bk} n^{[a}_{0} n^{b]}_{i} 
   h^{\dagger}_{-\bk}T^{\dagger]}_{i \bk}
          |0_{osc}\rangle \otimes |lm \rangle
+ |0_{osc} \rangle \otimes (L^{ab}_0 |lm \rangle ) ,
\ea
where we have used Equation~(\ref{eq:ch4.lab0}). Therefore we get
\be
\!\!\!\! \langle G,l'm'|1/2L^{ab}_0L^{ab}_0\omega_r| G,lm \rangle =
\left[l(l+N-2)\omega_r + (N-1)\omega_r \sum_{\bk \neq 0} g^2_{\bk}
\right] \delta_{ll'}\delta_{mm'} .
\ee
As expected, the degeneracy is lifted, and the ground states now
have a degeneracy of $l^2$, for a given $l$ value. The second term
in the above equation is an $l$ independent constant and can be absorbed
into the definition of the ground state energy. The first term is
$l$ dependent and is known as the rotator energy spectrum. This
formula was derived before by Leutwyler using
the rigid rotator approximation in the chiral Lagrangian formalism.
The significance of this energy spectrum in the Monte Carlo simulation
was also discussed \cite{leut4}.
The low energy excitations of the model exhibit a three hierarchy
of energy gaps. The largest energy gap is the mass gap of the Higgs
particle, whose energy is independent of the $3$-volume. The second
largest gap is the Goldstone particle, whose energy gap is typically
of order ${\cal O}(1/L)$, where $L$ is the size of the $3$ dimensional
cubic box. The smallest gap is the  rotator energy differences between
different $l$ values, whose energy is of order
${\cal O}(1/L^3)$. In a practical simulation, the size of the Higgs gap
and the Goldstone energy gap are of the same order, since the size of
the box is not large enough. However, the energy gap of the rotator
is usually much smaller compared with the Higgs and Goldstone. Therefore
we expect the Born-Oppenheimer picture should be a very good description
of the theory in the finite box. 

It is also possible to evaluate the second order correction to
the ground state energy. Let us first look at the following 
quantity
\ba
L^{ab}_0L^{ab}_0|G,lm\rangle&=&
(L^{ab}_0L^{ab}_0|0_{osc}\rangle )\otimes |lm\rangle
\nonumber \\
&+&2(L^{ab}_0|0_{osc}\rangle )\otimes(L^{ab}_0 |lm\rangle)+
|0_{osc}\rangle \otimes(L^{ab}_0L^{ab}_0 |lm\rangle) .
\ea
The first term in the above equation will contribute at the second order
but it is a term independent of $l$. If we only focus on the
$l$-dependent terms, we can forget about this term. The last term
is diagonal, it will not contribute to the second order correction of
the ground state energy. Therefore, only the second term will give
us $l$-dependent contribution to the ground state energy.
The state left over is
\be
2(L^{ab}_0|0_{osc}\rangle )\otimes(L^{ab}_0 |lm\rangle)=
2\sum_{\bk \neq 0,a,b,i} i g_{\bk}n^{[a}_0n^{b]}_i
h^{\dagger}_{\bk}T^{\dagger}_{i,-\bk} |0_{osc} \rangle
\otimes(L^{ab}_0 |lm\rangle) .
\ee
Therefore, it will contribute a second order energy correction that
looks like
\be
E^{(2)}_{0l}=-\sum_{\bk \neq 0,i}
{ \omega_r g^2_{\bk} \over \omega_{\bk}+\Omega_{\bk}}
\langle lm| L^{cd}_0n^{[c}_{0} n^{[d}_{i} n^{[a}_{0} n^{b]}_{0}L^{ab}_0
| lm \rangle .
\ee
The matrix element that appears in the above equation can be simplified
by noticing 
\be
L^{cd}_0n^{[c}_{0} n^{[d}_{i} n^{[a}_{0} n^{b]}_{0}L^{ab}_0=
4 L^{ab}_0 n^b_0 n^c_0 L^{ac}_0 .
\ee
We can pick our $O(N)$ axis such that the unit vector is in the 
direction $(0,0,\cdots,1)$. Then the above operator simplifies to
the difference of two Casimirs: the Casimir of the $O(N)$ and the
Casimir of the unbroken $O(N-1)$  
\be
E^{(2)}_{0l}=-\sum_{\bk \neq 0}
{ 2\omega^2_r g^2_{\bk} \over \omega_{\bk}+\Omega_{\bk}}
\left( l(l+N-2) -\langle lm|L^2_{O(N-1)}|lm\rangle \right) .
\ee
This correction is usually very small for practical simulation
parameters. However, there is another second order correction of the
rotator Hamiltonian. Recall that we can expand the rotator Hamiltonian
into the form
\be
{L^2+\Delta_N \over 2 V (v+\sigma)^2} =
  {L^2+\Delta_N \over 2 V v^2} (1 - {2\sigma \over v} 
           +{3\sigma^2 \over v^2}+\cdots ) .
\ee
Note that the operator 
$\sigma=(h_0+h^{\dagger}_0)/\sqrt{2Vm_0}$. therefore we
have a systematic expansion in inverse powers of
the $3$-volume. If we introduce the rotator energy
$\omega_r \equiv 1/2Vv^2$, then we have a expansion
in terms of the small quantity $\omega_r/m$. 
Therefore, there is another contribution from the
operator $L^2 \omega_r(\sigma/v)^2$ which is also
of the order of $\omega^2_r$. So we have
\ba 
E^{(2)}_{0l}&=&[l(l+N-2)+\Delta_N]\omega_r{ 3 \omega_r \over m} 
\nonumber \\
&-&\sum_{\bk \neq 0}
{ 2\omega^2_r g^2_{\bk} \over \omega_{\bk}+\Omega_{\bk}}
\left( l(l+N-2) -\langle lm|L^2_{O(N-1)}|lm\rangle \right) .
\ea 
The first term basically takes into account the nonrigid effects
of the rotator.

\subsection{$O(N)$ Model: The Zero Momentum Higgs States}

Let us consider the energy corrections to the state
$|n,0_{\bk \neq 0},lm \rangle$.
Since the zero momentum Higgs excitation is just the
radial excitation which commute with the angular
variables. Therefore the energy corrections to the state
are very much like the corrections for the ground states.
\ba
E^{(0)}_{nl}&=& (n+1/2) m , 
\nonumber \\
E^{(1)}_{nl}&=& \left[l(l+N-2)+\Delta_N \right]\omega_r ,
\nonumber \\
E^{(2)}_{nl}&=&(l(l+N-2)+\Delta_N)\omega_r(n+1/2){6\omega_r \over m}  
\nonumber \\
&-&\sum_{\bk \neq 0}
{ 2\omega^2_r g^2_{\bk} \over \omega_{\bk}+\Omega_{\bk}}
\left( l(l+N-2) -\langle lm|L^2_{O(N-1)}|lm\rangle \right) .
\ea 
The second order correction is down by an extra factor of $\omega_r/m$ 
compared with the first order correction. However, due
to the large numerical factor in front, the effect of the first term is
still quite significant.   
In a practical simulation, the correlation function
$\langle n^a(0)n^a(\tau)\rangle$ 
is measured and used as a way of extracting the
vacuum expectation value $v$. This correlation 
function will pick up the 
energy difference of $\Delta l=1$ states. 
In most of the old simulations on $O(4)$, the 
ratio $\omega_r/m$ is very small and
the rigid approximation of the rotator 
energy gives very reliable results. In our
recent simulations on the higher derivative
theories, this ratio is of the order of 
$10$ percent and the correction is noticeable.
We would find the wrong $v$ value if we did not
include this correction.

\subsection{$O(N)$ Model: Two Pion States}

We can perform the similar calculation for the two pion states. Let us
take the isospin zero channel states
$(N-1)^{-1/2} T^{\dagger}_{i,\bk}T^{\dagger}_{i,-\bk} |0\rangle
\otimes |lm \rangle$. We have
\ba
\!\!\!\!\!\!&&\langle l'm'| \otimes \langle 0| 
T_{i,\bk} T_{i,-\bk} L^2 \omega_r   
 T^{\dagger}_{i,\bk} T^{\dagger}_{i,-\bk} 
| 0 \rangle \otimes |l m \rangle/(N-1)
\nonumber \\
\!\!\!\!\!\!&=&\left( l(l+N-2) + (N-1) \sum_{\bp \neq 0} g^2_{\bp}
+2f^2_{\bk}+2g^2_{\bk} \right) \omega_r 
\delta_{ll'} \delta_{mm'}  ,
\ea
which implies that relative to the ground state the finite volume 
correction is 
\be
\Delta(2\omega_{\bk}) = 2(f^2_{\bk}+g^2_{\bk})\omega_r .
\ee
This correction is also very small when we extract the two pion 
energy. For the simulation points where we extract the two
pion energies, this correction is below $1$ percent and is 
therefore hidden in the statistical errors. 

\section{ Symmetry Breaking of the Higher Derivative $O(N)$ Model}
  
The similar analysis can be done with the higher derivative 
$O(N)$ model. Having discussed the ordinary $O(N)$ theory
we will be very brief and only point out the differences. 
Many steps are also similar to the quantization of the
higher derivative theory which was discussed in detail in  
Chapter~(\ref{ch:QUAN}).

One starts with the general higher derivative Lagrangian which 
has a global $O(N)$ symmetry
\be 
{\cal L} = {1 \over 2} \phi^a (-\rho_1 \Box
          -\rho_2 \Box^2-\rho_3 \Box^3 ) \phi^a 
          +{1 \over 2} \mu_0^2 \phi^a\phi^a 
           -{\lambda_0 } (\phi^a \phi^a)^2  ,
\ee
where $\Box=\partial^2_t -\nabla^2$ 
is the Minkowski space d'Alambert operator and 
the coefficients are parametrized as
\be
\rho_1=1+{m^2_0 \over {\cal M}^2}+{m^2_0 \over \bar{{\cal M}}^2},
\;\;\;\;
\rho_2={1 \over {\cal M}^2}+{1 \over \bar{{\cal M}}^2}
       +{m^2_0 \over {\cal M}^2\bar{{\cal M}}^2},
\;\;\;\;
\rho_3={1 \over {\cal M}^2\bar{{\cal M}}^2} .
\ee
After the usual steps of indefinite metric quantization, and
introduction of the Fourier modes, the
Hamiltonian has the form
( see Equation~(\ref{eq:ch2.hambro}) to
Equation~(\ref{eq:ch2.hambrof}) for detail )
%
\ba
H &=& {1 \over V}(i \pi^a_{10}\pi^a_{20}
      +{1 \over 2 \rho_3} \pi^a_{30}\pi^a_{30}
      +{\rho_1 \over 2} \pi^a_{20}\pi^a_{20} )
      +V ( {\rho_2 \over 2} \bar{\phi}^a_3\bar{\phi}^a_3
         + i \bar{\phi}^a_2\bar{\phi}^a_3 )
\nonumber \\
      && + 
        \sum_{\bk >0} i \pi^a_{1\bk}\pi^{a*}_{2\bk}
                    +i \pi^{a*}_{1\bk}\pi^{a}_{2\bk}
               +{1 \over \rho_3} \pi^{a}_{3\bk}\pi^{a*}_{3\bk}
               +(\rho_1+2\rho_2\bk^2+3\rho_3\bk^4)
                \pi^{a}_{2\bk}\pi^{a*}_{2\bk}
\nonumber \\
      &&   
      + (\rho_1\bk^2 + \rho_2\bk^4 + \rho_3\bk^6 ) 
                \phi^{a}_{1\bk}\phi^{a*}_{1\bk}
      + ( \rho_2\bk^4 +3\rho_3\bk^2 ) 
                \phi^{a}_{3\bk}\phi^{a*}_{3\bk}
       + i \phi^a_{2\bk}\phi^{a*}_{3\bk}
                    +i \phi^{a*}_{2\bk}\phi^{a}_{3\bk}
\nonumber \\
      &&   
          -\sum_{\bx}{1 \over 2} \mu_0^2 \phi^a_1\phi^a_1 
           +\sum_{\bx}{\lambda_0 } (\phi^a_1 \phi^a_1)^2  .
\ea 
Because we are now treating the system in a finite volume, we can
no longer neglect the motion of the zeromode. Instead, following
the idea of Born-Oppenheimer Approximation ( or Adiabatic 
Approximation ), 
we will single out the direction of the $\bar{\phi_1}^a$ variable
and make it the slow variable in our Born-Oppenheimer approximation.
We can then decompose
\be
\phi^a_1 =v n^a + h(\bx) n^a +\tilde{\phi}^a_{1T}(\bx) ,
\ee
and similarly for the $\phi_2$ and $\phi_3$ variables.
The Hamiltonian is then written as sum of three types of
terms
\ba
H &=& H_0+H_{\bk \neq 0}+H_{\rm int} ,
\nonumber \\
H_0&=&
      {1 \over V}(i \pi^a_{10}\pi^a_{20}
      +{1 \over 2 \rho_3} \pi^a_{30}\pi^a_{30}
      +{\rho_1 \over 2} \pi^a_{20}\pi^a_{20} )
      +V ( {\rho_2 \over 2} \bar{\phi}^a_3\bar{\phi}^a_3
         + i \bar{\phi}^a_2\bar{\phi}^a_3+{m^2_0 \over 2}\sigma^2 ) ,
\nonumber \\
H_{\bk \neq 0}&=&
        \sum_{\bk >0} i \pi^a_{1\bk}\pi^{a*}_{2\bk}
                    +i \pi^{a*}_{1\bk}\pi^{a}_{2\bk}
               +{1 \over \rho_3} \pi^{a}_{3\bk}\pi^{a*}_{3\bk}
               +(\rho_1+2\rho_2\bk^2+3\rho_3\bk^4)
                \pi^{a}_{2\bk}\pi^{a*}_{2\bk}
\nonumber \\
    &+&(\rho_1\bk^2+\rho_2\bk^4+\rho_3\bk^6+m^2_0)
       \phi^{a}_{1\bk L}\phi^{a*}_{1\bk L}
     + (\rho_1\bk^2+\rho_2\bk^4+\rho_3\bk^6)
       \phi^{a}_{1\bk T}\phi^{a*}_{1\bk T}
\nonumber \\
    &+&(\rho_2+3\rho_3\bk^2)
       \phi^{a}_{3\bk}\phi^{a*}_{3\bk}
     +i\phi^{a}_{2\bk}\phi^{a*}_{3\bk}
     +i\phi^{a*}_{2\bk}\phi^{a}_{3\bk} ,
\nonumber \\
H_{\rm int}&=& \sum_{\bx}
      {4 \lambda_0 v }h(h^2+
           \tilde{\phi}^{a}_{1T}\tilde{\phi}^{a}_{1T})
     +{\lambda_0  }(h^2+
           \tilde{\phi}^{a}_{1T}\tilde{\phi}^{a}_{1T})^2 .
\ea 
This Hamiltonian is identical to what we had in
Chapter~(\ref{ch:QUAN}),  except for  the $H_0$ piece
( see Equation~(\ref{eq:ch2.hampiece}) ).
For example,
the $\bk \neq 0$ piece can be diagonalized in the same way
as in Chapter~(\ref{ch:QUAN}). The interaction piece is also
expressed as the creation and annihilation operators through
the field variables. The $H_0$ piece can be decomposed as follows
in the finite volume. For convenience we use the rescaled
variables given by
\ba
p^a_1&=&(\rho_1 V)^{-1/2}\pi^a_{10},
\;\;\;\;\;
p^a_2=\sqrt{{\rho_1 \over V}}\pi^a_{20},
\;\;\;\;\;
p^a_3=(\rho_3 V)^{-1/2}\pi^a_{30} ,
\nonumber \\
q^a_1&=&(\rho_1 V)^{1/2}\bar{\phi}^a_{1},
\;\;\;\;\;
q^a_2=\sqrt{{V \over \rho_1}}\bar{\phi}^a_{2},
\;\;\;\;\;
q^a_3=(\rho_3 V)^{1/2}\bar{\phi}^a_{3} ,
\ea
and use the radial variables for $q^a_1$
\be
q^a_1=\sqrt{\rho_1 V}(v+\sigma) n^a= \rho n^a .
\ee
The derivatives for the $q^a_1$ are now substituted by
\be
{\partial \over \partial q^a_1}=n^a 
   {\partial \over \partial \rho} 
   +(\delta^{a\alpha}-n^an^{\alpha})
   {\partial \over \partial n^{\alpha}}  ,
\ee
where the index $a$ runs from $1$ to $N$ while the index
$\alpha$ only runs from $1$ to $N-1$. 
The main difference lies in the derivative term with respect
to the rotator variable $n^a$. In Chapter~(\ref{ch:QUAN}),
this was neglected because we were in the infinite volume.
This term practically serves as the kinetic energy of the
zeromode variable $n^a$. 
One can establish 
the following identity
\be
i p^a_1p^a_2 = i p_{2L}p_{1 \rho} -i p^a_{2T}        
         {n^b \over \rho} L^{ab}_0 ,
\ee
where $L^{ab}_0$ is the generator of the variable
$q^a_1$ only, i.e.,  
\be
L^{ab}_0= (-i)(q^a_1 {\partial \over \partial q^b_1}
            -q^b_1 {\partial \over \partial q^a_1} ) .
\ee
With these transformations, $H_0$ is further decomposed into
three parts
\ba
H_0&=& H_{0L}+H_{0T}+H_{0LT} 
\nonumber \\
H_{0L}&=& i p_{2L}p_{y} +{1 \over 2}p^2_{2L}
        +{1 \over 2}p^2_{3L}+{\rho_2 \over 2\rho_3}q^2_{3L}
        +i \sqrt{{\rho_1 \over \rho_3}}q_{2L}q_{3L}
        +{m^2_0 \over 2\rho_1} y^2 ,
\nonumber \\
H_{0T}&=& {1 \over 2}p^a_{2T}p^a_{2T}
        +{1 \over 2}p^a_{3T}p^a_{3T}
        +{\rho_2 \over 2\rho_3}q^a_{3T}q^a_{3T}
        +i \sqrt{{\rho_1 \over \rho_3}}q^a_{2T}q^a_{3T} ,
\nonumber \\
H_{0LT}&=& 
                   (-i)p^a_{2T}        
         {n^b \over \rho} L^{ab}_0 .
\ea 
The longitudinal part has the same form as the simple oscillator
and can be diagonalized easily. The transverse part can also be   
diagonalized  as shown in Chapter~(\ref{ch:QUAN})
\be
H_{0T}=\sum_{i \neq 0,a}a^{(+)a}_{i0T}a^{(-)a}_{i0T} \omega_{i0T} 
\ee
where the summation of $a$ is from $1$ to $N$ and the 
energy gap is $\omega_{10T}=\cm_g$ and 
$\omega_{20T}=\cmb_g$. 
In terms of these operators we can write out the explicit
form of $p^a_{2T}$ 
\be
p^a_{2T}=\sum_{i\neq 0} \sqrt{\omega_{i0T} \over 2} \epsilon_{i}
          (a^{(-)a}_{i0T}-a^{(+)a}_{i0T}) ,
\ee
where the polarization factor $\epsilon_i$ is given by
$\epsilon_1=\epsilon^{*}_2=i/(1-e^{-4i\theta_g})^{1/2}$.

To summarize, in the finite volume we would have the following
Hamiltonian
\ba
H&=& H_0 + H_{\rm int} +{(-i)\over2}
      \left( p^a_{2T}{n^b \over \rho}L^{ab}_0
           +{n^b \over \rho}L^{ab}_0 p^a_{2T} \right) ,
\nonumber \\
H_0&=& \sum_{i,\bk,\lambda} a^{(+)a}_{i\bk\lambda}a^{(-)a}_{i\bk\lambda}
                           \omega_{i\bk\lambda} ,
\nonumber \\
H_{\rm int}&=& \sum_{\bx}
      {4 \lambda_0 v }h(h^2+
           \tilde{\phi}^{a}_{1T}\tilde{\phi}^{a}_{1T})
     +{\lambda_0  }(h^2+
           \tilde{\phi}^{a}_{1T}\tilde{\phi}^{a}_{1T})^2 .
\ea
In this expression, the first two terms are just the Hamiltonian of
the model in the broken phase in the infinite volume. The third
term is a purely finite volume correction which describes the coupling
between the zeromode and the rest of degrees of freedom. The
index $\lambda$ takes the value $L$ and $T$ respectively. All the 
operators can be expressed in terms of the creation and annihilation
operators as
\ba
h(\bx)&=& \sum_{i\bk} {c_{iL} \over \sqrt{2\omega_{i\bk L}V}}
      \left( n^a a^{(-)a}_{i\bk L} e^{i\bk\cdot\bx}
           +n^a a^{(+)a}_{i\bk L} e^{-i\bk\cdot\bx} \right) ,
\nonumber \\
\tilde{\phi}^a_{T}(\bx)&=& \sum_{i\bk \neq 0} 
       {c_{iT} \over \sqrt{2\omega_{i\bk T}V}}
      \left(  a^{(-)a}_{i\bk T} e^{i\bk\cdot\bx}
           + a^{(+)a}_{i\bk T} e^{-i\bk\cdot\bx} \right) ,
\\
\rho&=&\sqrt{\rho_1 V}(v+\sigma)=\sqrt{\rho_1 V}\left( v
          +\sum_{i}{c_{iL} \over \sqrt{2\omega_{i0L}V}}
          (a^{(-)}_{i0L}+a^{(+)}_{i0L}) \right) ,
\nonumber
\ea
where the form factors $c_{i\lambda}$ are given by the following
table
\ba
c_{0L}&=& \sqrt{ \cm^2 \cmb^2 \over (m^2_0-\cm^2)(m^2_0-\cmb^2)},
\;\;\;c_{1L}=c^{*}_{2L}
=\sqrt{ \cm^2 \cmb^2 \over (\cm^2-m^2_0)(\cm^2-\cmb^2)},
\nonumber \\
c_{0L}&=& 1,
\;\;\;c_{1T}=c^{*}_{2T}=\sqrt{\cmb^2_g \over (\cm^2_g-\cmb^2_g)} .
\ea
The creation and annihilation operators enjoy the following
commutation realtions
\be
[ a^{(-)a}_{i\bk\lambda} , a^{(+)b}_{j\bp\lambda^{'}} ] 
= \delta_{ij}\delta_{\bk \bp}\delta_{\lambda \lambda^{'}}P^{ab}_{\lambda}
.
\ee
In the higher derivative model we have the similar relation
for the $O(N)$ generators acting on the ground state
\be 
L^{ab}_0 |0\rangle = i \sum_{i \bk \neq 0}
   g_{i \bk} a^{(+)[a}_{i\bk L}a^{(+)b]}_{i\bk T} |0 \rangle .
\ee
With these relations we can now calculate the rotator contribution
to the energy of the state. Due to the selection rule for the
operator $p^a_{2T}$, the first order correction vanishes. The lowest
order correction comes in at the second order in the perturbation 
Hamiltonian. Using the representations of the operators in terms
of the creation and annihilation operators, it is easy to show
that the first correction is simply the rotator energy,
\be
E^{(1)}_{0l} = [ l(l+N-2) + \Delta_N ] \omega_r .
\ee
Therefore, just like in the conventional $O(N)$ model in the
broken phase, the rotator energy spectrum is the most densely
spaced excitation  and dominates the invariant correlation
functions.

\vfill\eject

%% file: c5.tex
\chapter{Simulation Results and Discussions}  
\label{ch:SIMU}

\section{ Simulation Algorithms }

Finding the right algorithms for the higher derivative $O(N)$
model has been quite tricky \cite{hhiggs5,dallas5}.
In the beginning of this project, we ran many tests
on the existing algorithms for our model.  First we tried 
some conventional update algorithms, for example: metropolis, 
heatbath and hybrid Monte Carlo. But these type of 
algorithms had several serious problems.
One of these problems was that due to the next-next nearest neighbor
coupling terms in our model, the neighbor gathering process 
becomes a rather time consuming task. In four dimensions, with the
naive discretization, we would have had to collect the field
variables at $128$ neighbors for every lattice point. 
Compared with the ordinary theory, this is 
a factor of $16$ more.  
Another problem of such algorithms was the critical slowing
down when close to the criticality. 
This second problem was understandable
because, in our model, the spectrum of the Fourier modes is greatly 
broadened by the higher derivative term. 
To understand more about this issue, 
let us look at the autocorrelation
time in a standard hybrid Monte Carlo algorithm. 

Consider the higher derivative free field theory governed
by the Euclidean Lagrangian
\be
L_E= \sum_{\bp} \omega^2_{\bp} \phi(\bp) \tilde{\phi}(-\bp) ,
\ee
where the spectrum $\omega^2_{\bp}=\bp^2+\bp^6/M^4+m^2_0$. 
The acceptance and autocorrelationts in this Gaussian type of
hybrid  Monte Carlo has been  studied by A. D. Kennedy et. al.
 \cite{kenn5}. The autocorrelation time of the algorithm
was found to be: 
\be 
\tau= {2\tau_0 \over 1-\sqrt{1 -(2 \omega_{min}\tau_0)^2} } ,
\ee
where $\tau_0$ is the average length of each hybrid trajectory.
The quantity $\omega_{min}$ is the lowest frequency of the
Fourier modes, i.e. $\omega_{min}=\min_{\bp} \omega_{\bp}$.
The minimum of the autocorrelation time is obtained when
$\tau_0=1/(2\omega_{min})$ with the value $\tau=1/\omega_{min}$.
For the stability of the leapfrog integration scheme, the step
size cannot exceed $(1/\omega_{max})$, where 
$\omega_{max}=\max_{\bp}\omega_{\bp}$ is the highest frequency of
the Fourier modes. Therefore, the computer time that 
the algorithm consumes to generate an independent configuration
is given by
\be
T_{comp} \sim {\omega_{max} \over \omega_{min} } .
\ee
Thus, the computer time needed to generate an independent
configuration greatly depends on how broad the extent of the spectrum.
In the conventional model, the highest frequency is given by
$\omega_{max}=\sqrt{16+m^2}$. The lowest frequency is 
just $m$.   
With the higher derivative term added, the extension of 
this frequency is much broader than the former case. 
The highest frequency changes to
$\omega_{max}=\sqrt{16+ (16/M^2)^3+m^2}$  while
the lowest frequency remains unchanged. 
For the parameter range of $M$ where we perform our simulation,
this highest frequency is larger by a factor of $10$ or more. Therefore,
the autocorrelation time is enormous for the higher derivative
theory in standard hybrid Monte Carlo due to the broadening
effect of the frequency. 

For the Gaussian model, this effect can be overcome by
the so-called Fourier acceleration procedure 
\cite{parisi5,batrouni5,kogut5}, which is nothing
but noticing that the ideal
algorithm for the free Lagrangian above is to perform the simulation
in Fourier space by adding the momentum dependent kinetic 
energy part
\be
H = \sum_{\bp} {1 \over \omega^2_{\bp}}
       \tilde{\pi}(\bp)\tilde{\pi}(-\bp)
    +\omega^2_{\bp} \phi(\bp) \tilde{\phi}(-\bp) .
\ee
This $\bp$-dependent kinetic energy part will take into
account exactly the frequency differences of the modes and, in fact, the 
$\bp$-dependence for the step size then drops out completely 
from the Hamilton equation of motion, 
as one can easily check.
This hybrid algorithm is then equivalent to simulating $V$ independent
harmonic oscillators with frequency $1$ 
in lattice units. However, nobody would be impressed if
one can simulate a free theory effectively. When the interaction
terms are added, doing the simulation completely  in Fourier space is
sometimes hopeless. This is particularly true if the interaction is 
of the $\phi^4$ type, which is completely local in real
space, but highly nonlocal in Fourier space. Therefore, the 
hope is that we use a Fast Fourier Transformation program to
go back and forth between the real space and the Fourier space.
When the quadratic parts are evaluated, we go to the
Fourier space, and when the interaction part is needed, we go
to the real space. Obviously, this depends greatly on how fast
one can do the Fourier transform. It turns out the existing 
FFT package runs reasonably well on the cray with a speed of
$300-500$ Mflop on the C90-machine. Another complication is that
in the interacting theory we do not know what type of $\bp$-dependent     
kinetic energy term to add. The only clue is perturbation theory, 
however, one would expect that the low energy modes should be 
very well described by the renormalized parameters. It turns out
that the main effect is the broadening effect due to $M$, and
$M$ does not get renormalized very much. Therefore,  putting in the
bare value for $M$ basically overcomes most of the critical
slowing down. We are able to perform the simulation with an
autocorrelation time which is below $10$ hybrid Monte Carlo trajectories  
with each trajectory consisting of $15-20$ steps. Although this 
performance is not ideal, it works thousands of times better than
the old programs, for which the autocorrelation time was hopelessly
long. Also, in the Fourier accelerated Hybrid Monte Carlo,
it is trivial to extend the algorithm to the improved actions.
Since the quadratic part is evaluated in the Fourier space, it 
does not cost anything more for us to use the improved propagator
as compared with the naive one. If this were implemented in the
real space, it would require a lot more work. 

All of  our results were obtained with the appropriate Fourier
accelerated Hybrid Monte Carlo program. We currently have only 
the version for the finite bare coupling constant. Therefore, 
all results presented here are for some finite bare coupling constant.
However, some of our simulation points have a rather large
bare coupling constant  in continuum notation, therefore, 
we expect that most of the physically interesting results
will be quite similar in the nonlinear limit.

\section{ The Extraction of Physical Parameters }

We will now  extract some physical quantities from our 
simulation results \cite{dallas5}. 
One of the most interesting quantities is
the vacuum expectation value $v$. This is the quantity which
sets the energy scale of the simulation.
In the old simulations, this parameter was obtained by 
measuring the bare expectation value of the averaged 
field variable. The wave function renormalization constant
was then obtained from a linear fit to the momentum space
propagator. From these quantities, the 
renormalized vev is then obtained using
\be
v_R=Z^{-1/2} v_0 .
\ee
The crucial point is the measurement of the wave function
renormalization constant. But in our case, things are more
complicated. The momentum space propagator  will 
not only contain the usual $p^2$ term, but will also contain the higher
derivative terms. In general, the interaction will
 generate more terms which were not in
the bare free propagator. This makes it more difficult for
us to get a very accurate determination of the
wave function renormalization constant. 

Another way of extracting the renormalized vev is  
from the rotator correlation functions. Using the
theory discussed in Chapter~(\ref{ch:BOA}) , we can
write down an expression for the rotator correlation
function $n^a(0)n^a(\tau)$, where $n^a(\tau)$ is
the unit vector of the zeromode at a given time
slice $\tau$  
\be 
\langle n^{a}(\tau) n^{a}(0) \rangle = A
\sum_{l} l(l+1)e^{-\beta \omega_r (l(l+1)-1/2)}
\cosh [(2l+1)(\tau-\beta/2)\omega_r] ,  
\ee
where $\omega_r=(2L^3v^2_r)^{-1}$ is the
rotator energy unit.
This correlation function is dominated   
by the rotator energy spectrum in the finite volume.
All the other energy excitations are much higher 
than the rotator energy scale. Usually the lowest one
is the one Higgs contamination, whose energy 
scale is an order of magnitude higher. This
correction can be easily taken into account according to
the formula given in Chapter~(\ref{ch:BOA}) . Since the rotator
energy depends only on the renormalized vacuum 
expectation value (and the 3 volume), this is
a direct way of extracting the vev.
In our simulations, we have tried both methods and have obtained
compatible results.
\begin{figure}[htb]
\vspace{10mm}
\centerline{ \epsfysize=3.0cm
             \epsfxsize=5.0cm
             \epsfbox{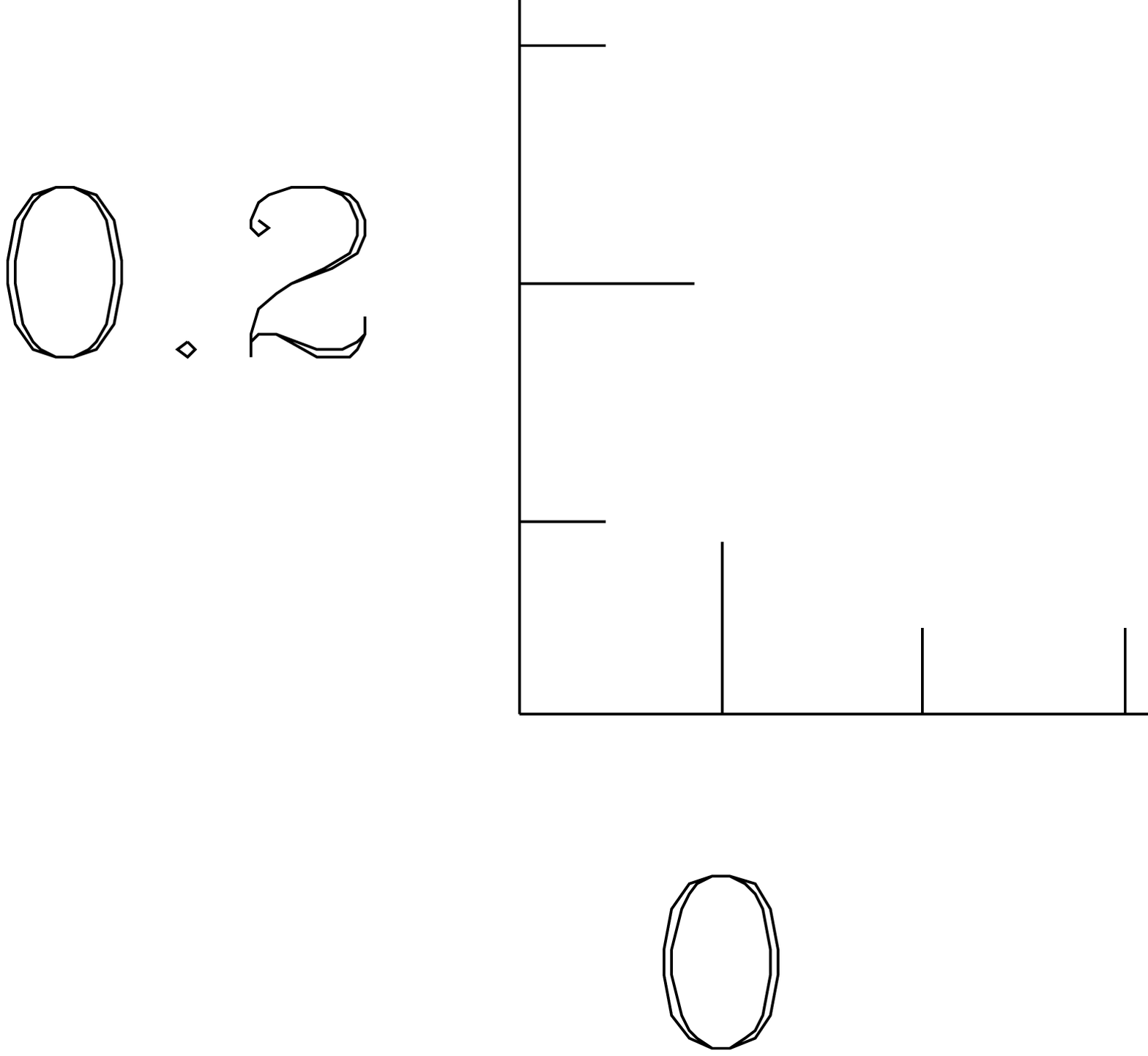} }
\vspace{17mm}
\caption{ The rotator correlation function is shown together with
the theoretical fit. The fit starts at $\tau=6$ and the quality of
the fit is good. The disagreement of the theoretical curve with the
data for small values of $\tau$ is because of the high energy contaminations. }
\label{fig:ch5.rotcor}
\end{figure}

In Figure~(\ref{fig:ch5.rotcor}), a typical rotator correlation function
is shown compared with the fit to the theoretical
form. The bare parameters are shown at the top
of the figure. The lattice size for this run is $16^3 \times 40$.
The output data has a total statistic of $32$k hybrid Monte Carlo 
trajectories. 
At very short distances, higher energy excitations
will contribute. Therefore, the fit was performed from $\tau=6$
all the way to the end. The fit is very stable if the starting
point is after $\tau=5$. The fit is also very stable with respect
to the number of rotator states ($nmax$ in the figure) that has been included.
It turns out that any number which is greater 
than $3$ would be adequate. 
In this fit, the correction of the single Higgs state is included 
using the formula described in Chapter~(\ref{ch:BOA}). 
This correction is about
$10$ percent even at large $\tau$ values. This is because
of the small vev value of our simulation. The corrections due to the
other states are all very small at large $\tau$ values. 
It is clear that we have found a very good agreement with
the theoretical formula.

\begin{figure}[htb]
\vspace{15mm}
\centerline{ \epsfysize=3.0cm
             \epsfxsize=5.0cm
             \epsfbox{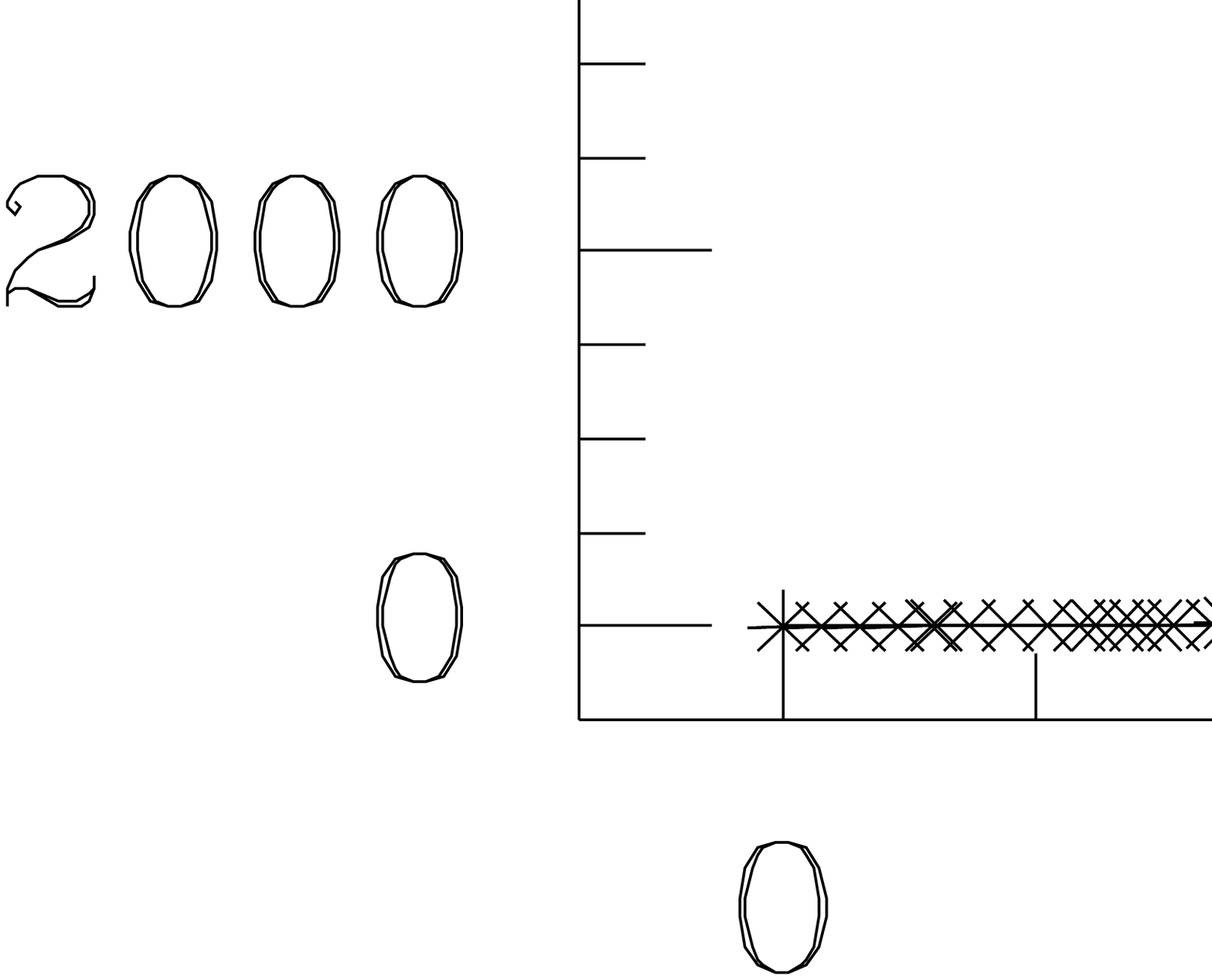} }
\vspace{12mm}
\caption{ The momentum space Higgs propagator is plotted as a function
of the lattice momentum squared for the bare parameters shown at the
top. The solid curve is a fit of the data to the polynomial form
up to order $\hat{p}^6$. The upper window is a magnified portion of
the lower window in the range $\hat{p}^2 < 4$. The quality of the
fit is reasonable, however, due to the 
ambiguity of the fitting functional form, the error in the
fitted wave function renormalization constant $Z$ is rather large. } 
\label{fig:ch5.momhig}
\end{figure}
For comparison, the momentum space Higgs propagator is shown in
Figure~(\ref{fig:ch5.momhig}).
 This momentum propagator was obtained from a run of the
same input bare parameters as in Figure~(\ref{fig:ch5.rotcor})
 except that it was on
a cubic geometry of $16^4$  with the statistic of $20$k trajectories.
The form of the fitting function is taken to be
$f(p^2)=Z^{-1}\hat{p}^2+Z^{-1}m^2+p_2\hat{p}^4+p_3\hat{p}^6$.
Note that the size of the coefficient of
$\hat{p}^4$ term is quite significant which is a signal of
strong interaction effects. We should keep in mind that the above
function has no justification if the interaction is strong.
In general, the interaction could introduce complicated functional
forms to the full Higgs propagator. It could generate
$\log \hat{p}^2$ terms, higher polynomial terms and even terms that
cannot  be written as functions  of $\hat{p}$ alone. Therefore, the
size of the interaction terms like $\hat{p}^4$ basically reflects
the ambiguity of the fit. If we had tried the same fit but setting
the coefficients of $\hat{p}^4$ to zero, we would have arrived at a
rather different value of $Z$ ( $Z^{-1}=1.33$ ).
From this we conclude that, due to the strong interaction, it would
be very difficult to extract the wave function renormalization constant
from the momentum space propagator.
Other methods are needed for the extraction of the physical
parameters and the momentum space propagator can only be used as an 
independent check.

Another important quantity is the mass of the
Higgs particle. In the old simulations, there were also
two ways of obtaining the Higgs mass. 
One way is to use a
fit to the momentum space propagator. The mass obtained 
this way has both advantages and disadvantages. The advantage 
is that the signal is very clean and we get a very
stable fit for the mass even with low statistics. 
We can fit the very low momentum portion of the  momentum 
space propagator where the effects of the interaction
terms are small and the mass values are rather stable.
The disadvantage is
that the mass obtained from the propagator is not yet the 
physical Higgs mass. We must use perturbation
theory to relate the two masses. This is legitimate in
the old $O(4)$ calculation because, in that case, the
theory is perturbative and the perturbative formula offers
us a rather accurate prediction. In a truly nonperturbative
theory, however, this could be misleading. The mass obtained
from the propagator fit, what we call the off-shell mass, 
could deviate significantly from the physical mass.

Another way of determining the Higgs mass is from the time
slice correlation function of the Higgs field. In this 
approach, the lowest energy gap of the Higgs excitation
is extracted and identified as the Higgs mass in the
finite volume. This, of course, should be closer to 
the physical mass than the off-shell mass and, in a
strongly interacting theory this is the only way to
get a good control of the Higgs mass. In our simulation of
the higher derivative theory, the interaction is much stronger
than the conventional $O(4)$ case, therefore, we used this method to
extract the Higgs mass. The off shell Higgs mass was also 
determined and only served as a comparison. 
\begin{figure}[htb]
\vspace{20mm}
\centerline{ \epsfysize=3.0cm
             \epsfxsize=5.0cm
             \epsfbox{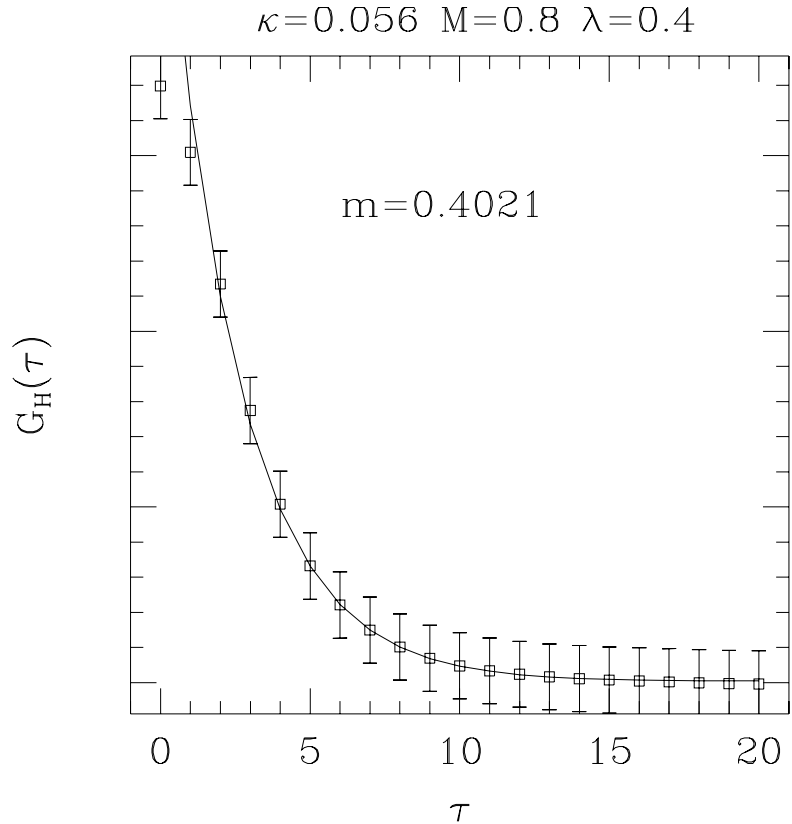} }
\vspace{17mm}
\caption{ The time sliced Higgs propagator is plotted as a function
of the Euclidean time separation $\tau$ for the bare parameters shown at the
top. The solid curve is a fit of the data to the  single Higgs
excitation. The fit was done in the range $5<\tau<17$ to ensure that
the higher energy excitations have died out.
The quality of the fit is reasonable, but the error for the mass
parameter remains to be determined. }
\label{fig:ch5.higcor}
\end{figure}

In Figure~(\ref{fig:ch5.higcor}), 
we have shown the time sliced Higgs correlation function 
as a function of the Euclidean time separation $\tau$. The bare
parameters are also shown at the top of the figure. At small distances,
all higher energy excitations contribute, including the ghost states.
Therefore, to ensure that we extract the lowest radial excitation, we
started the fit from some $\tau$ values so that the fit was stable 
from there on. The functional form that we used is the standard 
hyperbolic cosine function for a single excitation.
The data of the correlation function is derived from
a blocking analysis of $32$k hybrid Monte Carlo trajectories. If we 
compare this fitted mass value  with the off shell mass, we find
that the difference is very significant, which means the interaction
is really much stronger when compared with the conventional $O(4)$ case.
The data points of the correlation function are highly correlated.
Therefore, we should develop a method to determine the error of
the fitted mass value.

To determine the error of the mass parameter, we performed the following
blocking procedure. The output data is originally divided into small
blocks. For this particular example, we had $80$ blocks available.
Due the large fluctuation, a single block is not enough to give stable
mass values. Therefore, the small blocks are first grouped together to
form $N_b$ larger blocks, large enough so that we can extract stable mass
values from them. For each large block $i$, the following ratio is formed 
\be
R_i(\tau) \equiv { G_i(\tau +1)-G_i(\tau) \over
                 G_i(\tau)-G_i(\tau -1) } ,
\ee
where $i$ runs from $1$ to the total number of large blocks $N_b$.
If we have only a single excitation that dominates the correlation
function, then
the correlation function should be of the form
\be
G^{theo}(\tau)=A \cosh[m(\tau-L_t/2)] + B .
\ee
Therefore, the ratio should only depend on the mass $m$ and
the Euclidean time separation $\tau$,
\be
R^{theo}(\tau) ={\cosh[m(\tau+1-L_t/2)]-\cosh[m(\tau-L_t/2)]
                \over
                 \cosh[m(\tau-L_t/2)]-\cosh[m(\tau-1-L_t/2)] } .
\ee
Then the blocked values $R_i(\tau)$ are set to the theoretical value
and we can solve for the mass numerically for each $\tau$. The outcome of this
procedure is called the ``effective mass'', denoted as $m^i_{eff}(\tau)$.
Then, the averaged effective mass is obtained by
\be
m_{eff}(\tau)={1 \over N_b} \sum^{N_b}_{i=1} m^i_{eff}(\tau) .
\ee
We can also obtain an error for the effective mass by
\be
\Delta m_{eff}(\tau)=\sqrt{ {1 \over N_b (N_b-1)} \sum^{N_b}_{i=1} 
     [m^i_{eff}(\tau)-m_{eff}(\tau)]^2 } .
\ee
\begin{figure}[htb]
\vspace{13mm}
\centerline{ \epsfysize=3.0cm
             \epsfxsize=5.0cm
             \epsfbox{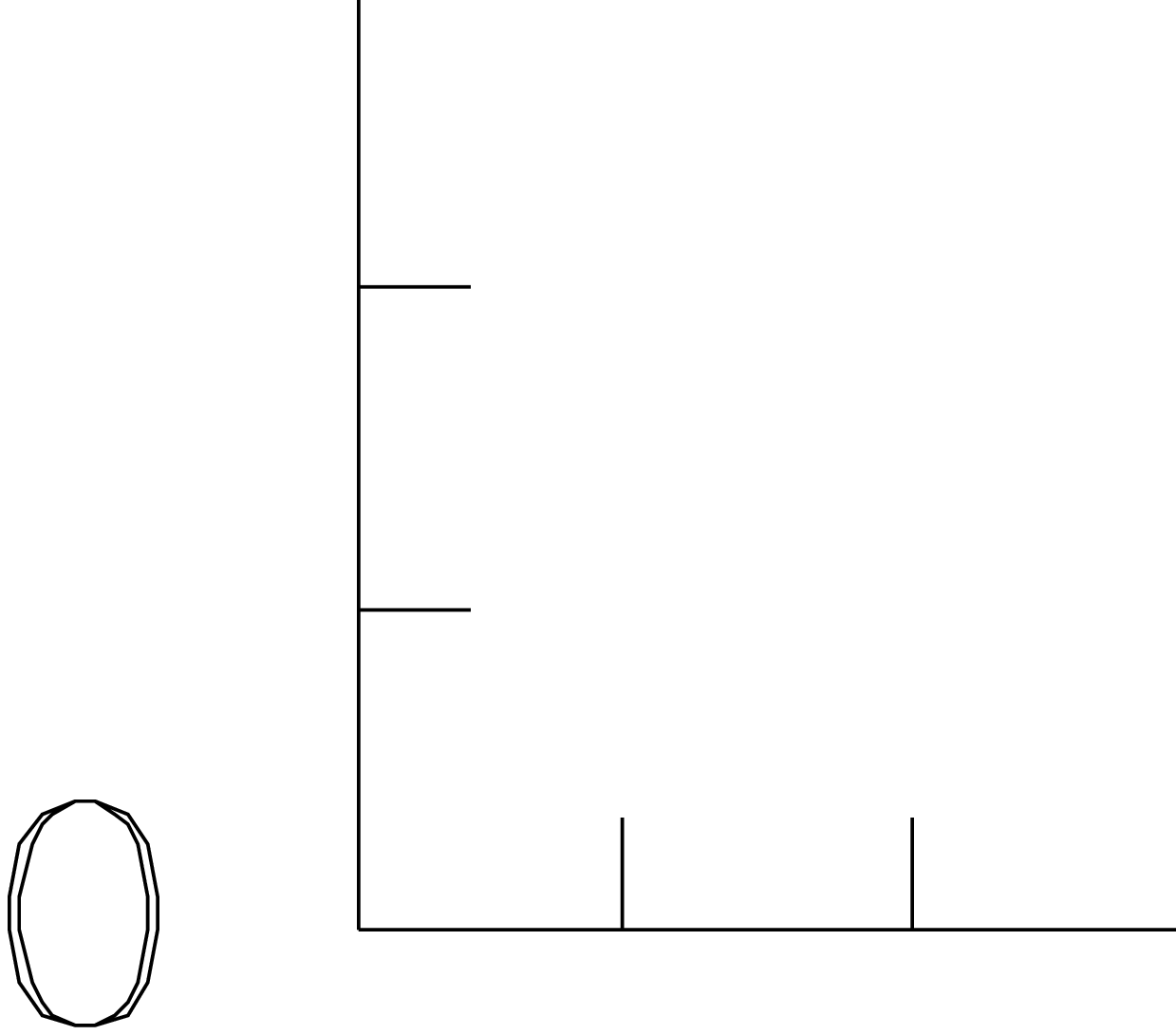} }
\vspace{17mm}
\caption{ The effective mass plot for the time slice Higgs correlation
 function for the bare parameters listed at the top.
 The Higgs mass value is obtained from the $\chi^2$ fit to the plateau
 starting at $\tau=7$. The dashed line tick marks denote the range of
 the fit. The horizontal solid line is the fitted mass value which is
 also labeled in the figure. The horizontal dashed lines denotes the
 error of the fitted mass value. The mass value from the effective mass
 plot is consistent with the value 
 from the exponential fit.}
\label{fig:ch5.higmeff}
\end{figure}
We can then plot the effective mass  as a function of the time
separation $\tau$, together with the appropriate errors. 
This is shown in Figure~(\ref{fig:ch5.higmeff}).
We  found
that, 
since many states contribute for small values of $\tau$, the 
effective mass is varying with $\tau$. However, if we go to $\tau$
values that are large enough, all the higher energy excitations die 
out exponentially and the lowest energy excitation dominates. Therefore
starting from some $\tau$ value, we should see a plateau behavior of
the effective mass. The value of the plateau should basically be
the energy of the lowest energy excitation. 
Since the signal is getting exponentially small with $\tau$, the error
of the effective mass function will grow significantly with $\tau$. 
Usually near the endpoint ($\tau=L_t/2$), the errors become so
large that effective mass value is no longer meaningful. We 
can then perform a $\chi^2$ fit to the effective mass, giving higher
weight to the more accurate points. From this fit, we can determine 
the mass and its error.

But this is not the whole story yet. In fact even in the second
approach, what we extract is not the infinite volume Higgs mass.
The reason for this is very simple. All the simulations are
done in a finite volume, and finite size effects must 
be taken into account. Among all the finite effects, there is
one effect that is most disturbing. In the infinite volume, the
Goldstone particles are exactly massless. Therefore, the Higgs
particle can decay into two Goldstone particles, thus the Higgs 
has a finite lifetime. In the simulation, however, because the
volume is finite, the lowest Goldstone pair is not at zero
energy, but is equal to $4\pi/L$. This number is rather
large for most of our simulations. In fact, it is larger 
than the Higgs mass itself. So the situation that we have in
our simulation is that the Higgs is lighter than the Goldstone
pair, and it therefore cannot decay. Of course, when the volume 
is increased, the Higgs mass energy level will meet the two
Goldstone levels and the so-called level crossing phenomenon occurs. This 
was  noticed quite some time ago. In fact, many groups have 
used this picture to get both the physical Higgs mass and its
width from the measurement of the two Goldstone levels. In this
picture, the Higgs is viewed as a resonance of the Goldstone
Goldstone scattering process. L\"uscher derived a formula which
relates the Goldstone pair energy level in the finite volume
to the infinite volume Goldstone-Goldstone scattering phase
shift. By measuring the two Goldstone energy levels as accurately 
as possible for various volumes, one gets the continuum
scattering phase shift profile in an energy range. If all the
parameters are well chosen, one would be able to see a phase
shift stepping from almost zero to almost $\pi$ exactly at
the threshold energy which is equal to the physical Higgs mass.
One would also be able to get the physical width of the Higgs
by fitting it to the Breit-Wigner shape near the resonance.
So, instead of fighting against the finite
volume effects, one could utilize it to gain precious information
about the continuum theory. 

To carry out a similar calculation in our model is more difficult
than the usual $O(N)$ model. First of all, we must establish
an equivalent formula in the higher derivative theory which can
relate the energy levels in the finite volume to the phase shift
in the infinite volume. Secondly, we have extra particles in our
model, namely the ghost pairs. We have to control their 
contribution to the correlation functions in order to get
reliable results for the two Goldstone energy levels. Thirdly,
our model requires much more computing power to get good 
stable results for the time sliced correlation functions.
The detailed analysis of this problem is given in the
next chapter.

The simulation results we have obtained belong to one of  
the following two categories. One is performed with the naive
discretization action and the other category is performed by using
the improved action. We have done simulations in both 
phases of the theory. The following table summarizes the 
bare parameter and extracted physical quantities of the points.
\begin{table}[htb]
\begin{tabular}{@{\hspace{2mm}}c|c@{\hspace{3.2mm}}
                  c@{\hspace{3.2mm}}
                  c@{\hspace{3.2mm}}
                  c@{\hspace{3.2mm}}
                  c@{\hspace{3.2mm}}|
                  c@{\hspace{3.2mm}} 
                  c@{\hspace{3.2mm}} 
                  c@{\hspace{3.2mm}}|
                  c@{\hspace{3.2mm}}}
\hline
$P$ & $\kappa$ & $M$ & $\lambda$ & $V$ 
      & Stat & $v_0$ & $v_r$ & $m_H$ & $m_H/v_R$\\
\hline
\hline
$A$ & $0.056$ & $0.8$ & $0.4$ & $16^3*40$ 
     & 32k & $0.0478(1)$ & $0.057(2)$ & $0.40(2)$ & $7.0(4)$ \\
$B$ & $0.056$ & $0.8$ & $0.4$ & $20^3*40$ 
     & 20k & $0.0365(1)$ & $0.045(1)$ & $0.33(2)$ & $7.3(5)$ \\
$C$ & $0.105$ & $0.8$ & $0.1$ & $16^3*40$ 
     & 40k & $0.0607(1)$ & $0.065(2)$ & $0.31(2)$ & $4.8(4)$ \\
$D$ & $0.115$ & $0.8$ & $0.05$ & $16^3*40$ 
     & 60k & $0.0798(1)$ & $0.082(1)$ & $0.24(1)$ & $2.9(1)$ \\
$E$ & $0.081$ & $1.0$ & $0.3 $ & $16^3*40$ 
     & 28k & $0.0878(1)$ & $0.093(1)$ & $0.42(2)$ & $4.5(2)$ \\
$F$ & $0.081$ & $1.0$ & $0.3 $ & $20^3*40$ 
     & 24k & $0.0826(1)$ & $0.088(1)$ & $0.38(2)$ & $4.3(3)$ \\
$G$ & $0.053$ & $0.8$ & $0.4 $ & $16^3*16$ 
     & 70k & $---$ & $----$ & $0.434$ & $4.1(7)$ \\
\hline
\hline
$H$ & $0.088$ & $2.0$ & $0.99 $ & $16^3*40$ 
     &108k & $0.0477(1)$ & $0.058(1)$ & $0.351(5)$ & $6.1(1)$ \\
$I$ & $0.088$ & $2.0$ & $0.99 $ & $20^3*40$ 
     & 64k & $0.0354(1)$ & $0.045(1)$ & $0.29(1)$ & $6.4(3)$ \\
$J$ & $0.104$ & $2.0$ & $0.4 $ & $16^3*16$ 
     &100k & $---$ & $----$ & $0.352(5)$ & $2.2(2)$ \\
\hline
\hline
\end{tabular}
\end{table}
In this table, points $A$ through $G$ are  the results for 
the naive action while points $H$ through
$J$ are for the improved action.  Point
$G$ and point $J$ are in the symmetric phase, while 
all other points are in the broken phase.

In the symmetric phase, the important physical quantity
is the renormalized coupling constant, which could be defined
to be the connected $4$-point function at zero external
momenta. In order to get this quantity, the propagator mass is
measured. The renormalized coupling
constant 
is directly measured by forming the connected 
$4$-point function. The
measurement of the  renormalized coupling 
constant is very noisy, which requires large
statistics of the data. We used the following
formula to extract the connected four point function
\be
\lambda_R= {\Omega m^4_R \over 24} ({3 N \over N+2})
           \left(
           { {N+2 \over N} \langle \bar{\phi}^2 \rangle^2 
             -\langle \bar{\phi}^4 \rangle 
           \over
            \langle \bar{\phi}^2 \rangle^2 } 
           \right) ,
\ee
where $N$ is the number of components of the field,  
$m_R$ is the propagator mass and $\Omega$ is the
$4$-volume of the system. The quantity 
$\bar{\phi}^2$ is defined to be    
$\sum^{N}_{a=1} \bar{\phi}^a \bar{\phi}^a $
where $\bar{\phi}^a $ is the $4$-volume average of
the field $\phi^a(x)$. 
The quantity $\bar{\phi}^4$ is just a short hand notation for
$(\bar{\phi}^2)^2$,  and the bracket means the
Monte Carlo ensemble average.
It is the subtraction in the bracket which causes most of 
the noise. Therefore, in order to get sensible results
we have accumulated large statistics for the two points
in the symmetric phase (Point G and J in the table). 
\begin{figure}[htb]
\vspace{15mm}
\centerline{\epsfysize=3.0cm  \epsfxsize=5.0cm  
            \epsfbox{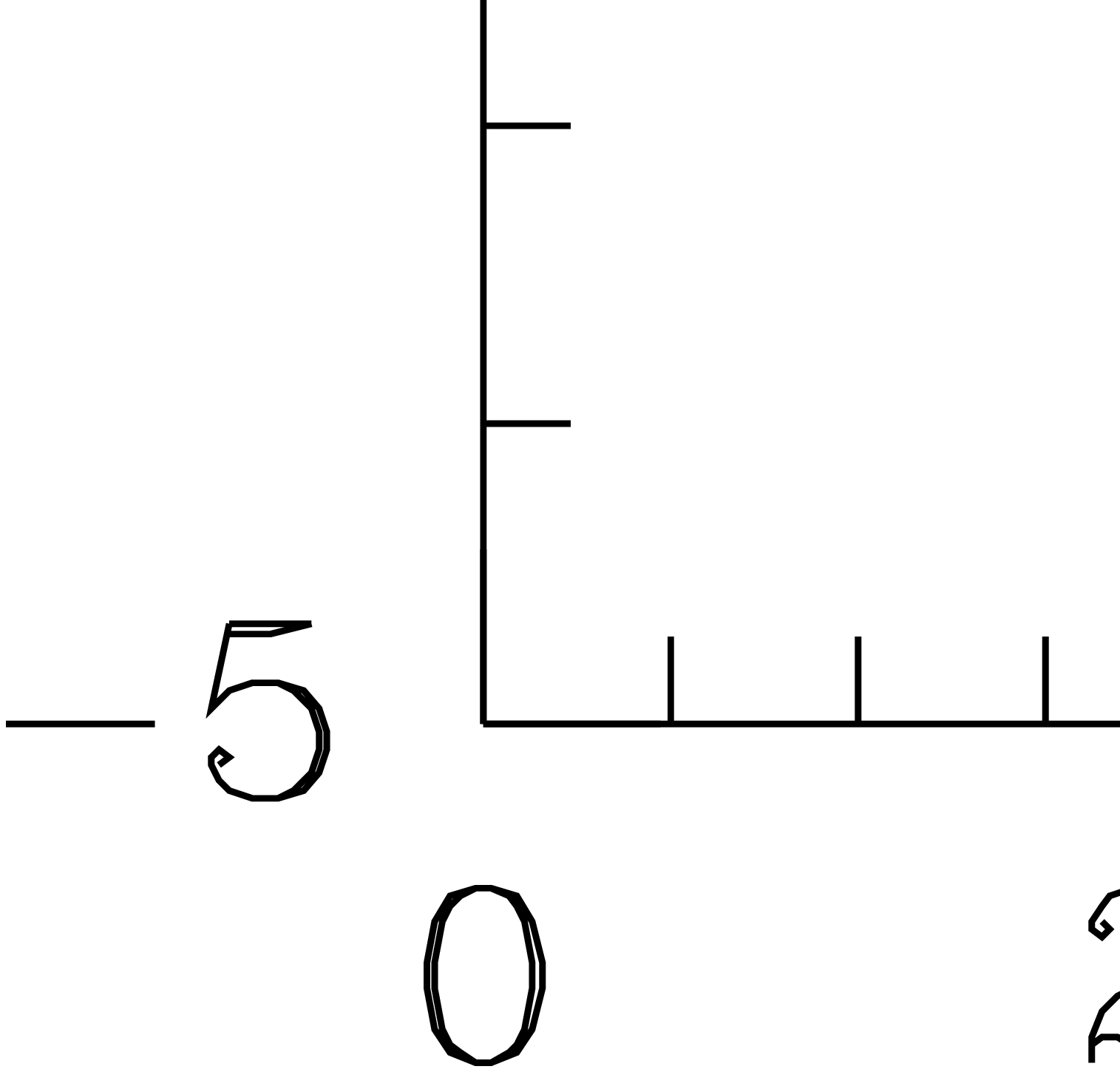}
             }
\vspace{15mm}
\caption{ The renormalized coupling constant (connected four
point function at zero external momenta) is plotted for individual
runs. Due to the subtraction the signal is quite noisy and a 
large statistic is needed to get a sensible accuracy for this 
quantity. }
\label{fig:ch5.symlam}
\end{figure}
In Figure~(\ref{fig:ch5.symlam}),  
we have shown the renormalized coupling constant for
individual runs for the higher derivative $O(4)$ model.
It can be seen that the result is
quite noisy and the points scatter a lot around the 
average. Usually $100K$ is needed for an error of
about $10$ percent. 
 
We can now compare the renormalized coupling constant that
we measured for the higher derivative $O(4)$ model with that
of the conventional $O(4)$ model \cite{kuti5}.
In the symmetric phase (point G and J), 
we found that the renormalized coupling constants
were much larger than in the
old lattice simulation results of $O(4)$ model.
In the conventional
$O(4)$ model, when the correlation length was about $2-3$ 
the renormalized coupling constant $\lambda_R$ was typically
of the order of $0.6 - 0.8$. In our model, however,  we saw
a huge jump (about a factor of $2$ to $3$) 
of the renormalized coupling constant.  This is a signal that
the higher derivative model is much more strongly coupled 
than the conventional $O(4)$ model. 
Recall that, from the large $N$ calculations in Chapter~(\ref{ch:NUNI}), large
$N$ also predicts a jump in the renormalized coupling constant in
the symmetric phase. Therefore, our simulation results agree
with the large $N$ results qualitatively. 

In the broken phase, the renormalized vacuum expectation
values are obtained using the rotator correlation functions 
as described above. The errors are estimated from
a blocking analysis of the data. 

The Higgs mass is  taken to be 
lowest radial energy excitation in the finite volume.
As described above, we tried two ways of extracting
this energy gap. One by fitting the time sliced correlation
function to the hyperbolic cosine function, the other from
the effective mass plot. Both methods gave compatible results
and the errors are determined from the $\chi^2$ fit of the
effective mass plateau in the appropriate range.

Identifying this energy gap with the infinite volume
Higgs mass is of course a rather crude approximation
and is subject to finite volume corrections. However, as
shown in the table, we did not see a  significant change
in the $m_H/v_R$ ratio when the lattice volume was increased.
In fact, they are compatible with each other within errors.
We also tested this within the framework of the large
$N$ approximation. We found that the ratio $E(L)/v(L)$ was 
rather stable when the size of the box was changed, as long
as the box size was not too small and the energy crossing
phenomenon had not occurred. And the value of the ratio was
in agreement with the infinite volume large $N$ value.
Therefore, we expect that this ratio represents the feature
of the continuum higher derivative theory. The correct way
of extracting the Higgs mass has to come from the finite
volume resonance picture, which we will discuss  in the
next chapter.
 
Another issue in the Higgs mass bound problem is to determine
how much scaling violations (cutoff effects) are present
in our results. This turns out to be a rather subtle 
issue.  To study this problem, we have to answer the following
two questions: (1) what is the nature of the scaling violations
in our model and, (2) how can we calculate the scaling violations
once the Higgs mass and the ghost parameters are known.

First we will review how the above two questions are answered in
the conventional $O(N)$ model simulations. In the conventional
$O(N)$ model, the scaling violation is due to the hypercubic 
lattice that violates Euclidean (or rotational) invariance. 
This scaling violation can be defined both perturbatively and
nonperturbatively. To calculate this scaling violation, we can 
check the rotational invariance of some quantity, for example, the
free propagator of the field \cite{lang5}, or
evaluate the Goldstone scattering amplitude and compare with 
perturbation theory \cite{lusc5,hase5,zimm5,neub5}. The second method   
seems to be more closely related to measurable quantities, but
it relies on the perturbative nature of the problem. 
It worked out nicely for the conventional $O(4)$ simply
because even at the highest bound, the theory is still perturbative.
The first method 
offers us an unambiguous result
without using perturbation theory.

Now, let us look at the situation for the higher derivative 
lattice theory.
People tend to think that in the higher derivative $O(4)$ theory
there exist two types of scaling violations. One is the effect due
to the lattice; the other one is what is usually called the
Pauli-Villar cutoff (or ghost) effects. However, such a statement
is very misleading. In fact, as we have shown in the
previous chapters, this should not be the view, at least not the 
only view,  of the higher
derivative theory. This theory is a well defined field theory which
has a unitary $S$-matrix and the ghost effects can easily evade the 
experimental tests. It is also
a well-defined theory free of divergences.
Therefore, if we could do the simulation
in the continuum, we would have had no cutoff effects
at all. It is only because the computer cannot handle 
infinite number of variables that we have to introduce
the underline lattice to the theory. As long as we
can constrain our lattice effects to be small, our 
simulation results should represent the higher derivative
$O(4)$ model in the continuum. In other words, there are no
``ghost effects'' if the ghosts are well
hidden from any experiment.   

As stated previously, in analyzing the lattice effects, 
perturbation theory should only be taken as a hint.
There have been ways of doing
 nonperturbative analysis of the lattice 
effects, though none of them is really 
sophisticated. One of the things that could be
done is to analyze the breaking of the Euclidean
invariance of the free propagator at some given parameters.
This was first discussed by Lang et. al. in 1988 \cite{lang5}. 
Although it only uses the tree level propagator, 
 it is still a very good measurement
of the amount of lattice violations in the theory. 
Obviously, going beyond this using perturbation 
theory is hopeless if the
theory is strongly interacting. One can try to
carry out the same analysis for the propagator
in the large $N$ approximation, but again, the 
justification for the large $N$ approximation at
$N=4$ is not very promising either.

Let us now review some of the basic ideas of how this
procedure is carried out for the propagator.
On the lattice, the propagator in momentum space
is, in  general, a function 
of every individual momentum component.
In the continuum, however, it should only depend on
the combination $p^2=\sum^{4}_{\mu=1}p_{\mu}p_{\mu}$
due to Euclidean invariance. This symmetry is violated on
the lattice and we can define a quantity 
${\cal N}_G$ which represents the 
amount of violation due to the lattice.
For the inverse momentum space propagator, 
the quantity ${\cal N}_G$
is defined in the following way.
Let us pick some prescribed momentum scale  $p_{cut}$ in lattice
units, and pick our reference momentum to be
${\bp}_0=(p_{cut},0,0,0)$. Then we can form all the momenta that
have the same magnitude as this reference momentum in
the form 
${\cal R}{\bp}_0=p_{cut}(\cos\theta_1, \sin\theta_1\cos\theta_2, 
 \cos\theta_1\sin\theta_2 \cos\theta_3, 
\cos\theta_1\sin\theta_2 \sin\theta_3) $. 
We can then define the rotational invariance violation by
${\cal N}_G$ by
\be
\label{eq:ch5.ri}
{\cal N}_G(p_{cut}) = \int d{\cal R} \sqrt{
             {(G({\cal R}{\bp}_0)-G({\bp}_0))^2 
              \over
               G({\bp}_0)^2 } 
              } ,
\ee
where $d{\cal R}$ is the invariant measure for the rotational
group normalize in such a way that $\int d{\cal R}=1$. Obviously
this quantity is identically zero in the continuum where the
rotational invariance is restored. On the lattice, the size
of this quantity is a measure of the lattice effects in the
discretized theory. In principle we can define
similar quantities for other functions. 

\begin{figure}[htb]
\vspace{9mm}
\centerline{\epsfysize=3.0cm  \epsfxsize=5.0cm  
            \epsfbox{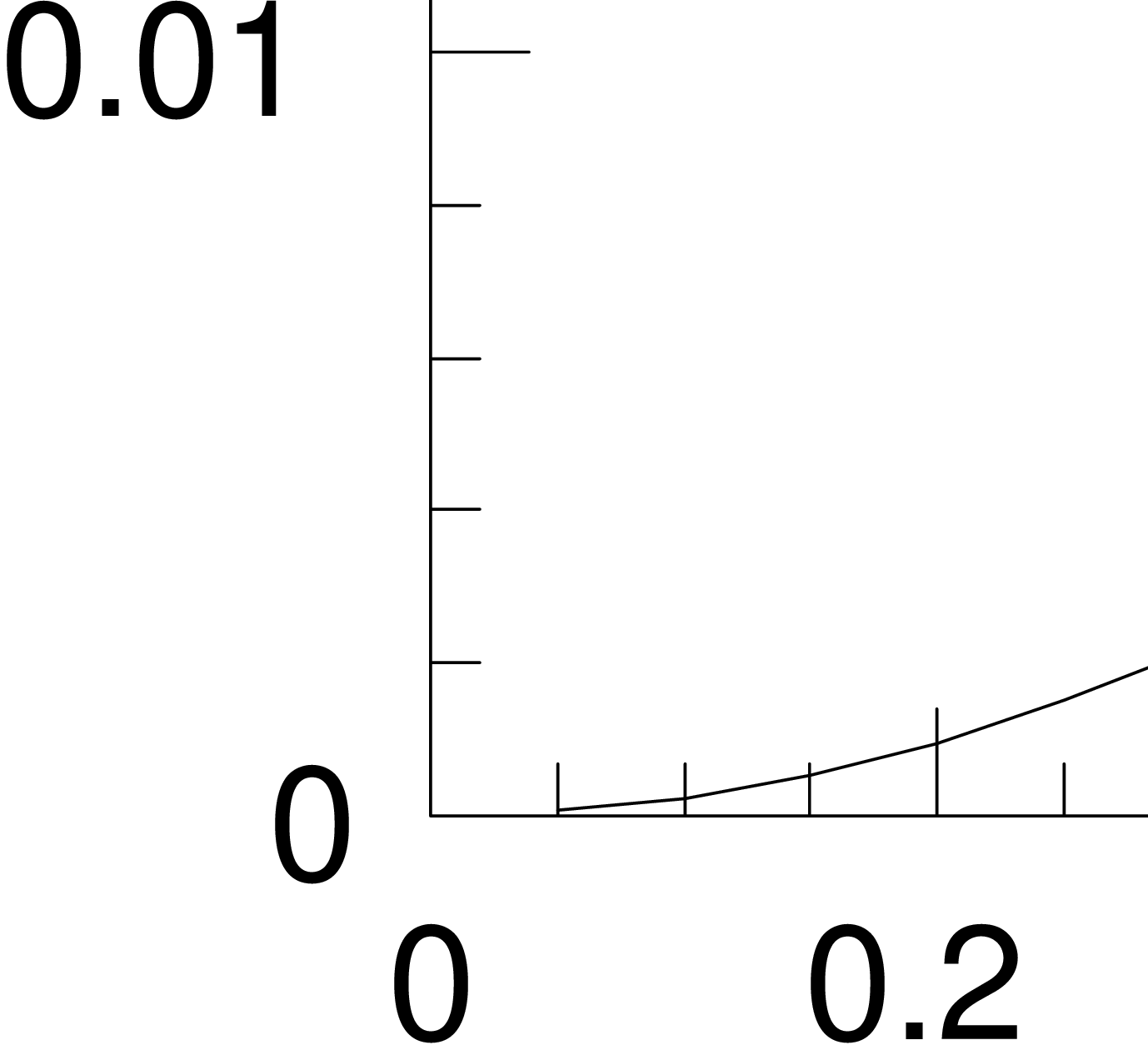}
             }
\vspace{3mm}
\caption{ The rotational invariance violation for 
  the free inverse propagator on an
   infinite hypercubic lattice
   is plotted 
 for various cases. The bottom
  two curves are the naive propagator and the one using 
  the improvement up to 14th order. The upper two boxes
  show the Pauli-Villars case for $M=2$ (the solid lines)
  and $M=0.8$ (the dashed lines) when using the naive and
  improved action. It is clearly seen that for the parameters
 that our simulation are performed , the rotational 
 invariance violation is very small. }
\label{fig:ch5.ri}
\end{figure}

We have performed the rotation invariance analysis for our
simulation points using both the tree level and 
large $N$ approximation. 
In Figure~(\ref{fig:ch5.ri}),  we have shown some of the rotational
invariance violations 
for the tree level propagator of our simulation points 
We found that all our 
simulation points have very small lattice effects.
For example, even with the naive propagator, in 
the Higgs mass range where we did our simulation, 
the rotational invariance violation is not larger
than the old $O(4)$ simulation points with correlation
length of $2-3$.

We can also calculate the finite volume lattice violation
in both the free propagator and in the large $N$ approximation.
In a finite box with lattice structure the lattice momenta 
are discrete and can only be multiples of $2\pi/L$. For each
integer $n_{cut}$ there will be more than one set of solution
$(n^{(i)}_1,n^{(i)}_2,n^{(i)}_3,n^{(i)}_4)$
to the equation $n_{cut}=n_1^2+n_2^2+n_3^2+n_4^2$.
Denoting the total number of solutions by $D(n_{cut})$,
we can then define the counterpart of ${\cal N}_G$ in 
the finite lattice 
\ba
\label{eq:ch5.calgfi}
\bar{G}(n_{cut})&=&{1 \over D(n_{cut})}\sum^{D(n_{cut})}_{i=1}
                 G(({2\pi \over L})n^{(i)}_1,
                   ({2\pi \over L})n^{(i)}_2,
                   ({2\pi \over L})n^{(i)}_3,
                   ({2\pi \over L})n^{(i)}_4) ,
\\
{\cal N}_G(n_{cut})&=&\sqrt{ 
       {1 \over D(n_{cut})}\sum^{D(n_{cut})}_{i=1}
                 { (G(({2\pi \over L})n^{(i)}_1,
                   ({2\pi \over L})n^{(i)}_2,
                   ({2\pi \over L})n^{(i)}_3,
                   ({2\pi \over L})n^{(i)}_4) -\bar{G}(n_{cut}) )^2
                \over 
                \bar{G}(n_{cut})^2 }
                  } .
\nonumber 
\ea
Due the finite size effects, the momentum lattice is coarse grained.
This will result in some zigzag behavior of the function
${\cal N}_G(n_{cut})$, as $n_{cut}$ is increasing.
However, for a  reasonably large lattice, we will recover the 
infinite lattice results. 
\begin{figure}[htb]
\vspace{16mm}
\centerline{\epsfysize=3.0cm  \epsfxsize=5.0cm  
            \epsfbox{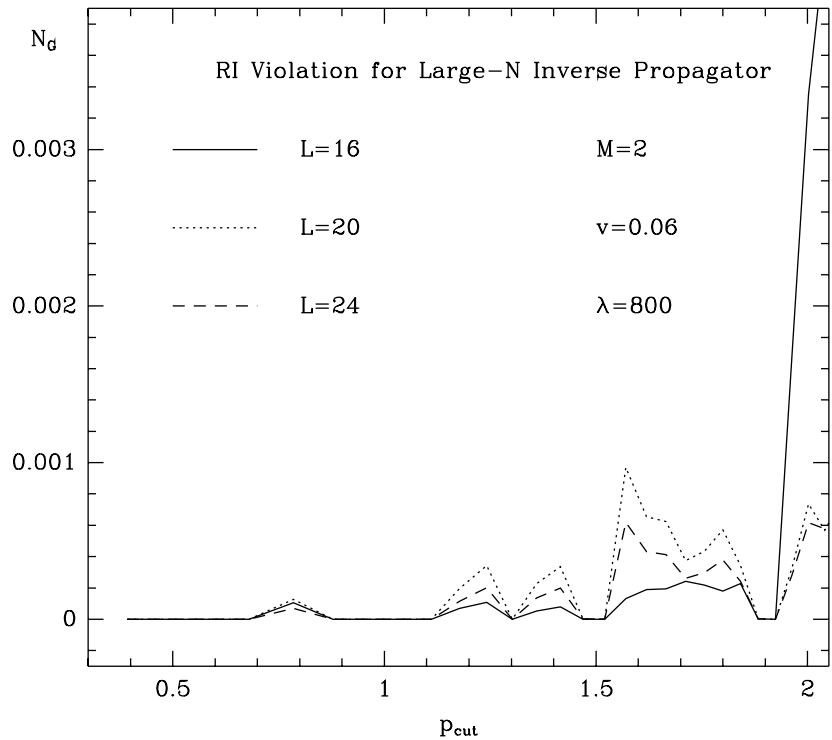}
             }
\vspace{10mm}
\caption{ The rotational invariance violation for the large $N$
  propagator is plotted different lattice sizes.
  The bare parameters are chosen to be close to the ones
  in our simulation. For small lattices, because the momentum is
discrete, the function is not smooth. But for the larger lattices
the function approaches the infinite volume result. }
\label{fig:ch5.rin}
\end{figure}
In Figure~(\ref{fig:ch5.rin}), this rotational violation
is shown for one of our simulation points for 
the large $N$ propagator. All the rotational invariance violation
are well under one percent level. 
We are therefore confident that our results should 
represent the features of the higher derivative
theory in the continuum.

\vfill\eject

%% file: c6.tex
\newcommand {\bn} {{\bf n}}
\chapter{ Extracting Scattering Phase Shift Using Finite Size Techniques}  
\label{ch:LUSC}

\section{ Resonance in Finite Volume }

In the previous chapter we argued that extracting the mass
parameter of an unstable particle in a finite volume is not 
a trivial task. For the volume that people usually perform
their Monte Carlo simulations, the lowest two Goldstone particle state has
an energy eigenvalue which is higher than the Higgs mass parameter.
This means that in such volumes the Higgs cannot decay into the
Goldstone pair as it should in the infinite volume, even
if the interaction between the Higgs and two Goldstone state
is turned on. This problem
can be solved in two ways if the theory is
only weakly interacting. In the first conventional
way, one tries to extract the propagator mass in the
finite volume, then the finite volume corrections are added
to get the propagator mass in the continuum infinite volume.
After that the perturbation theory is used again to relate
the propagator mass to the on-shell physical mass of the
Higgs particle. The width of the Higgs can also
be calculated using perturbation theory. This method
heavily utilizes perturbation theory. The second method is 
to extract the infinite volume continuum results directly  by measuring
some quantity in the finite volume. With this method, one needs 
a general formula which will relate the infinite volume
quantities to the finite volume quantities without using
perturbation theory. In the conventional $O(4)$ simulations
both methods have been tried  and they give compatible 
results. It is obvious that for our higher derivative 
$O(N)$ model, due to its strong interaction, only the
second one can be used to analyze the finite size effects.
In fact, the basic idea of the second approach is to 
make use of the finite size effects 
instead of fighting them. Let us now review some of the
basic ideas of this approach.

We start with the conventional $O(N)$ model without the
higher derivatives. The basic particle excitations in the
broken phase in a finite box consist of Higgs excitations and Goldstone
excitations. Consider the eigenstate of one Higgs 
excitation and the eigenstates of two
Goldstone excitations.
\begin{figure}[htb]
\vspace{5mm}
\centerline{ \epsfysize=3.0cm
             \epsfxsize=5.0cm
             \epsfbox{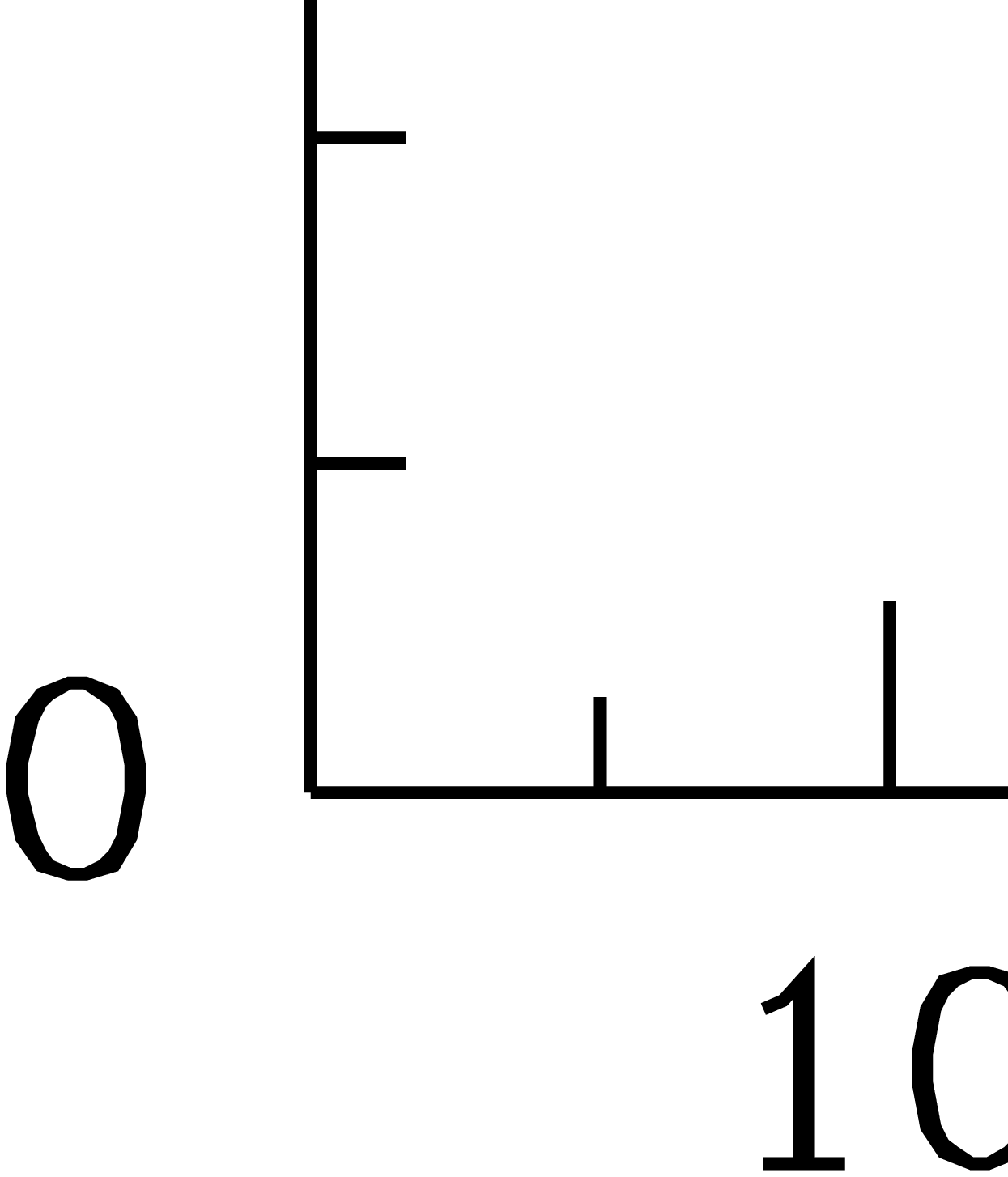} }
\vspace{10mm}
\caption{ The zeroth order of the level crossing is shown
schematically as a function of box size $L$. The Higgs excitation is $L$ 
independent while the two Goldstone excitations is decreasing with $L$.
At some value of $L$, the Higgs energy level will cross the Goldstone
energy levels and if the interaction between the two is turned on, the
two levels will repel each other  and split. }
\label{fig:ch6.cross}
\end{figure}
Due to the Euclidean invariance, we can 
take the rest frame of the Higgs particle. 
There is a major difference between the energy of the 
Higgs excitation and the Goldstone excitations.
The Goldstone pair with opposite three
momenta has an energy eigenvalue which is
dependent of the box size. The lowest one
is $4\pi/L$ where $L$ is cubic box size.
The Higgs excitation, however, will not  
depend on the box size and is just a constant.
To the lowest order, these  particles behave just like
free particles. In Figure~(\ref{fig:ch6.cross}), the dependence of these
eigenvalues are shown.
After the interaction is turned on, 
the one Higgs excitation mixes with the 
Goldstone pair excitation. For small box 
sizes, when the Higgs mass is below the
lowest Goldstone pair excitation energy
for that box size, ordinary perturbation theory
will give us the correction of these 
energy levels if the interaction is not too 
strong. For some larger box size, however, the Higgs
excitation will cross the Goldstone pair excitation
and for that particular box size, degenerate perturbation
theory should be used to calculate the level crossing. 
If the interaction is weak, one would expect a plateau
in a range of $L$ which should be identified as the 
physical Higgs mass and the splitting at the crossing
point basically gives you the width of the Higgs particle.
In order to use this picture of resonance in a finite box,
a nonperturbative relation must be established from which
one can get the relation between infinite volume quantities 
and the finite volume quantities. Finally, if the finite volume
quantities are measured in the Monte Carlo simulations, we
can use this relation to deduce the infinite volume results
nonperturbatively.

\section{ L\"uscher's Formula}

In the infinite volume, the Higgs particle is identified as
a resonance in the isospin $0$ channel in Goldstone-Goldstone
scattering. As in any two particle scattering process, the 
scattering cross section is characterized by the scattering 
phase shift $\delta(E)$ at a given center of mass energy.
When the center of mass energy is at the physical Higgs mass, 
we see a peak in the scattering cross section and the 
scattering phase shift rises dramatically from almost
zero to almost $\pi$.  In this case, there is
a resonance in the scattering process and a Higgs particle 
is produced. The mass of the Higgs particle is identified
by the position of the peak or 
equivalently by the energy at which $\delta(E)$ crosses
$\pi/2$. The width of the particle is given
by the range in which $\delta(E)$ steps from almost $0$
to almost $\pi$.
For an ideal resonance, that is   
the resonance which is infinitely narrow,
the scattering phase shift will step up
exactly $\pi$. But for wide resonances  the sharpness
and the height of the step is greatly reduced.
Therefore, the scattering phase shift in the infinite volume
fully describes the properties of the Higgs particle.

People have derived a relation which relates the infinite volume
phase shift to the two Goldstone particle energy eigenvalues 
in the finite volume. This relation, with the name L\"uscher's
formula, was derived first by DeWitt 
in a different form \cite{dewitt6}. 
Later L\"uscher rederived it and expreseed in a form suitable
for nonperturbative Monte Carlo simulations 
\cite{luscf6,luscwolf6}. It was used by
Zimmermann et. al. to study the conventional $O(4)$ model
and proved to work very well \cite{zimmph16,zimmph26}. 
We now derive this formula with
a method that is based on DeWitt, since this can be easily generalized
to the higher derivative case.

Consider a quantum mechanical system governed by the Hamiltonian
$H=H_0+V$ in a three dimensional box whose side is $L$. We can 
define the resolvent operators $G(z)$ and $G^{0}(z)$ as
\ba
G(z)&=&(z-H)^{-1} \;\;\;\; G^{0}(z)=(z-H_0)^{-1} ,
\nonumber \\
G(z)&=&G^{0}(z) + G^{0}(z) V G(z) ,
\ea 
where $z$ is just an arbitrary complex number.
Consider the matrix elements of $G(z)$ 
between two states  $|\alpha \rangle$ and $|\beta \rangle$. 
where $|\alpha \rangle$ and $|\beta \rangle$ 
are  eigenstates of the free
Hamiltonian $H_0$, i.e. 
$H_0 |\alpha \rangle = E_{\alpha} |\alpha \rangle$,
$H_0 |\beta \rangle = E_{\beta} |\beta \rangle$.
We have
\be
\langle \alpha|G(z)|\beta \rangle
={1 \over z-E_{\alpha}} (\delta_{\alpha\beta}
+\langle \alpha | V G(z) | \beta \rangle ) .
\ee
Let us now define the self energy operator such that
\be
\langle \alpha|\Sigma(z)|\beta \rangle
={ \langle \alpha|V G(z)|\beta \rangle
  \over
  \langle \beta| G(z)|\beta \rangle } .
\ee
Then the matrix element of $G(z)$ may be written as
\be
\label{eq:ch6.gsigma}
\langle \alpha|G(z)|\beta \rangle
={1 \over z-E_{\alpha}} (\delta_{\alpha\beta}
+\langle \alpha|\Sigma(z)|\beta \rangle
\langle \beta|G(z)|\beta \rangle ) .
\ee
The self energy operator defined above satisfies the following
integral equations
\be
\langle \alpha|\Sigma(z)|\beta \rangle
=\langle \alpha|V|\beta \rangle
+\sum_{\gamma \neq \beta}
\langle \alpha|V|\gamma \rangle
{ 1 \over z-E_{\gamma}}
\langle \gamma|\Sigma(z)|\beta \rangle .
\ee
Setting $\alpha=\beta$ in the above equation we get
\be 
\langle \alpha|G(z)|\beta \rangle
=( z - E_{\alpha} - \langle \alpha|\Sigma(z)|\alpha \rangle)^{-1} .
\ee
Note that the pole of $G(z)$ in the complex $z$ plain should be
the exact eigenvalue of state $|\alpha \rangle$, and we get
\be
\label{eq:ch6.eshift}
\epsilon_{\alpha}-E_{\alpha} = 
\langle \alpha|\Sigma(\epsilon_{\alpha})|\alpha \rangle ,
\ee
where $\epsilon_{\alpha}$ is the eigenvalue of the
full Hamiltonian for the state that is perturbed from $|\alpha\rangle$.
This formula tells us that the expectation value of the
self energy operator in some state gives us 
the so-called  energy shift which is the energy
difference between the exact eigenvalues and the free 
eigenvalues. 

Let us now look at this integral equation in a very large
box, where the intermediate states are very dense and we
would expect to be able to go to the continuum limit. In fact, 
we can write down an expression
\be
+\sum_{\gamma}
{ |\gamma \rangle \langle \gamma |
\over
z-E_{\gamma} }= {\cal P} {1 \over z-H_0}
+\delta(z-H_0) \Phi(z) ,
\ee
where the function $\Phi(z)$ is given by the energy shell sum
\be
\Phi(z)= \sum { dE(z) \over z-E_{\gamma} } .
\ee
This relation is obtained in the following way.  Imagine that
$z$ is some real number and we take some small real positive
number $\epsilon$ and divide the real axis into a small interval
$(z-\epsilon, z+\epsilon)$ and the rest. For a very large box, the 
eigenvalues $E_{\gamma}$ will be very dense and they are treated 
separately, depending on whether they fall in the interval or not.
For those states whose eigenvalues fall outside the small interval,
the sum will better approximate the principal valued expression 
 if we take smaller $\epsilon$ values. For any fixed $\epsilon$
there will be infinite eigenvalues which fall into the small interval, as
long as the box size is going to infinity. For these states, if the 
operator is inserted in some smooth function of the energy, 
they are equivalent to the delta function which selects out 
the specific energy. The function $\Phi(z)$ is basically the 
degeneracy sum of all the states that has almost the same energy
in the infinite volume.

With this relation we can rewrite Equation~(\ref{eq:ch6.gsigma}) 
in the following way
\be 
\langle \alpha|G(z)|\beta \rangle 
=\langle \alpha| V | \beta \rangle +
\sum_{ \gamma } \langle \alpha | V {|\gamma\rangle \langle \gamma |
                                \over z-E_{\gamma }   }
\Sigma (z)|\beta \rangle -\langle \alpha|V|\beta \rangle 
{ \langle \beta|\Sigma(z)|\beta \rangle 
\over z-E_{\beta} } .
\ee
If we now take $z=\epsilon_{\beta}$ and make use of 
Equation~(\ref{eq:ch6.eshift}) we get
\be
\langle \alpha|\Sigma(z)|\beta \rangle 
=\langle \alpha| (1-V{\cal P}{1 \over z-H_0})^{-1}
V \delta(z-H_0) \Sigma(z) |\beta \rangle \Phi(z) .
\ee
We can then multiply both sides by a factor of
$2\pi \delta(E_{\alpha}-E_{\beta})$. 
Note that the scattering phase shift operator
is given by
\be
\langle \alpha|-2 \tan\delta|\beta \rangle 
=2\pi\delta(E_{\alpha}-E_{\beta})
\langle \alpha| (1-V{\cal P}{1 \over E_{\alpha}-H_0})^{-1}
V |\beta \rangle ,
\ee
also we can take our states $|\alpha\rangle$ to
be diagonal in angular momentum, we then get
\be
\Phi_{\lambda_{\alpha}}(\epsilon_{\alpha})
=-\pi \cot \delta_{\lambda_{\alpha}}(\epsilon_{\alpha}) .
\ee
This is the basic formula which relates the scattering 
phase shift in the infinite volume to the exact
energy eigenvalues in the finite volume.
To be specific with the function $\Phi$, note that
in the isospin $0$ channel of two Goldstone
particles with opposite momenta $\bk$, we have the relation
\be
1={L^3 \over (2\pi)^3}(4\pi) k^2 dk
={L^3 \over 4 \pi^2} kE_1dE ,
\ee
where $E_1$ is the energy of one Goldstone particle. We 
get $dE=4\pi^2/(L^3kE_1)$ ,so
\ba
\Phi_{0}(z)&=& \sum {2 \pi^2 \over L^3}{ 1 
\over kE_1( z/2 - E_1(k)) }
\nonumber \\
&=&\sum {4 \pi^2 \over L^3}{ 1 
\over k( k^2 - ({2\pi \over L})^2 {\bf n}^2) }
\nonumber \\
&=&- {1 \over \sqrt{\pi} q} {\cal Z}_{00}(1, q^2) ,
\ea
where we have used the dispersion relation for the
Goldstone $z/2=k$ and $q=kL/(2\pi)$. The zeta function
is defined to be
\be
{\cal Z}_{lm}(1,q^2) = \sum_{{\bf n}} 
{ {\cal Y}_{lm}({\bf n}) \over {\bf n}^2 -q^2} ,
\ee
where the function ${\cal Y}_{lm}({\bf n})$ is the usual
spherical harmonics. When this expression is substituted
into the general formula, we get
\ba
\cot \delta_{0}(E) &=& {1 \over \pi^{3/2} q} {\cal Z}_{00}(1, q^2) ,
\nonumber \\
E^2/4&=& k^2 + m^2_{\pi}, \;\;\;\;\; m_{\pi}=0  ,
\nonumber \\
q&=& {kL \over 2\pi} .
\ea
This is exactly L\"uscher's formula that has been used by
Zimmermann et. al. in their simulation except that they
were working with the nonzero mass case for the Goldstone
particle. It is clear from the above derivation that
the condition of the massive Goldstone is not necessary.
In fact, as we will see below, 
we have tested the massless case in the conventional
$O(4)$ case and got consistent results with Zimmermann et. al..

This problem  can be understood in the following way. Recall that
in L\"uscher's derivation of the formula, he assumes that
the pion (Goldstone) has a finite mass due to 
the non-vanishing external source. 
This external source tilts the potential and makes the
potential lower in the direction of the external source.
Therefore in the potential valley it is not flat but rather  
has a slope. This is why the Goldstone particles become  massive.
Then the Higgs field is defined to be the 4-volume
average of the field variable along that particular
direction. The Goldstone field is defined to be the field along the
directions that are orthogonal to the Higgs field.
However, when the external source is gets smaller,
the tilting in the potential becomes weaker. As a
result, the fluctuation around the Higgs direction becomes
stronger. In the limit of a vanishing external source, the 
special direction is not defined at all and the  potential
becomes totally $O(N)$ invariant.
It is clear that in this limit, the Higgs field and the pion
field is not well-defined.
This is also seen in the finite volume correction of the
pion mass.
Therefore, if the
box is too small, we will not be able to disentangle the
energy correction to the one pion energy and the interaction
between the two. In this case, the correction to the single
pion mass depends on the quantity $m_{\pi}L$ exponentially.
When the pion mass goes to zero, the finite size correction
to the single pion energy will be very large. This is the 
main reason that one has to take the nonzero pion mass. 

In fact, the situation in the $O(N)$ model is more subtle.
First of all, the massless pion dispersion relation is
protected by the symmetry in the very large volume limit. 
Therefore, the energy of a   
single pion would be exactly multiple of $2\pi/L$, even if
the interaction is turned on. 
Exactly at the vanishing external source, we know that the
above picture is not a good picture of the symmetry breaking
mechanism in the finite volume. Instead, we should use
the Born-Oppenheimer picture discussed in
Chapter~(\ref{ch:BOA}) . In the Born-Oppenheimer picture, the pions are
massless, and the zero momentum pion is replaced by the 
rotator excitations. The Higgs field can also be meaningfully 
defined. As we have seen in Chapter~(\ref{ch:BOA}) ,
there will be no large finite volume corrections
to the two pion energy hence the energy shift in the
finite box totally reflects the interaction between the
two pions. Note that there is no contradiction to the finite
external source case. If the external source is present and
significantly different from $0$, then the conventional picture of
the massive pion works very well and  
the Born-Oppenheimer picture
would be a very bad approximation since the potential is so tilted.
On the other hand, the Born-Oppenheimer picture is valid for
very small external source 
where the conventional picture breaks down. 
Therefore, our conclusion is that L\"usher's formula will still
work even in the massless pion case, as long as the 
field definitions are adjusted according to the 
Born-Oppenheimer picture described in detail in Chapter~(\ref{ch:BOA}).

\section{ Integral Representation of the Zeta Function }

In the L\"uscher's formula, the zeta function needs to be dealt with
carefully. For the cubic geometry, it turns out that a useful integral
representation of the function 
exists which we will now discuss \cite{luscf6}.

In general, the zeta function is defined to be
\be
{\cal Z}_{lm}(s,q^2)=\sum_{ \bn \in Z } {\cal Y}_{lm}(\bn)
   (\bn^2 -q^2)^{-s} ,
\ee
where the symbol ${\cal Y}_{lm}(\bn)$ stands for the usual spherical
harmonic functions and the summation is over all the three dimensional
integers. In order to derive the integral representation, let us also
define the heat kernel by
\ba
\label{eq:ch6.heat}
{\cal K}(t,\bx) &=&{1 \over (2\pi)^3}\sum_{\bn \in Z}
   e^{i\bn \cdot \bx -t \bn^2}
\nonumber \\
   &=&{1 \over (4\pi t)^{3/2}}\sum_{\bn \in Z}
   e^{{1 \over 4t}(\bx-2\pi\bn)^2} .
\ea
We will also need the truncated heat kernel
\be
{\cal K}^{\lambda}_{lm}(t,\bx)={1 \over (2\pi)^3}
\sum_{|\bn| > \lambda} {\cal Y}_{lm}(\bn) 
e^{i \bn \cdot \bx -t \bn^2} .
\ee
Then the zeta function has the following representation
\ba
{\cal Z}_{lm}(s,q^2)&=&\sum_{|\bn| < \lambda} {\cal Y}_{lm}(\bn)
       (\bn^2 -q^2 )^{-s}
\nonumber \\
  &+&{(2\pi)^3 \over \Gamma(s)} \int^{\infty}_{0} dt
   t^{s-1} e^{tq^2} {\cal K}^{\lambda}_{lm}(t,0) ,
\ea
as long as $s$ satisfies the condition $Re(s) >l/2+3/2$. Note that
the combination $\exp(tq^2){\cal K}^{\lambda}_{lm}(t,0)$  
has the following asymptotic behavior
\ba
e^{tq^2}{\cal K}^{\lambda}_{lm}(t,0) &\sim& e^{-t(\lambda^2-q^2)},  
\;\;\;\; t \rightarrow +\infty ,
\nonumber \\
e^{tq^2}{\cal K}^{\lambda}_{lm}(t,0) &\sim& 
{\delta_{l0}\delta_{m0} \over (4\pi)^2} t^{-3/2}+{\cal O}(t^{-1/2}),
\;\;\;\; t \rightarrow 0 .
\ea
Therefore we immediately have the following analytic continuation for
the zeta function in the range $Re(s)>1/2$,
\ba 
{\cal Z}_{lm}(s,q^2)&=& \sum_{|\bn| < \lambda} {\cal Y}_{lm}(\bn)
(\bn^2 -q^2)^{-s} 
+ {(2\pi)^3 \over \Gamma(s)} \left(
{\delta_{l0}\delta_{m0} \over (4\pi)^2(s-3/2)} \right.
\nonumber \\
&& +\int^{1}_{0} dt t^{s-1}(e^{tq^2} {\cal
K}^{\lambda}_{lm}(t,0) -{\delta_{l0}\delta_{m0} \over (4\pi)^2t^{-3/2}})
\nonumber \\
&&\left. +\int^{\infty}_{1} dt t^{s-1} e^{tq^2} {\cal K}^{\lambda}_{lm}(t,0)
\right) .
\ea
In particular for $s=1$ we have
\ba
{\cal Z}_{lm}(1,q^2)&=& \sum_{|\bn| < \lambda} {\cal Y}_{lm}(\bn)
(\bn^2 -q^2 )^{-1} 
\nonumber \\
&+&(2\pi)^3 \int^{\infty}_{0} dt \left( e^{tq^2}{\cal
K}^{\lambda}_{lm}(t,0) - 
{ \delta_{l0}\delta_{m0} \over (4\pi)^2 t^{3/2}} \right) .
\ea 
The above integral representation is suitable for numerical evaluation
of the zeta function. The integrand is evaluated for any $t$ value 
and the integrals are performed numerically using the standard
integration subroutines (e.g. IMSL). When evaluating the integrand one
has to distinguish the case for $t>1$ and $t<1$. The first 
line in the representation~(\ref{eq:ch6.heat})
 is used for the case $t>1$ while the
second line is used for the case $s<1$ for better convergence. It
turns out that an accurate numerical answer can be obtained for
$q^2$ values not larger than $10$, which is the case in the practical
applications. 

\section{ Simulation Results on the Conventional $O(4)$ Model }

As mentioned in the previous section, we performed a test simulation
first on the conventional $O(4)$ model  {\em{without}} the external source
term. Therefore, our pion dispersion is the massless dispersion. The
simulation was done using a cluster update program which runs very 
efficiently on the alpha AXP workstation. In the simulation, we made
measurements after every $10$ cluster updates and for each lattice size
a 100,000 to 200,000 measurements were accumulated. The operators that we
took into account were the radial Higgs field, and the four lowest pion
pair states. We chose our 
simulation point so that the Higgs mass would 
come out around $0.6$. We were also working in the nonlinear limit of the
$O(4)$ model and the input bare parameter was the hopping parameter
$\kappa$ which we fixed to be $0.315$. Old simulation results indicate
\begin{figure}[htb]
\vspace{12mm}
\centerline{ \epsfysize=3.0cm
             \epsfxsize=5.0cm 
             \epsfbox{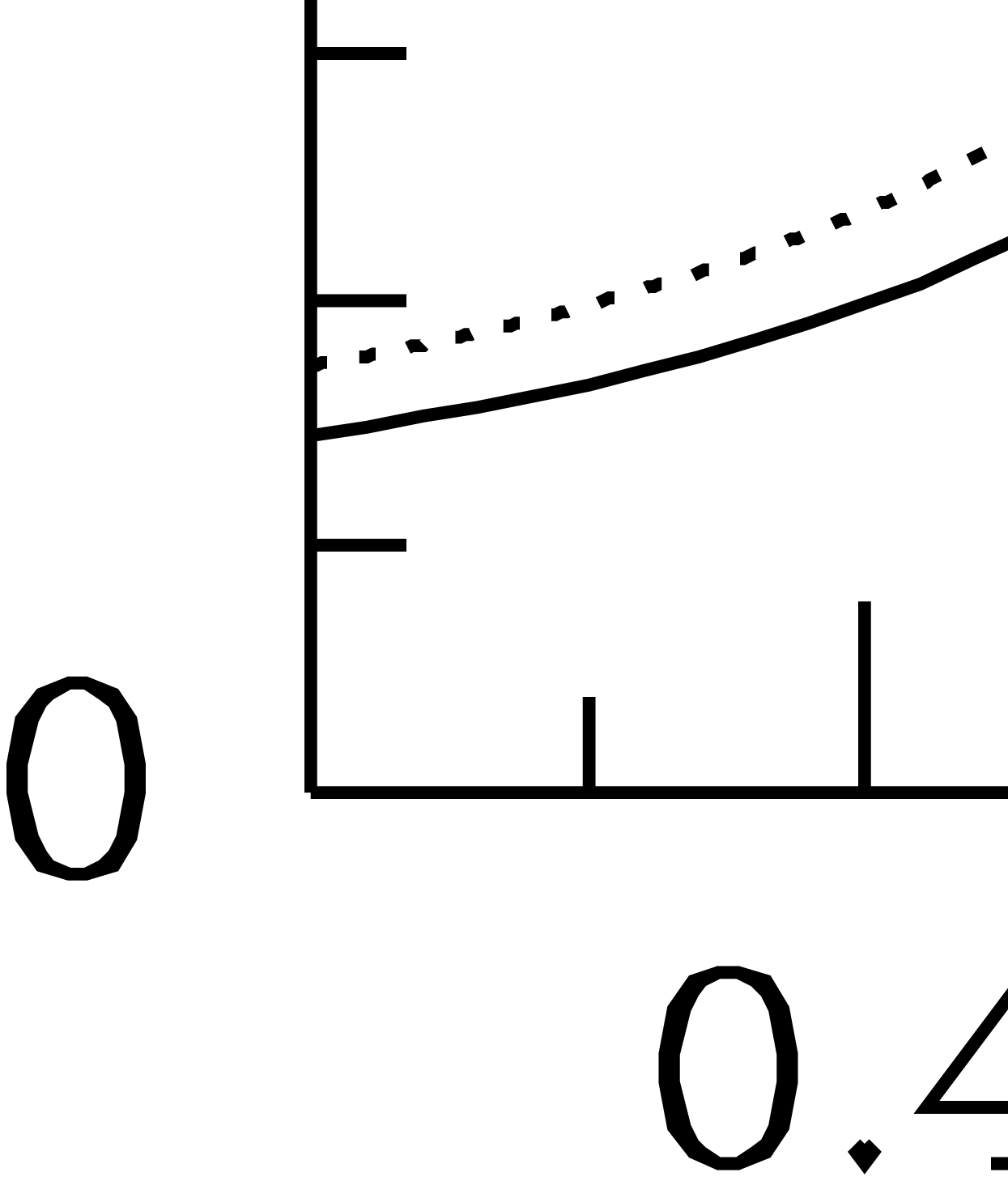} }
\vspace{20mm}
\caption{ The scattering phase shift is extracted using L\"uscher's
formula in the isospin $0$ channel with zero Goldstone mass. 
The solid line is the fit
to the relativistic Breit-Wigner shape and the fitted values of Higgs mass and
width are also shown. This is in good agreement with the perturbative
prediction (dashed line). }
\label{fig:ch6.phaseo4}
\end{figure}
that the Higgs mass for this point should be $m_H=0.581$. We scanned the
size of our 3-volume from 8 to 24 with a step of 2. The correlation
functions were then analyzed to extract the energy levels in this isospin
zero channel. The correlation matrix was diagonalized and the eigenvalues
were used to extract the energy levels. The errors of the energy levels 
were obtained by blocking the data. These errors in the energy levels
then translate into the errors in the phase shift when using 
L\"uscher's formula.
The final results can be summarized in Figure~(\ref{fig:ch6.phaseo4}),  
where we
have plotted the the scattering phase shift as a function of the center
of mass energy obtained from the application of L\"uscher's formula 
(the data points). The solid line is a perturbative fit to the data
which yields a mass and width compatible with the expected results. The
dashed line represent the 
perturbative results when the old values of $m_H$ are
substituted in. 
The highest point is for the lattice size $8^3 \times 32$ and
this point overshoots the expected values. This could be
because of the lattice effects and  the subleading finite
volume effects. The rest of the points agree nicely with the perturbative
results. The large error bars for the larger lattices is
purely due to the lack of the statistics.
This plot is a clear indication that the formula also works in
the massless case, as long as we define our Higgs field properly.

\section{ L\"uscher's Fomula for Higher Derivative Theory }

The relation in the higher derivative theory is quite similar to that
in the conventional theory. We just have to repeat most parts
of the previous derivation and make adjustments accordingly.
In the case of the higher derivative theory, the Hamiltonian is still
self adjoint, i.e.
\be
\eta H^{\dagger} \eta = H \;\;, \;\;\;\; 
\eta H^{\dagger}_0 \eta = H_0 \;\;. 
\ee
We will still define the resolvent operators as
\ba
G(z)&=&(z-H)^{-1}\;\;, \;\;\;\;G_0(z)=(z-H_0)^{-1}\;\;, 
\nonumber \\
G(z)&=&G_0(z)+G_0(z)H_1G(z) .
\ea
We will set up the basis ${|\alpha \rangle}$ such that
\be
H_0 |\alpha \rangle = E_{\alpha} |\alpha \rangle , \;\;\;\;
\sum_{\alpha} | \alpha \rangle \langle \bar{\alpha} | \eta  = 1 ,
\ee 
where the eigenvalue $E_{\alpha}$ could be complex. The state
are chosen to satisfy $\langle \bar{\beta}|\eta | \alpha \rangle 
=\delta_{\alpha \beta}$.  We can now define the self energy operator
according to
\be
\langle \bar{\alpha} |\eta \Sigma(z)| \beta \rangle =
{ \langle \bar{\alpha} |\eta H_1 G(z)| \beta \rangle \over  
  \langle \bar{\alpha} |\eta  G(z)| \beta \rangle }  ,
\ee
and it satisfies the following integral equation
\be
\langle \bar{\alpha} |\eta \Sigma(z)| \beta \rangle =
 \langle \bar{\alpha} |\eta H_1 | \beta \rangle  + 
 \sum_{\gamma \neq \beta} (z-E_{\gamma})^{-1}
\langle \bar{\alpha} |\eta  H_1| \gamma \rangle   
\langle \bar{\gamma} |\eta  \Sigma(z)| \beta \rangle  . 
\ee
Similarly the energy shift is given by the diagonal matrix element of
the self energy operator.
\be
\Delta E_{\alpha} = 
\langle \bar{\alpha} |\eta  
\Sigma(E_{\alpha}+\Delta E_{\alpha}| \alpha \rangle .  
\ee
In the large 3-volume limit,
the intermediate state summation in the integral equation of the self
energy reduce to the following
\ba
\sum_{\gamma} 
{ | \gamma \rangle \langle \bar{\gamma} | \eta \over z-E_{\gamma} }
&=&
\sum_{\gamma,E_{\gamma} \in \Re } 
{ | \gamma \rangle \langle \bar{\gamma} | \eta \over z-E_{\gamma} }+
\sum_{\gamma,E_{\gamma} \ni \Re } 
{ | \gamma \rangle \langle \bar{\gamma} | \eta \over z-E_{\gamma} } 
\nonumber \\
&=&
\sum_{\gamma,|E_{\gamma}-z|<\epsilon  } 
{ | \gamma \rangle \langle \bar{\gamma} | \eta \over z-E_{\gamma} }+
{\cal P} {1 \over z-H_0} 
\nonumber \\
&=&
\delta(z-H_0) \Phi(z) +
{\cal P} {1 \over z-H_0} .
\ea
Taking $z=E_{\beta}+\Delta E_{\beta}$ we again arrive at
\be
\langle \bar{\alpha} |\eta \Sigma(z)| \beta \rangle =
\langle \bar{\alpha} |\eta 
(1-H_1{\cal P}{1 \over z-H_0})^{-1}H_1
\delta(z-H_0)\Sigma(z)| \beta \rangle \Phi(z) .
\ee 
The delta function in the above equation restricts the intermediate
states summation to take only 
the real energy eigenstates. The matrix element
of the operator $(1-H_1 {\cal P} {1 \over z-H_0})^{-1}H_1$ between
the physical states is nothing but the phase shift as can be verified
from the general formula established in Chapter~(\ref{ch:NUNI}). 
Therefore, we would
conclude that the L\"uscher's formula will still work, even in the
case of the higher derivative theory with ghost states.

\section{ Phase Shift for Higher Derivative Theory in $1/N$ Expansion }

As we have established the validity for the L\"uscher's formula for
the higher derivative, it is very instructive to show that this 
indeed will work out in the large $N$ expansion. Recall that the
continuum scattering phase shift for the higher derivative theory has
been calculated in Chapter~(\ref{ch:NUNI}) in the large $N$ expansion. There the
\begin{figure}[htb]
\vspace{14mm}
\centerline{ \epsfysize=3.0cm   
             \epsfxsize=5.0cm 
             \epsfbox{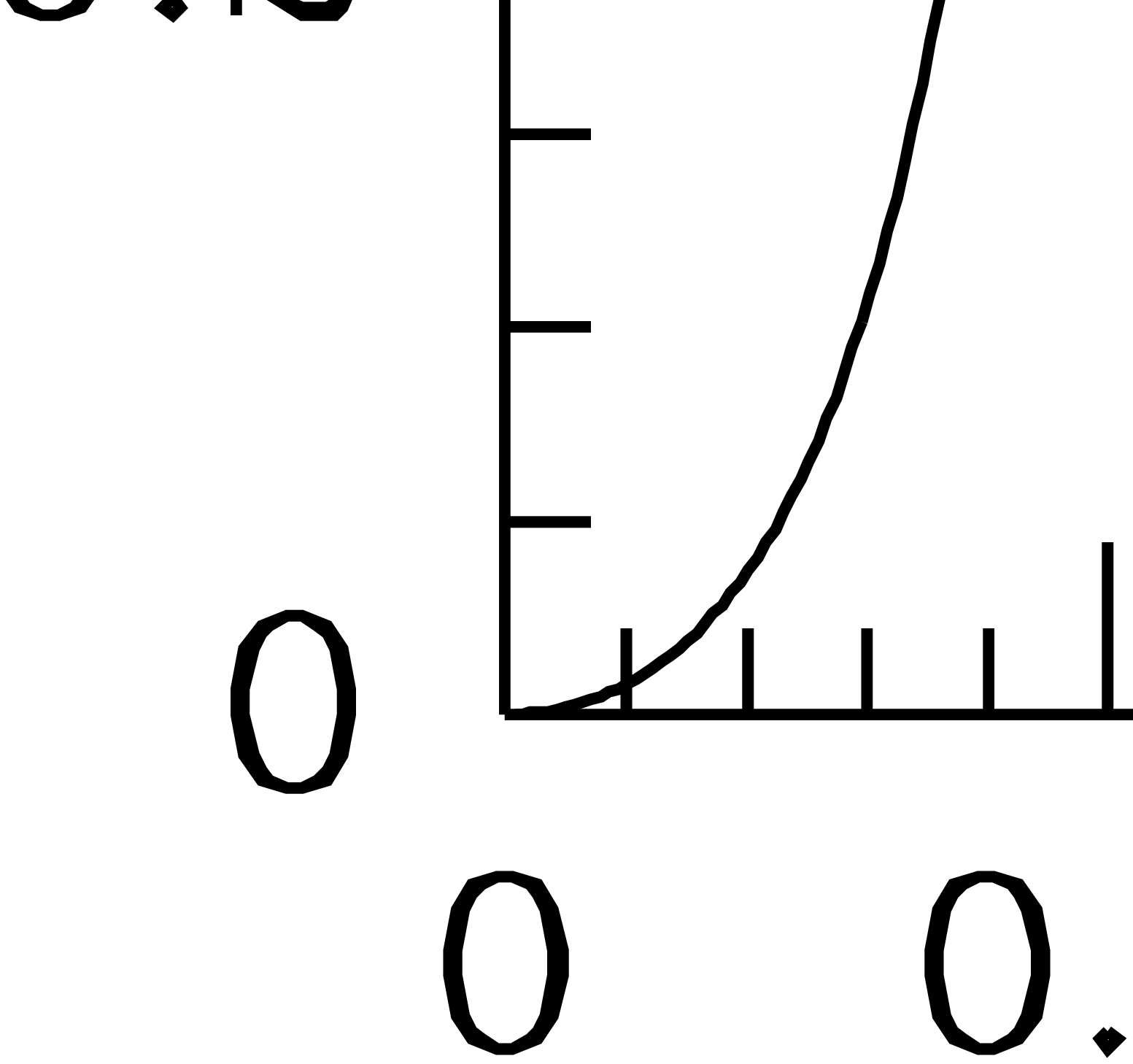} }
\vspace{18mm}
\caption{ The result of the scattering phase shift for the higher
derivative theory in the large $N$ limit is shown as a function of
the center of mass energy. The data points are obtained from 
the two Goldstone particle energy eigenvalues in the finite cubic box
in the large $N$ limit by applying L\"uscher's formula.
The solid line is the continuum large $N$ calculation for the same set
of parameters as described in Chapter (3). 
The corresponding cross section
is also shown. The agreement of the two methods is clearly seen. }
\label{fig:ch6.phasen}
\end{figure}
unitarity is maintained in the large $N$ expansion and the phase 
shift could have some microscopic acausal effects. To verify that 
L\"uscher's formula in the higher derivative theory, we have evaluated
the energy eigenvalues for the two Goldstone particle in the isospin
$0$ channel in the large $N$ limit. 
This is done by solving for the real roots of the matrix element
$\langle h|(z^2 -H^2)^{-1}|h \rangle$. In the leading order of
the $1/N$ expansion this matrix element reduces to the geometric
summation of the Goldstone bubbles. 
Then these eigenvalues are
substituted into L\"uscher's formula to extract the  phase shift, which
is then plotted against the center of mass energy of the scattering.
In Figure~(\ref{fig:ch6.phasen}),
  the result of this calculation is shown. The points are
the phase shift values obtained from L\"uscher's formula. The solid line
is the corresponding continuum calculation of the phase shift as 
described in Chapter~(\ref{ch:NUNI}). 
It is evident that the formula is working very
well. We have tried the same procedure for other set of parameters 
and they all give good agreement with the continuum calculations. 

\section{ Phase Shift Simulations for Higher Derivative Theory }

The result from the large $N$ expansion makes us confident 
that the same procedure could be carried out for the higher derivative 
field theory just as it was done for the conventional $O(N)$ model.
Owing to the improved action, we can now select our Higgs mass
value to be around $0.7$ and the ghost mass parameter at
$M=2.0$ and still keep the lattice effects small. This makes it
possible to perform such simulations on the higher derivative
$O(N)$ model.
However, there are quite a number of technical difficulties. 

One of the
main difficulties is that some efficient 
algorithms that are available for the
conventional theory break down miserably for the higher derivative 
theory. For example, the over relaxation algorithm 
is very slow for the higher derivative theory due to the neighbor
gathering. The cluster algorithm is simply not working at all (
the whole lattice tends to become one huge cluster).
In fact the only usable algorithm is the Fourier accelerated hybrid
Monte Carlo algorithm, which only 
works for finite bare coupling. Also, it
is not as efficient as the algorithms mentioned above for the
conventional theory. This means that to really get the stable
energy levels, we would have to run a rather long time.

Another difficulty is the understanding of the shape of the
phase shift as a function of the center of mass energy. If the
theory is strongly interacting, we can no longer hope to fit the
simulation data to the perturbative results.
A scheme to extract the physical parameters like the Higgs mass
and its width is needed. If the Higgs resonance is well separated 
from the ghost, we could try the Breit-Wigner shape near the
resonance, neglecting the effects of the ghosts. 

The simulation of this project is still in progress and we hope
to release the results in the near future.

\vfill\eject

%% file: c7.tex
\chapter{Conclusions}
\label{ch:CONC}

Our project was first motivated by the study of the Higgs
mass bound problem in a Pauli-Villars regulated theory
\cite{hhiggs7,dallas7,neub7}.
This theory
can be viewed as a limiting case of the higher derivative
$O(N)$ scalar field theory.
The study of the higher derivative theory goes
beyond the scope of the Higgs mass bound problem. 

The higher derivative $O(N)$ model that we have studied is
obtained from the conventional $O(N)$ scalar field theory by
adding higher derivative terms to the Higgs kinetic energy
\cite{ghost7,hhiggs7,dallas7}.
We have established 
the consistent quantization procedure of the higher derivative
scalar field theory, and have shown this theory  to
be finite and unitary  with possible violations of microscopic
causality \cite{scat7}.  Therefore, the ghost states in the theory
can easily evade experimental tests.
We have also studied 
the model  nonperturbatively in 
computer simulations by introducing an underlying lattice structure. 

In the continuum,
the higher derivative $O(N)$ model can be viewed as the
Pauli-Villars regulated conventional $O(N)$ model in the
small $m_H/M$ limit, where $m_H$ is the Higgs mass and
the $M$ is the Pauli-Villars mass parameter. It can also
be viewed as a finite, well-defined 
and unitary theory with ghost excitations.
The continuum large $N$ study of our model shows that
this theory can incorporate a heavy Higgs particle in the
TeV range, with the ghost pair well hidden at a few times
heavier than the Higgs particle \cite{hhiggs7}. 

On the lattice, our model can represent different universality
classes of models, depending on how the criticality is approached.
It could represent the conventional 
trivial $O(N)$ model at criticality, 
in which case, the higher derivative terms indeed become irrelevant. 
However, in another limit, it 
could also represent the higher derivative $O(N)$ theory in the
continuum, in which case, the theory is not trivial and 
the higher derivative terms cannot be viewed as irrelevant
operators in the Lagrangian. 

From our simulation results of the model, it is evident that
any attempt to perform a systematic search of higher dimensional 
operators to determine the Higgs mass bound would not make
any sense \cite{neub7}, since, 
as far as  the Higgs mass  bound is concerned,
one cannot tell whether a higher dimensional operator is
relevant for the problem or not.

In our nonperturbative simulation of the model, we find:

(1) Our model can generate a much heavier Higgs particle than
the conventional $O(N)$ model, which is in agreement with the
large $N$ result qualitatively. 
Without introducing the more complicated structures like
technicolor, 
it is possible in our model to
have a strongly interacting 
Higgs sector, which was excluded
by earlier lattice studies of the conventional model.

(2) It is difficult to establish a bound for Higgs particle
in our model, because by the time the Higgs is heavy enough, it
would be impossible for us to define the scaling violations in
our model. In fact, in our model, we believe the notion of the
Higgs mass bound loses its meaning, unless some new nonperturbative
definition is provided. 

Many interesting theoretical issues remains unsolved for our
higher derivative $O(N)$ model. For example, can this theory
incorporate a techni-rho-like resonance in the isospin $1$ channel?
This is obviously a nonperturbative problem. To answer it, we have
to extract the phase shift in the isospin $1$ channel for the higher
derivative theory.  We are still working on this issue.

\vfill\eject